\documentclass{article}

\usepackage{geometry}
\usepackage{natbib}
\usepackage{amsmath}
\usepackage{amssymb}
\usepackage{amsthm}
\usepackage{float}
\usepackage{graphicx}
\usepackage{bm}
\usepackage{setspace}
\usepackage[subrefformat=parens,labelformat=parens]{subcaption}
\usepackage[inline]{enumitem}
\usepackage{authblk}

\usepackage[compact]{titlesec}

\usepackage[capitalize,noabbrev]{cleveref}
\usepackage{color,xcolor}
\usepackage{caption}

\def\bSig\bm{\Sigma}

\newtheorem{prop}{Proposition}
\newtheorem{assumption}{Assumption}

\title{Effective treatment allocation strategies under partial interference}

\author{Samantha G. Dean$^{1}$,
Georgia Papadogeorgou$^{2}$, and
Laura Forastiere$^{1}$ \\
$^{1}$Department of Biostatistics, Yale University, New Haven, CT, U.S.A. \\
$^{2}$Department of Statistics, University of Florida, Gainesville, Florida, U.S.A.}

\date{\today}

\begin{document}

\maketitle

\label{firstpage}

%  put the summary for your paper here

\begin{abstract}
Interference occurs when the potential outcomes of a unit depend on the treatment of others. Interference can be highly heterogeneous, where treating certain individuals might have a larger effect on the population's overall outcome. A better understanding of how covariates explain this heterogeneity may lead to more effective interventions. In the presence of clusters of units, we assume that interference occurs within clusters but not across them. 
We define novel causal estimands under hypothetical, stochastic treatment allocation strategies that fix the marginal treatment probability in a cluster and vary how the treatment probability depends on covariates, such as a unit's network position and characteristics. We illustrate how these causal estimands can shed light on the heterogeneity of interference and on the network and covariate profile of influential individuals. %For example, we could evaluate a strategy assigning treatment to units more central in a network, a common strategy in settings with contagion or diffusion processes across a network. 
For experimental settings, we develop standardized weighting estimators for our novel estimands and derive their asymptotic distribution. We design an inferential procedure for testing the null hypothesis of interference homogeneity with respect to covariates. We validate the performance of the estimator and inferential procedure through simulations.
%and we develop a statistical test for detecting the key characteristics driving the heterogeneity of the interference mechanism and that could be used to design an effective targeting allocation strategy.
We then apply the novel estimators to a clustered experiment in China to identify the important characteristics that drive heterogeneity in the effect of providing information sessions on insurance uptake. 
\end{abstract}

%  Please place your key words in alphabetical order, separated
%  by semicolons, with the first letter of the first word capitalized,
%  and a period at the end of the list.
%

Keywords: Networks; Inverse Probability Weighting; Optimal treatment allocation; Partial Interference; Policy Evaluation; Spillover.

%  As usual, the \maketitle command creates the title and author/affiliations
%  display 

%\maketitle

%  If you are using the referee option, a new page, numbered page 1, will
%  start after the summary and keywords.  The page numbers thus count the
%  number of pages of your manuscript in the preferred submission style.
%  Remember, ``Normally, regular papers exceeding 25 pages and Reader Reaction 
%  papers exceeding 12 pages in (the preferred style) will be returned to 
%  the authors without review. The page limit includes acknowledgements, 
%  references, and appendices, but not tables and figures. The page count does 
%  not include the title page and abstract. A maximum of six (6) tables or 
%  figures combined is often required.''

%  You may now place the substance of your manuscript here.  Please use
%  the \section, \subsection, etc commands as described in the user guide.
%  Please use \label and \ref commands to cross-reference sections, equations,
%  tables, figures, etc.
%
%  Please DO NOT attempt to reformat the style of equation numbering!
%  For that matter, please do not attempt to redefine anything!

\doublespacing

\section{Introduction}
\label{s:intro}

% \subsection{Background and Motivation}
Most causal inference methods rely on the stable unit treatment value assumption (SUTVA), which requires that there is no  interference between units 
%and that a unit's potential outcomes depend only on their own treatment 
\citep{rubin1980sutva}. However, in health, social, and educational settings, among others, interactions among individuals might lead to interference, and one unit's potential outcomes might depend on their own treatment as well as the treatment of others. 

%Defining and estimating causal effects in the context of interference requires special attention. 
In the presence of interference, a common approach is to replace SUTVA with a partial interference assumption \citep{sobel2006housing}, which partitions units into clusters (e.g., villages, households) and allows treatment effects to ``spillover'' within a cluster, but not across them. 
%A cluster can represent a socially defined group such as a village, classroom, or family. 
%
%\cite{sobel2006housing} introduced causal inference with partial interference, where potential outcomes are defined depending on a neighborhood's treatment allocation scheme rather than solely on the individual's treatment. 
%
Under partial interference, \cite{hudgens2008causal} defined causal estimands and derived the properties of difference-in-means estimators under %considered 
%the setting of causal inference under partial interference with 
a two-stage randomization scheme.
%, where the treatment vector in a cluster is assigned according to a completely randomized experiment.
%
\cite{tchetgen2012causal} and \cite{perez2014interference} extended this work to observational settings, and proposed inverse probability weighting estimators
for causal estimands 
%that represent average outcomes 
under an independent Bernoulli allocation. 
%for the units in the cluster. 
% The aforementioned work considers estimands under counterfactual treatment allocations that do not depend on individual characteristics. 
%
\cite{papado2019interference} and \cite{Barkley2020causal} introduced estimands under realistic treatment allocations according to which a unit's treatment status depends on covariates while varying the average treatment probability in a cluster.
%counterfactual treatment allocations depend on covariates in the same way as the treatment propensity in the observed data, while the proportion of treated units varies within each group\citep{papado2019interference}.
For example, these estimands may represent the mean difference in average potential outcomes when the average probability of treatment is 10\% versus 20\%, but treatment depends on the covariates in the same manner as in the observed data, e.g., women are twice as likely to be treated compared to men. 
There exists much work in causal inference with interference that does not rely on a partial interference assumption \citep[e.g.,][]{aronow2017estimating, athey2018exact, forastiere2021identification}, though we refrain from delving into it here.

% Papadogeorgou et al. account for the fact that spillover may be affected by covariates, but only allows for comparisons between scenarios with the same distribution of covariates. Additional methods are needed to evaluate interventions in the setting of heterogeneous interference, where counterfactual potential outcomes depend not only on how many individuals are treated but also who specifically is treated. 

Methods for interference have commonly relied on the so-called stratified interference assumption \citep[e.g.,][]{hudgens2008causal, forastiere2021identification}, which states that interference only depends on the proportion of treated units, regardless of who is treated.  
However, within a cluster, which units receive the treatment can impact the overall outcome \citep{lee2023influencers, zhang2024policylearning}. This can be due to four mechanisms: a) heterogeneity in the individual response to treatment (direct effect) with respect to individual characteristics, b) heterogeneity in the influence on the outcome of others (spillover effect), c) the direct effect of receiving the treatment may vary depending on who else is treated in the cluster, and d) 
%network interference, that is, spillover effects occur 
interference occurring through network connections, resulting in a heterogeneous effect of an individual's treatment on the cluster's outcome depending on their network position.
% The first three heterogeneities in the direct and spillover effects can depend on
% %both of which can vary based on 
% observed and unobserved covariates, while the heterogeneity due to network interference is defined with respect to the individual's network properties. 
Therefore, 
%under heterogeneous interference, 
treatment strategies targeting units that benefit from the treatment and have a strong spillover on others, either because of their influence or because of their network position, can lead to improvements in outcomes over treatment strategies that are agnostic to a unit's ability to influence others \citep{banerjee2020influence, kim2015influence}. 
\cite{qu2022interference} allow for heterogeneous spillover effects (b) by replacing the stratified interference assumption with a conditional exchangeability assumption, allowing a unit's potential outcome to depend on the proportion of treated cluster members within each category defined by covariates.
%propose an approach for measuring direct and spillover effects in settings of heterogeneous partial interference. In clustered data, they assign individuals into different categories based on covariates (e.g. parents versus children) and allow for heterogeneity in interference between different categories of individuals. 
%
% This partition approach is in contrast to defining individual treatment to be stochastic, where treatment propensities are conditional on covariates which can be individual-level or group-level, such as in Papadogeorgou et al. and the work we present here \citep{papado2019interference}. 
Additionally, \cite{zhang2024policylearning} present a policy learning approach for finding optimal assignments under partial interference.
%
%\cite{chattopadhyay2023design} propose causal estimands for settings where treatment is stochastically assigned to one group of units while outcomes are measured in a separate group of units. 
%
%Principled approaches for characterizing interference heterogeneity would lead to improved design of treatment allocation strategies to improve a population outcome. 

% The literature for homogeneous partial interference is well developed, but there is room to better understand heterogeneous interference and in particular to estimate and compare counterfactual outcomes under complex interventions in the presence of heterogeneous interference. 

% Our Aims
Principled approaches for characterizing interference heterogeneity are required to improve the design of treatment allocation strategies.
In this paper, we present novel causal estimands for partial interference settings %that target interference heterogeneity. 
%
%Our estimands represent average potential outcomes 
under stochastic interventions that fix the average treatment probability in a cluster and vary which individuals are targeted for receiving the treatment based on their covariate profile (\cref{s:estimands}).
%
% is larger for units more central in a cluster conditional on covariates and in the second approach the treatment probability is the same for all units. Similarly, we can compare strategies where we target individuals with different socio-demographic characteristics: for example a strategy where women have higher treatment propensity versus a strategy where men have higher treatment propensity. These socio-demographic characteristics may or may not be correlated with network characteristics like centrality. 
%
In a randomized setting, we introduce identifying assumptions, propose standardized weighting estimators, and derive their asymptotic distributions (\cref{s:estimation}). Based on the estimators' asymptotic distributions, we present a statistical test to determine whether the causal estimands vary with the covariates that drive treatment assignment, 
%which allows us to detect characteristics that drive the heterogeneity of interference. 
allowing to detect the characteristics that should be used to design heterogeneous allocation strategies.
Even though we develop our estimators and theory for data from randomized experiments, all the results extend straightforwardly to the case of observational data.
We evaluate the performance of the estimator and the inferential procedure in simulations (\cref{s:simstudy}).
% We close by applying the novel estimators to a dataset from the disaster insurance intervention just described \citep{cai2015insurance}.

Finally, we apply our approach to data from
%evaluate interference heterogeneity for the effect of information sessions on insurance uptake using data from 
%the following motivating example, which we use as an application for the proposed methods later in this paper: A 
a randomized experiment designed to evaluate the effect of information sessions on a weather insurance uptake in rural China \citep{cai2015insurance} (\cref{s:cai}).
%In this experiment, the population is organized into villages, and some individuals in each village are randomized to attend an information session on disaster insurance. The outcome of interest is whether individuals in the cluster, whether they attended the information session or not, purchase disaster insurance. 
Social interactions of individuals in the same village might lead to interference, and heterogeneous interference is possible if certain individuals are more receptive to the information sessions or have more influence over others.
% because individuals who do not attend the information session may still purchase or decline to purchase insurance at the recommendation or discouragement of those who do attend the information session, and some who attended the information session may be more convincing than others. 
% The partial interference assumption is plausible if we assume people mostly interact with others who live in the same village as them. After this randomized trial has occurred, we want to estimate what would have happened if we had treated a different set of individuals. Specifically, we want to know 
We study what would have happened 
%in disaster insurance uptake when targeting 
if we targeted individuals with different characteristics, without changing the average probability of treatment in each village. %Therefore, our work provides information for understanding the characteristics of influential individuals in this context.

\section{Causal Estimands for Interference Under Heterogeneous Stochastic Interventions}
\label{s:estimands}

\subsection{The setup}

Let $N$ be the total number of units organized in $I$ clusters, where $n_i$ is the number of units in cluster $i$, for $i \in \{ 1, 2, \cdots , I \}$,  and $\sum_i n_i = N$.
We use $j \in \{ 1, 2, \cdots, n_i\}$ to denote a unit in the cluster, such that $ij$ denotes unit $j$ in cluster $i$. 
%Let us define the treatment vector as $\bm{A} \in \{0,1\}^N$, the outcome vector as $\bm{Y} \in \mathbb{R}^N$, and the covariate matrix as $\bm{X} \in \mathbb{R}^{N \times k}$ where k is the number of covariates. 
Let $A_{ij}\in \{0,1\}$ be the treatment status of unit $j$ in cluster $i$.
We use $\bm{A}_i \in \mathcal{A}(n_i) = \{0,1\}^{n_i}$ to denote the vector of treatment for all individuals in cluster $i$, and $\bm{A}_{i, -j} \in \mathcal{A}(n_i - 1)$ to denote the treatment of all individuals in cluster $i$, excluding unit $j$. For each unit, we 
measure $K$ covariates $\bm{X}_{ij}$ and an outcome $Y_{ij}$. We use $\text{X}_i = (\bm X_{i1} \ \bm X_{i2} \ \cdots \ \bm X_{i_{n_i}})^\top$ to denote the $n_i \times K$ matrix of covariates, and $\bm Y_i = (Y_{i1}, Y_{i2}, \dots, Y_{in_i})$ to denote the vector of observed outcomes in cluster $i$.
%in a randomized trial with a known treatment randomization schema, we observe the vector $(A_{ij}, Y_{ij}, \bm{X}_{ij})$. % Notation is adapted from that used by Tchetgen Tchetgen and VanderWeele and Papadogeorgou et al. \citep{tchetgen2012causal, papado2019interference}.

In full generality, a unit's potential outcomes can depend on everyone's treatment. Let $Y_{ij}(\bm a_1, \bm a_2, \dots, \bm a_I)$ denote the potential outcome for unit $j$ in cluster $i$ when the treatment of units across all clusters is equal to $\bm a_1, \bm a_2, \dots, \bm a_I$ for $\bm a_i \in \mathcal{A}(n_i)$. 
%We assume partial interference.
However, we limit the number of potential outcomes by imposing the partial interference assumption \citep{sobel2006housing}. 

%Let $Y_{ij}(\bm{a})$ be the potential outcome of unit $j$ in cluster $i$ under treatment vector $\bm{a} = (\bm{a}_1, \cdots, \bm{a}_{I})$, where $\bm{a_i} = (a_{i1}, \cdots a_{i n_i})$ and $a_{ij}$ is the treatment of unit $j$ in cluster $i$. Here, 
%
\begin{assumption}[Partial interference]
For vectors of treatment assignments $(\bm a_1, \bm a_2, \cdots, \bm a_I)$ and $(\bm a_1', \bm a_2', \cdots, \bm a_I')$ such that $\bm a_i = \bm a_i'$, we have that $Y_{ij}(\bm a_1, \bm a_2, \dots, \bm a_I) = Y_{ij}(\bm a_1', \bm a_2', \dots, \bm a_I')$.
\end{assumption}
This assumption states that the potential outcomes of unit $j$ in cluster $i$ can only depend on the treatment of individuals in cluster $i$, and it allows us to denote potential outcomes for each individual as $Y_{ij}(\bm{a}_i) = Y_{ij}(a_{ij}, \bm{a}_{i, -j})$.

\subsection{Covariate-Dependent Hypothetical Treatment Allocations} \label{counterfactual}

To evaluate the effect of targeting individuals with specific characteristics, we consider stochastic, covariate-dependent \textit{hypothetical} treatment allocation strategies. These allocation strategies are hypothetical in that they can differ from the observed treatment assignment mechanism. Similarly to \cite{tchetgen2012causal} and \cite{perez2014interference}, we consider hypothetical allocations under Bernoulli assignment mechanisms. %where the probability of treatment for each unit depends on the overall target prevalence of treatment. However,
At the same time, similarly to \cite{papado2019interference} and \cite{Barkley2020causal}, the probability of treatment for each unit depends on its characteristics. However, in contrast to previous work, we consider a class of hypothetical stochastic allocations that vary the extent to which covariates drive the treatment while maintaining a fixed cluster-average probability of treatment.

Specifically, a hypothetical treatment allocation denoted by $P_{\alpha, \bm \gamma, \text{X}}$ specifies a conditional probability of treatment for unit $j$ in cluster $i$ with covariates $\bm{X}_{ij}$ as
\begin{equation}
\label{eq: propensity}
    \mathrm{logit}(P_{\alpha, \bm \gamma, \text{X}}(A_{ij}=1 \mid \bm{X}_{ij})) = \xi_i^{(\alpha, \bm \gamma, \text{X})} + \bm \gamma^\top \textbf{X}_{ij}.
\end{equation}
Therefore, a unit's probability of treatment depends on its characteristics based on the vector of coefficients $\bm \gamma \in \mathbb{R}^K$. 
The cluster-specific intercept $\xi_i^{(\alpha,\bm \gamma, \text{X})}$ is chosen such that the cluster average treatment propensity is equal to $\alpha \in (0, 1)$, that is,
\(
\frac1{n_i} \sum_{j = 1}^{n_i} P_{\alpha, \gamma, \text{X}}(A_{ij}=1 \mid \textbf{X}_{ij}) = \alpha.
\)
%As a consequence, the intercept $\xi_i^{(\alpha,\bm \gamma, \text{X})}$ also depends on the choice of $\bm \gamma$, but we refrain from including this dependence in the notation for simplicity of presentation.
%
When $\gamma_k= 0$ for all $k=1, \dots, K$, the treatment assignment does not depend on covariates and $P_{\alpha, \gamma, \text{X}}(A_{ij}=1 \mid \textbf{X}_{ij}) = \alpha$. 
% In this case we have $\xi_i^{(\alpha, \bm \gamma, \text{X})}=\mathrm{logit}(\alpha)$.
When the coefficient corresponding to the covariate ${X}^{(k)}$ is not zero, $\gamma_k\neq 0$, the hypothetical strategy gives a higher or lower probability of treatment to those with a larger ${X}^{(k)}$ if $\gamma_k> 0$ or $\gamma_k< 0$, respectively. When $\gamma_k\neq 0$ only for one covariate, the hypothetical treatment allocation assigns treatment only based on that covariate. On the other hand, when $\gamma_k\neq 0$ for multiple covariates 
$k\in\mathcal{K}_P\subseteq [1,\dots, K]$, the treatment allocation assigns treatment based on the value of the $P=|\mathcal{K}_P|\leq K$ covariates, $X^{(k)}$ with $k\in\mathcal{K}_P$. 
Finally, since the hypothetical assignment assumes independent assignment across units, the probability of a treatment vector $\bm a_i$ in cluster $i$ is given by:
 \[
 P_{\alpha, \bm \gamma, \text{X}}(\bm A_i = \bm a_i \mid \bm{X}_i) = \prod_{j = 1}^{n_i} P_{\alpha, \gamma, \text{X}}(A_{ij} = 1 \mid \bm{X}_{ij})^{a_{ij}}\left(1- P_{\alpha, \gamma, \text{X}}(A_{ij} = 1 \mid \bm{X}_{ij})\right)^{1-a_{ij}}.
 \]

In Equation \ref{eq: propensity} we adopted a logistic link function with additive dependence between the treatment and covariates. Alternative specifications can be easily incorporated.

\subsection{Causal Estimands} \label{sec: targets}

Under stochastic allocation $P_{\alpha, \bm \gamma, \text X}$, we can define individual average potential outcomes as 
\[\overline{Y}_{ij}(\alpha, \bm \gamma) = \sum_{\bm s \in \mathcal{A}(n_i)} Y_{ij}(\bm{A_i} = \bm s) P_{\alpha, \bm \gamma, \text{X}}(\bm{A_i} = \bm s  \mid  \bm{X}_i).
\]
Here, $\overline Y_{ij}(\alpha, \bm \gamma)$ represents the average potential outcome for a unit $j$ in cluster $i$ when all units in the cluster are assigned to treatment according to $P_{\alpha, \bm \gamma, \text X}$. %Then $\bm s$ indexes all possible treatment vectors for the $n_i$ units in the cluster. 

We define the cluster average potential outcomes as averages of the individual average potential outcomes within the cluster: \( % \displaystyle
\overline{Y}_{i}(\alpha, \bm \gamma) = \frac{1}{n_i} \sum_{j=1}^{n_i} Y_{ij}(\alpha, \bm \gamma)
\). 
We assume that clusters are drawn from a super-population of clusters with distribution $G_0$, and we define the population average potential outcomes as
\( % \displaystyle
\overline{Y}(\alpha, \bm \gamma) = \mathbb{E}_{G_0} \{\overline{Y}_{i}(\alpha, \bm \gamma)\}.
\)

We define overall effects (OE) as contrasts of average potential outcomes for two different treatment strategies that prioritize units differently for treatment based on their characteristics.
Specifically, given two vectors of coefficients, $\bm \gamma, \bm \gamma' \in \mathbb{R}^K$, representing two different covariate-dependent allocation strategies $P_{\alpha, \bm \gamma, \text X}$ and $P_{\alpha, \bm \gamma', \text X}$, defined under the same cluster-average treatment propensity $\alpha$, the individual-level overall effect is defined as 
\begin{equation}\label{eq:overall_effect}
    {OE_{ij}(\alpha, \bm \gamma, \bm \gamma') = \overline{Y}_{ij}(\alpha, \bm \gamma) - \overline{Y}_{ij}(\alpha, \bm \gamma')}.
\end{equation}
Cluster average overall effects are defined as $OE_i(\alpha, \bm \gamma, \bm \gamma') = \overline{Y}_i(\alpha, \bm \gamma) - \overline{Y}_i(\alpha, \bm \gamma')$ and population average overall effects as $OE(\alpha, \bm \gamma, \bm \gamma') = \overline{Y}(\alpha, \bm \gamma) - \overline{Y}(\alpha, \bm \gamma')$. 
Therefore, the overall effects describe changes in average potential outcomes when the average probability of treatment in the cluster is fixed at $\alpha$, but units are prioritized differently for treatment based on their characteristics under $\bm \gamma$ and $\bm \gamma'$. 
% Therefore, these contrasts quantify interference heterogeneity based on the covariates targeted for treatment.
We consider overall effects of the form  $OE(\alpha, \bm \gamma, \bm 0)$, where the comparison baseline is the strategy where treatment is independent of covariates, $\bm \gamma' = \bm{0}$. All other overall effects can be derived from these.
%Then $OE(\alpha, \bm \gamma, \bm 0)$ gives the change in average potential outcome when treatment is assigned according to $\bm \gamma$ versus when treatment is independent of covariates.  

%When there is no interference or homogeneous interference, the average potential outcome is the same for all treatment strategies with a given $\alpha$, even as $\bm \gamma$ varies. Then, any contrast of average potential outcomes under two different treatment allocation strategies which differ only by their respective $\bm \gamma$s is 0. Accordingly, $OE(\alpha, \bm \gamma, \bm 0) = 0$ for all $\bm \gamma \in \mathbb{R}$ under no interference or homogeneous interference. When there is heterogeneous interference, where the treatment of different units has different effects on cluster average outcomes, the average potential outcomes, and consequently the OEs, depend on the choice of $\bm \gamma$. 

%I moved this paragraph here and added a few things
\cite{papado2019interference} and \cite{Barkley2020causal} defined overall effects by contrasting treatment allocation strategies as in \cref{eq: propensity} holding $\bm \gamma$ fixed, i.e., keeping the  dependence of the treatment assignment on covariates fixed, while   varying the overall prevalence of treatment $\alpha$. %Therefore, they considered allocation strategies for which the dependence of the treatment assignment on covariates was fixed, while they varied the overall prevalence of treatment. 
In contrast, the OEs in \cref{eq:overall_effect} reflect the impact of prioritizing units with certain characteristics for receiving the treatment while holding the prevalence of treatment fixed. 
Therefore, our OEs capture four  
distinct mechanisms of heterogeneity which may contribute to the impact that prioritizing different individuals for treatment might have on the population:
\begin{enumerate*}[label=(\alph*)]
\item units may respond to their own treatment differently based on their characteristics,
\item they might have different spillover effects on the outcome of other units in the cluster, 
\item the effect of receiving the treatment may vary depending on who else is treated in the cluster, and
\item spillover effects occur through a network of connections and the overall cluster outcome depends on the network position of treated units.
\end{enumerate*}
When any of these mechanisms are at play, the characteristics of the units that receive the treatment under a given choice of $\bm \gamma$ will impact the population average potential outcomes, and, consequently, the OEs. Note that the first mechanism (a) will result in non-zero OEs even when interference is not present. Therefore, these OEs encapsule a comprehensive notion of heterogeneity across units, including but not limited to heterogeneous interference.
% The OEs can be non-zero even in the absence of interference, or if interference is homogeneous, the average potential outcomes may vary if the effect of the treatment on those who receive it is heterogeneous.

%DIRECT AND INDIRECT EFFECTS
We also define average potential outcomes when a unit's treatment is fixed to a value $a$ and the rest of the cluster is assigned to treatment  under a stochastic allocation $P_{\alpha, \bm \gamma, \text X}$ as
%, the individual average potential outcome is
\begin{align*}
    \overline{Y}_{ij}(a, \alpha, \bm \gamma) &= \sum _{\bm{s} \in \mathcal{A}(n_i - 1)} Y_{ij}(A_{ij} = a, \bm{A}_{i, -j} = \bm{s})  P_{\alpha, \bm \gamma, \text{X}}(\bm{A}_{i, -j} = \bm{s} \mid \bm{X}_i).
\end{align*}  
%representing the average outcome for unit $j$ in cluster $i$ when its own treatment is $a$ and all the other units in the cluster are assigned to treatment according to $P_{\alpha, \bm \gamma, \text X}$. 
%Then $\bm s$ indexes the possible treatment vectors of the $n_i - 1$ units other than $j$.
%
% Equation \eqref{eq: indavg} shows that the individual average potential outcome is defined by averaging over all the potential outcomes for each of the possible treatment allocations under the specified counterfactual strategy defined by the parameters $\alpha$ and $\bm \gamma$ and the set of covariates in the cluster $\text{X}_i$.
%
Corresponding cluster and population average potential outcomes are defined as
%Indexed by treatment, we define the cluster average potential outcomes as averages of the individual average potential outcomes within the cluster:
\( %\displaystyle
\overline{Y}_i(a, \alpha, \bm \gamma) = \frac{1}{n_i}\sum_{j=1}^{n_i} Y_{ij}(a, \alpha, \bm \gamma)\)
%. Again, we assume that clusters are drawn from a super-population with distribution $G_0$, and we define the population average potential outcome as
and
\( % \displaystyle
\overline{Y}(a, \alpha, \bm \gamma) = \mathbb{E}_{G_0}\{\overline{Y}_i(a, \alpha, \bm \gamma)\} \),
respectively.

These average potential outcomes allow us to study the extent to which a unit's outcome is driven by its own treatment or the treatment of others in its cluster. To do so, we define direct and indirect effects, respectively.
Direct effects (DE) contrast average potential outcomes when a unit is treated or not, holding the treatment strategy of others (determined by $\alpha$ and $\bm \gamma$) fixed. The individual average direct effect is defined as:
\[
%\label{eq:direct_effect}
{DE_{ij}(\alpha, \bm \gamma) = \overline{Y}_{ij}(1, \alpha, \bm \gamma) - \overline{Y}_{ij}(0, \alpha, \bm \gamma)}.
\]
Cluster and population direct effects, $DE_{i}(\alpha, \bm \gamma)$ and $DE(\alpha, \bm \gamma)$, are defined similarly.
%It follows that average group direct effects are defined as $DE_{i}(\alpha, \bm \gamma) = \overline{Y}_{i}(1, \alpha, \bm \gamma) - \overline{Y}_{i}(0, \alpha, \bm \gamma)$, and the average population direct effect is defined as $DE(\alpha, \bm \gamma) = \overline{Y}(1, \alpha, \bm \gamma) - \overline{Y}(0, \alpha, \bm \gamma)$. 
%Direct effects are designed to evaluate the effect of a unit's treatment on its own outcome when the rest of the cluster is assigned to treatment under the assignment strategy $P_{\alpha, \bm \gamma, \text X}$. 
Direct effects are non-zero 
%for any $\alpha$ and  $\bm \gamma$ 
if the  treatment received by one individual has an effect on their own outcome. Here, we  focus on the extent to which $DE(\alpha, \bm \gamma)$ varies by $\bm \gamma$.
%In the presence of interference, the effect of the individual treatment may depend on the treatment prevalence in the cluster and, thus, $DE_{i}(\alpha, \bm \gamma)$ may vary by $\alpha$. 
This occurs when the effect of receiving the treatment depends on who else is treated in the cluster, that is, when mechanism (c) is at play and depends on covariates $X^{(k)}$ %corresponding to the change in $\bm \gamma$, i.e., 
for which $\gamma_k\neq\gamma_k'$. 
%potentially in addition to other mechanisms of heterogeneous interference.
%Since the effect of the individual treatment may depend on the treatment distribution in the rest of the cluster, $DE(\alpha, \bm \gamma)$ may vary by $\bm \gamma$ in the presence of heterogeneous interference. 
This mechanism can potentially coexist with other mechanisms of heterogeneous interference.
For instance, the effect of receiving the treatment may decrease when, in the cluster, treated individuals are those who have a higher influence on their cluster members' outcomes.
Under no interference or homogeneous interference, DEs may be non-zero, but do not vary with $\bm \gamma$.
Note also that the heterogeneity in the direct effect with respect to individual characteristics, i.e., mechanism (a), is not captured by the variation of $DE(\alpha, \bm \gamma)$ with $\bm \gamma$.
%Here, DEs are defined as average direct effects in the population. Hence 
%The DEs, as opposed to OEs, are net of the effect of assigning the treatment to different people who may themselves respond differently.

Indirect effects (IEs) contrast average potential outcomes when the unit's own treatment is fixed at some level $a \in \{0, 1\}$, while changing the cluster's treatment assignment strategy. We consider indirect effects under treatment strategies that fix $\alpha$ and vary $\bm \gamma$. For vectors $\bm \gamma, \bm \gamma' \in \mathbb{R}^K$, we define average individual indirect effects as: 
\begin{equation}
\label{eq:indirect_effect}
{IE_{ij}(a, \alpha, \bm \gamma, \bm \gamma') = \overline{Y}_{ij}(a, \alpha, \bm \gamma) - \overline{Y}_{ij}(a, \alpha, \bm \gamma')}.
\end{equation}
Cluster and population indirect effects, $IE_{i}(a, \alpha, \bm \gamma, \bm \gamma')$ and $IE(a, \alpha, \bm \gamma, \bm \gamma')$, are defined similarly. Depending on the value of 
%the individual treatment 
$a$, IEs represent average indirect effects 
%in the population 
when a unit is treated ($a=1$) or not ($a=0$). As for OE, we consider IEs where $\bm \gamma' = \bm 0$, i.e., $IE(a, \alpha, \bm \gamma, \bm 0)$.
%IEs, like OEs and in contrast to DEs, are driven by the structure of spillover effects.
Under no interference or homogeneous interference, IE is constant at 0. Under heterogeneous interference, IEs would be non-zero for stochastic allocations that treat the same number of individuals, but one treats high-spillover individuals with higher probability.
%and one treats low-spillover individuals. 
Note that IEs would be non-zero even if there is interference but only mechanism (a) is at play.
In addition, mechanism (c), i.e., when the effect of receiving the treatment depends on who else is treated in the cluster, 
%that is, when we have mechanism (c),  heterogeneous interference may also have a different impact on treated and untreated individuals, making 
would make
$IE_i(1, \alpha, \bm \gamma, \bm \gamma')$ different from $IE_i(0, \alpha, \bm \gamma, \bm \gamma')$. %Therefore, when interference is homogeneous, $IE_i(a, \alpha, \bm \gamma, \bm \gamma')$ is zero.

In principle, any $\bm \gamma \in \mathbb{R}^K$  represents a potential strategy. However, to ensure that the observed data are informative of causal estimands under these strategies, we consider values of $\bm \gamma$s that are reflected in the observed data
%. We define this space 
as follows. For each covariate $k\in \mathcal{K}_P$ of interest and each cluster $i$, let $\delta_i^{(k)}$ be the estimated coefficient from a regression of $A_{ij}$ on $\bm X_{ij}^{(k)}$ within cluster $i$. Across the $I$ clusters, we estimate coefficients $\bm \delta^{(k)} = (\delta_1^{(k)}, \delta_2^{(k)}, \ldots, \delta_I^{(k)})$ for covariate $k$. Then, we set $\Gamma_k$ to be the range of values from the 10th percentile to the 90th percentile of the vector $\bm \delta^{(k)}$, and we consider values $\bm \gamma \in \bm \Gamma(\mathcal{K}_P) = \otimes_{k \in \mathcal{K}_P} \Gamma_k \subset \mathbb{R}^K$.
 
\section{Estimation}
\label{s:estimation}

\subsection{Randomized Experiment and Identifying Assumptions}
We assume that data are acquired from a randomized trial with a treatment assignment mechanism defined by a known cluster-level propensity score $f(\bm{A}_i \mid \bm{X}_i)$, where covariates might drive the assignment. The estimands and theory presented here extend straightforwardly to the case where the propensity score is estimated, as in the case of observational data \citep{tchetgen2012causal, perez2014interference, papado2019interference, Barkley2020causal}. 
%\cite{perez2014interference} and \cite{tchetgen2012causal} explore the case of an unknown propensity score in more detail. 

%\subsection{Causal Assumptions and Identification}
For identifiability of causal estimands, we assume consistency of potential outcomes and cluster-level positivity and ignorability. 

\begin{assumption}
    \textit{Consistency of potential outcomes.} A unit's observed outcome is equal to its potential outcome under the observed treatment for its cluster:
    $Y_{ij} = Y_{ij}(\bm{A}_{i})$
\end{assumption}

\begin{assumption}
    \textit{Positivity.} There is a non-zero probability of observing any treatment vector that is possible under the hypothetical stochastic allocation, i.e. for $\bm a_i$ such that $ P_{\alpha, \bm \gamma, \text{X}}(\bm{A_i} = \bm a_i  \mid  \bm{X}_i) > 0 $ it holds that $f(\bm{A}_i = \bm a_i \mid \bm{X}_i) > 0$. 
    %where $f_{A \mid \mathbb{X}}$ is the cluster level propensity score in the observed data \citep{papado2019interference}. 
\end{assumption}

\begin{assumption}
    \textit{Conditional Ignorability.} The cluster treatment assignment is independent of potential outcomes given covariates: 
     $\bm{A}_i \perp \bm{Y}_i(\bm{a}_i) \mid \bm{X}_i$
\end{assumption}

In a randomized experiment, conditional ignorability holds by design, and, in order to satisfy positivity, researchers should consider hypothetical treatment allocation strategies that give non-zero probability to treatment vectors that are possible under the study design. Consistency must be satisfied by the treatment implementation.
Identification of OEs, DEs, and IEs under these assumptions have been demonstrated in the literature \citep{hudgens2008causal, tchetgen2012causal, perez2014interference}, and those results extend to the covariate-dependent treatment strategies we propose here.

\subsection{Standardized Weighting Estimators}

%\subsection{Standardized weighting estimators}
\label{sec: estimator}
We develop standardized weighting estimators for average potential outcomes in the setting of partial interference. This estimator reweights clusters according to the relative probability of the observed treatment vector under the hypothetical stochastic allocation and the cluster-level propensity score, while standardizing the weights across clusters to improve efficiency. 
For each unit $j$ in cluster $i$,  we define a weight $w_{ij}$ as
$$w_{ij}(\alpha, \bm \gamma) = \frac{P_{\alpha, \gamma, \text{X}}(\bm{A}_{i}  \mid \mathbf{X}_i)}{f(\bm{A}_i  \mid  \text{X}_i)}$$
where $f(\bm{A}_i  \mid  \text{X}_i)$ is the probability of observing the cluster treatment vector $\bm{A}_i$  under the realized randomization experiment, and 
$P_{\alpha, \gamma, \text{X}}(\bm{A}_{i} \mid 
%\alpha, \gamma, 
\mathbf{X}_i)$ is the probability of the cluster treatment vector $\bm{A}_i$ under the hypothetical allocation (as described in Section~\ref{counterfactual}).
The weight $w_{ij}(\alpha, \bm \gamma)$ is equal for all units in a given cluster.
Then, the standardized weighting estimator for population average potential outcomes $\overline{Y}(\alpha, \bm \gamma)$ is given by:  
\vspace{0.3cm}
\[{\widehat{\bm{Y}}(\alpha, \bm \gamma) = \frac{\sum_{i=1}^I \sum_{j=1}^{n_i} w_{ij}(\alpha, \bm \gamma) Y_{ij}}{\sum_{i=1}^I \sum_{j=1}^{n_i} w_{ij}(\alpha, \bm \gamma)}}\]
%
%\vspace{0.3cm}

\noindent Similar weights have been previously employed in the interference literature \citep{papado2019interference, Barkley2020causal}, and the weight standardization has been shown to lead to efficiency gains \citep{hajek1971, liu2016hajek}.

To estimate the average potential outcome when we fix the individual treatment to $a$, $\overline Y(a, \alpha, \bm \gamma)$, we define weights as
$$w_{ij}(a, \alpha, \bm \gamma) = \dfrac{I(A_{ij} = a) P_{\alpha, \gamma, \text{X}}(\bm{A}_{i,-j}  \mid  \text{X}_i)}{f(\bm{A}_i  \mid  \bm{X}_i)}$$ where here
$P_{\alpha, \gamma, \text{X}}(\bm{A}_{i,-j} \mid 
%\alpha, \gamma, 
\bm{X}_i)$ is the probability of the cluster treatment vector $\bm{A}_{i,-j}$ under the hypothetical allocation, excluding unit $ij$. 
%and conditional on the covariate matrix $\mathbf{X}_i$. 
%, with individual j's treatment fixed  
%For $P_{\alpha, \gamma, \text{X}}(\bm{A}_{i,-j} \mid \alpha, \bm \gamma, \text{X}_i)$, we do not fit a model, but generate individual predicted probabilities of treatment for each counterfactual scenario with a chosen $\bm \gamma$.
%
% We then define the standardized weighting estimator for cluster-level average potential outcomes $\overline{Y}_i(a, \alpha, \bm \gamma))$: 
% %
% \[{\widehat{\bm{Y}}_i(a, \alpha, \bm \gamma) = \dfrac{ \sum_{j=1}^{n_i} w_{ij}(a, \alpha, \bm \gamma) Y_{ij}}{ \sum_{j=1}^{n_i} w_{ij}(a, \alpha, \bm \gamma)}}\] 
% %
The standardized weighting estimator %for population-level potential outcomes $\overline{Y}(a, \alpha, \bm \gamma))$ 
is \[{\widehat{\bm{Y}}(a, \alpha, \bm \gamma) = \frac{\sum_{i=1}^I \sum_{j=1}^{n_i} w_{ij}(a, \alpha, \bm \gamma) Y_{ij}}{\sum_{i=1}^I \sum_{j=1}^{n_i} w_{ij}(a, \alpha, \bm \gamma)}}. \]

% Marginalized over treatment, we define weights $w_{ij}(\alpha, \bm \gamma) = \frac{P_{\alpha, \gamma, \text{X}}(\bm{A}_{i}  \mid  \alpha, \gamma, \text{X}_i)}{f(\bm{A}_i  \mid  \text{X}_i)}$. and the population-level standardized weighting estimator for potential outcomes is:  \[{\widehat{\bm{Y}}(\alpha, \bm \gamma) = \frac{\sum_{i=1}^I \sum_{j=1}^{n_i} w_{ij}(\alpha, \bm \gamma) Y_{ij}}{\sum_{i=1}^I \sum_{j=1}^{n_i} w_{ij}(\alpha, \bm \gamma)}}\]
% %
% Note that $w_{ij}(\alpha, \bm \gamma)$ is equal for all units in a given cluster. 

%These estimators use a novel representation of counterfactual treatment distributions to calculate potential outcomes for counterfactual stochastic interventions which depend on individual covariates. 

\subsection{Asymptotic Results} \label{sec:asymptotics}
We acquire the asymptotic properties of the estimators as the number of clusters grows using M-estimation \citep{%tsiatis2006mest, 
stefanski2002mestimation, perez2014interference}. Standard regularity conditions are included in Web Appendix A. 

\begin{prop}\label{prop: mest}
    Consider R distinct coefficient vectors $\{\bm \gamma_1, \bm \gamma_2, \dots, \bm \gamma_R\} = \bm \Gamma^* \subset \bm \Gamma$
    %$\bm \gamma_1, ..., \bm \gamma_R$ 
    for the stochastic allocation $P_{\alpha, \bm \gamma, \text{X}}$ and a fixed value $\alpha \in (0, 1)$.
    % Let $\mu_{a, \alpha, \bm \gamma}=\overline{Y}(a, \alpha, \bm \gamma)$ and $\mu_{\alpha, \bm \gamma}=\overline{Y}(\alpha, \bm \gamma)$ denote population average potential outcomes.
    %
    Define the collection of the true causal estimands $\bm \mu_{\alpha, \bm\Gamma^*} = \big\{ \overline{Y}(0, \alpha, \bm \gamma), \overline{Y}(1, \alpha, \bm \gamma), \overline{Y}(\alpha, \bm \gamma)\big\}_{\bm \gamma \in \bm\Gamma^*}$ and the corresponding estimators
    %
    %Define $\widehat{\bm \mu} = \{ \widehat{Y}(0, \alpha, \bm \gamma_1), \allowbreak \widehat{Y}(0, \alpha, \bm \gamma_2), \allowbreak \ldots, \widehat{Y}(0, \alpha, \bm \gamma_R), \widehat{Y}(1, \alpha, \bm \gamma_1), \allowbreak \widehat{Y}(1, \alpha, \bm \gamma_2), \ldots, \widehat{Y}(1, \alpha, \bm \gamma_R) , \widehat{Y}(\alpha, \bm \gamma_1), \widehat{Y}(\alpha, \bm \gamma_2), \ldots, \widehat{Y}(\alpha, \bm \gamma_R)\}$
    $\widehat{\bm \mu}_{\alpha, \bm\Gamma^*} = \big\{ \widehat{Y}(0, \alpha, \bm \gamma), \widehat{Y}(1, \alpha, \bm \gamma), \widehat{Y}(\alpha, \bm \gamma)\big\}_{\bm \gamma \in \bm\Gamma^*}$,  across the values of $\bm \gamma \in \bm \Gamma^*$.
    Then $\sqrt{I}(\hat{ \bm \mu}_{\alpha, \bm\Gamma^*} - \bm \mu_{\alpha, \bm\Gamma^*}) \rightarrow N(0, \bm \Sigma_{\alpha, \bm\Gamma^*})$ when $I \rightarrow \infty$, where:
    \begin{equation*}
            \bm \Sigma^{-1}_{\alpha, \bm\Gamma^*}  = \mathbb{E} \{\bm{\psi}(\bm{Y}_i, \bm{X}_i, \bm{A}_i; \bm{\mu}_{\alpha, \bm\Gamma^*}) \bm{\psi}(\bm{Y}_i, \bm{X}_i, \bm{A}_i; \bm{\mu}_{\alpha, \bm\Gamma^*})^T \}
    \end{equation*}
    and $\bm{\psi}$ is a vector of estimating functions, given in Web Appendix A. 
    
\end{prop}

We estimate the asymptotic variance $\bm \Sigma_{\alpha, \bm\Gamma^*}$
by taking empirical expectations over the estimating functions, i.e.,
\( \displaystyle 
    \widehat{\bm \Sigma}^{-1}_{\alpha, \bm\Gamma^*} =  \frac{1}{I} \sum_{i=1}^I \left\{\bm{\psi}(\bm{Y}_i, \bm{X}_i, \bm{A}_i; \bm{\mu}_{\alpha, \bm\Gamma^*}) \bm{\psi}(\bm{Y}_i, \bm{X}_i, \bm{A}_i; \bm{\mu}_{\alpha, \bm\Gamma^*})^T \right\}.
\)
This estimator is consistent under standard regularity conditions \citep{iverson1989convergence}.

All the causal estimands that we are interested in, DEs, IEs, or OEs, are contrasts of average potential outcomes included in $\bm \mu_{\alpha, \bm \Gamma^*}$. Therefore, we obtain the asymptotic distribution of the estimators of DEs, IEs, and OEs  using the delta method and design 95\% confidence intervals. 

%Asymptotic 95\% confidence intervals for each estimator can be calculated using their respective covariance matrix and applying the central limit theorem to estimate a standard error. 
%Let $\sigma^2_{OE, \gamma_r}$ be the $r^{th}$ diagonal element of the OE covariance matrix $\bm \Sigma_{OE}$ Then the 95\% CI for an estimated $OE(\alpha, \bm \gamma_r, \bm \gamma_R)$ is $OE(\alpha, \bm \gamma_r, \bm \gamma_R) \pm 1.96 \sigma_{OE, \gamma_r}$

\subsection{Statistical Test for Heterogeneity} \label{sec:test}
We propose a statistical test to determine whether there is 
%heterogeneous interference by a 
heterogeneity with respect to a certain covariate $k$ or set of covariates $\mathcal{K}_P$ that could be used for targeting units for treatment. 
In particular, to test heterogeneity with respect to a set of covariates $\mathcal{K}_P$, consider all the vectors of coefficients $\bm \gamma \in \bm\Gamma$ such that $\gamma_k\neq 0$ for at least one covariate $k \in \mathcal{K}_P$. Let $\bm\Gamma(\mathcal{K}_P)$ be such a subspace of interest. Then, the null hypothesis to be tested is as follows:
%the null hypothesis tested is as follows, where $\bm \gamma_a$ and $\bm \gamma_b$ differ only for the covariate that we are interested in testing for heterogeneous interference and $\bm \gamma_R$ is the reference intervention that all counterfactuals are compared to:
%
\begin{equation}\label{null}
\begin{aligned}
&\mathcal{H}_0: OE(\alpha, \bm \gamma, \bm 0)-OE(\alpha, \bm \gamma', \bm 0) = 0 \textrm{ for all } \bm \gamma, \bm \gamma' \in \bm \Gamma(\mathcal{K}_P),
    % &\mathcal{H}_0: OE(\alpha, \bm \gamma_a, \bm \gamma_R) = OE(\alpha, \bm \gamma_b, \bm \gamma_R) \textrm{ for all } a, b \\
    % &\mathcal{H}_A: \exists \bm \gamma_a, \bm \gamma_b \text{ such that } OE(\alpha, \bm \gamma_a, \gamma_R) \neq OE(\alpha, \bm \gamma_b, \bm \gamma_R).
\end{aligned} 
\end{equation}
against the alternative that there exist two   $\bm \gamma$ and $\bm \gamma'$ for which the OE differs. As discussed in Section \ref{s:estimands}, if the null hypothesis $\mathcal{H}_0$ for OEs is rejected, then one of the mechanisms of heterogeneity  (a)-(d) is at play for at least one of the covariates $k\in \mathcal{K}_P$, and the overall cluster outcome will depend on 
the extent to which these covariates are targeted through $\bm \gamma$.
%the treatment allocation strategy  $P_{\alpha, \gamma, X}$.

For testing $\mathcal{H}_0$, we propose the following procedure. We consider the test statistic that compares overall effects under different values of $\bm \gamma$, as
$T = \max_{\bm \gamma_1, \bm \gamma_2 \in \bm \Gamma(\mathcal{K}_P)} \{OE(\alpha, \bm \gamma_1, \bm 0) - OE(\alpha, \bm \gamma_2, \bm 0)\}$.
If $T$ is large, it would imply that treatment assignment mechanisms that prioritize the covariates in $\mathcal{K}_P$ differently lead to different outcomes. 
We acquire the distribution of $T$ under the null hypothesis using the asymptotic distribution of our causal estimators derived in \cref{prop: mest}. Let $\Sigma_{OE}$ denote the asymptotic covariance matrix of the OE estimators. We generate $B$ draws from the multivariate normal distribution $N(\bm 0, \Sigma_{OE})$. For each draw $b = 1, 2, \dots, B$, we calculate the value of the test statistic $T_b$. 
Then, we reject the null hypothesis at level $\alpha$ if the observed value of the test statistic is larger than the $(1-\alpha)$ quantile of the values $\{T_b\}_{b = 1}^B$.
Since the test statistic contrasts overall effects, setting the mean of the normal distribution to $\bm 0$ does not affect the performance of the test. 

We can also consider a setting where we want to compare a strategy that targets individuals using a set of covariates $\mathcal{K}_P$ to a strategy that targets individuals using a smaller set of  covariates $\mathcal{K}'_{P'}\subset \mathcal{K}_P$, with $P'< P$. We can test whether there is additional heterogeneity in OEs introduced by the additional covariates in $\mathcal{K}_P\backslash\mathcal{K}'_{P'}$. 
In this case, the null hypothesis in \cref{null} would be defined with $\bm \gamma \in \bm \Gamma(\mathcal{K}_P)$ and $\bm \gamma' \in \bm \Gamma(\mathcal{K}'_{P'})$
%to scenarios where $\bm \gamma$ is a vector of covariates where two terms are non-zero and $\bm \gamma'$ is a vector of covariates where only one term is nonzero. Similarly, we calculate a test statistic where considered $\bm \gamma_1$ has two nonzero terms and considered $\bm \gamma_2$ has one nonzero term. 

The same procedure can be used to test if the DE or IE is constant across intervention strategies, substituting the corresponding estimators, covariance matrix, and test statistic. Testing whether DEs or IEs vary with $\bm \gamma$ can provide us with insights on the existence of difference mechanisms of heterogeneous interference, as discussed in Section \ref{s:estimands}.
%Validation of the proposed statistical test is described in Web Appendix G.

%Rather than a single test that incorporates information from all estimated OE's, we could generate a series of tests for each contrast of overall effects, where we calculate $T = \{OE(\alpha, \bm \gamma, \bm 0) - OE(\alpha, \bm \gamma', \bm 0)\}$ for each $\bm \gamma, \bm \gamma' \in \Gamma(\mathcal{K}_P)$. This would be a less powerful approach, however, because we would need to adjust for multiple testing, and we would not be able to take advantage of correlations between test statistics.

\section{Simulation Study}
\label{s:simstudy}

We conducted a simulation study to validate the performance of the estimators. We considered scenarios with no interference, homogeneous interference, and heterogeneous interference driven by one or two covariates. Data sets consist of 200 clusters. We considered scenarios where interference is imposed through a linear outcome model %, where the outcome directly depends on the cluster treatment, 
or a social network diffusion process. Scenarios are detailed in Web Appendix B. We generate 600 data sets per scenario. 
% and average results over replicates. We chose 600 iterations because of computational limitations. %Bias is consistently minimal and coverage levels of the 95\% confidence intervals are close to the expected level.

%section 1: simulated dataset - linear model
\subsection{Data generating process}
\subsubsection{Scenario 1: Outcome linear model}

In this scenario, we consider clusters that consist of 15 units each. For each unit, we generate two binary covariates ${X}^{(1)}$ and ${X}^{(2)}$, where $X^{(1)}_{ij} \sim$ Bernoulli(0.5) and $X^{(2)}_{ij}$ is generated to be independent of ${X}^{(1)}$ or correlated with $X^{(1)}$, with $P(X^{(2)} = 1 ) = 0.5$. The correlation of the two covariates depends on the concordance parameter $\rho = \mathbb{P}(X^{(1)}_{ij} = X^{(2)}_{ij})$, that we vary over $\{0.5, 0.65\}$. 
We generate a randomized treatment $A_{ij} \sim$ Bernoulli(0.5). We use the term {\it neighbors} to refer to the other units in the cluster. For each unit, we define the proportion of treated neighbors, $T_{ij} = \frac{\sum_{h \ne j} A_{ih}}{n_i - 1}$; the proportion of treated neighbors with $X^{(1)} = 1$, $T^{(1)}_{ij} = \frac{\sum_{h \ne j} A_{ih}X^{(1)}_{ih}}{\sum_{h \ne j}A_{ih}}$; and the proportion of treated neighbors with $X^{(2)} = 1$, $T^{(2)}_{ij} = \frac{\sum_{h \ne j} A_{ih}X^{(2)}_{ih}}{\sum_{h \ne j}A_{ih}}$. The values $(T_{ij}, T^{(1)}_{ij}, T^{(2)}_{ij})$ play a role in defining the type of interference, described below.
Finally, we simulate the observed outcome as
%Simulated outcomes depend on cluster treatment characteristics using 
\[{Y_{ij} = \beta_0 + \beta_1 A_{ij} + \bm \beta_2 \bm X_{ij} + \beta_3 T_{ij} + \beta_4 T^{(1)}_{ij} + \beta_5 T^{(2)}_{ij} + \epsilon_{ij}}\] where $\epsilon_{ij} \sim N(0,1)$.

Depending on the values of the parameters, this model will impose 
no interference, homogeneous interference, or heterogeneous interference. 
The intercept $\beta_0$ is fixed at 1. 
The parameter $\beta_1$ represents the direct effect of the treatment and is fixed at 3. Here, because there are no interactions between individual treatment and covariates, the direct effect is homogeneous with respect to covariates and does not depend on who else is treated in the cluster, that is, mechanisms (a) and (c) are not present. We set $\bm \beta_2$ to zero for these simulations. The parameter $\beta_3$ controls the magnitude of homogeneous interference, whereas the parameters $\beta_4$ and $\beta_5$ control the magnitude of heterogeneous interference through $ X^{(1)}$ and $ X^{(2)}$, respectively. That is, when $\beta_4\neq 0$ or $\beta_5\neq 0$
individuals with different values of $X^{(1)}$ or $ X^{(2)}$ have a different spillover effect on their neighbors. This corresponds to mechanism (b) described earlier.
We consider scenarios where  $\beta_3 \in (0,1)$, $\beta_4 \in (0,1,2)$, and $\beta_5 \in (0,1)$. When $\beta_3 = \beta_4 = \beta_5 = 0$, there is no interference present. When $\beta_3\neq 0$, with $\beta_4=\beta_5=0$, then there is only homogeneous interference.
Under this model specification, when there is no interference or only homogeneous interference, $OE(\alpha, \bm \gamma, \bm 0)=0$ for any $\alpha$ and $\bm \gamma$.
When instead $\beta_4 > 0$ or $\beta_5 > 0$, there is heterogeneous interference with respect to $X^{(1)}$ or $X^{(2)}$, respectively. This means that strategies with $\gamma_k>0$ giving a higher probability of treatment to those with $X^{(k)}=1$, with $k=1,2$ depending on the heterogeneity, will result in a higher average potential outcome, i.e., $OE(\alpha, \bm \gamma, \bm 0)>0$ for any $\alpha$ and $\bm \gamma$ with $\gamma_k>0$. 

%section: simulated dataset - diffusion
\subsubsection{Scenario 2: diffusion process}
Scenario 2 considers the presence of network interference, where spillover effects occur through a network of connections. In particular, here we consider a setting with an outcome diffusion process, where treating central units ensures a larger diffusion of the outcome. Mechanism (d) is then at play. 
%We consider cases where a covariate $X^{(2)}_{ih}$ is more or less correlated with network centrality.
Here, a cluster consists of five units: one central unit and 4 alters connected only to the central unit, forming an undirected star network.
We let $X^{(1)}_{ij}$ indicate whether a unit is the central unit of its cluster. We generate  $\bm X^{(2)}$ to be independent of $\bm X^{(1)}$, or correlated with $\bm X ^{(1)}$ depending on the concordance parameter with values $\rho \in \{ 0.5, 0.65 \}$. 
We generate treatment under a Bernoulli assignment with constant probability: $A_{ij} \sim$ Bernoulli(0.25).
We denote by 
%Here, a unit's neighbors, denoted by 
$\mathcal{N}_{ij}$
%, are the other units in the cluster with which it is 
unit $ij$'s neighbors, that is, the set of units 
connected to $ij$ by a network edge. The neighbors of a central node $ij$ are all other nodes in the cluster, i.e., 
$\mathcal{N}_{ij}=\{ih: X^{(1)}_{ih}=0\}$, whereas for a non-central node $ij$, the only neighbor is the cluster's central node, i.e., 
$\mathcal{N}_{ij}=\{ih: X^{(1)}_{ih}=1\}$.
Given this network structure and the treatment vector $\bm A$, we generate the outcomes $\bm Y$ by a diffusion process \citep{kempe2003influence}. In particular, we let $Y_{ij}=1$ if the unit is treated, i.e., $A_{ij} = 1$. If the unit $ij$ is untreated, i.e., $A_{ij}=0$, we assume that the outcome can diffuse from the treated neighbors (whose outcome is equal to 1) under an independent cascade model. That is, each treated neighbor can diffuse the outcome independently with a constant probability $p_d$: 
\[
Y_{ij}
%\begin{cases}
%=1 \qquad \text{if} \quad A_{ij}=1\\
\sim\text{Bernoulli}\biggl(1 - (1-p_d)^{5 - \sum_{ih\in \mathcal{N}_{ij}} A_{ih}}\biggr) 
%\qquad \text{if} \quad A_{ij}=0
%\end{cases}
\]
We consider $p_d \in (0,0.2,0.5,0.8)$.
Treating a central node would give rise to a possible outcome diffusion to the four non-central nodes, whereas if we treat a non-central node only the central node can receive the outcome by diffusion. Therefore, given the graph structure and the diffusion process, we expect that treating central nodes will result in an increased population average outcome than treating non-central ones, with the increase depending on the diffusion parameter $p_d$. 

%section 2: counterfactual treatment allocaton
\subsection{Counterfactual treatment allocation}
Causal estimands are defined under a hypothetical treatment allocation based on both covariates $X^{(1)}$ and $X^{(2)}$
with probability 
\[ \mathrm{logit}(P_{\alpha, \bm \gamma, \text{X}}(A_{ij}=1 \mid  X^{(1)}_{ij},  X^{(2)}_{ij})) = \xi_i^{(\alpha, \bm \gamma, \text{X})} + \gamma_1 X^{(1)}_{ij} + \gamma_2 X^{(2)}_{ij}\]
as in \cref{eq: propensity}. By varying $\gamma_1$, $\gamma_2$ or both, we evaluate the average potential outcome that would result from the corresponding covariate-dependent treatment allocation based on $X^{(1)}$, $X^{(2)}$, or both, respectively.
Calculation of true average potential outcomes for counterfactual treatment allocations is detailed in Web Appendices G and H.

%section 4: simulation results
\subsection{Simulation results}
\label{s:simresults}
In each scenario and for each dataset, we apply our estimator presented in Section \ref{sec: estimator} for overall, indirect, and direct effects, for a range of $\bm \gamma = (\gamma_1, \gamma_2)$  such that $\gamma_1, \gamma_2 \in (-1.3, 1.3)$ (Scenario 1) or $\gamma_1, \gamma_2 \in (-0.5, 0.5)$ (Scenario 2). The range of $\bm \gamma$s was chosen to reflect the space $\Gamma$, described in Section \ref{sec:test}. We also consider confidence intervals based on the estimators' asymptotic distribution and a bootstrap procedure that resamples clusters \citep{papado2019interference}.
Here, we present results for the overall effects, while results for the direct and indirect effects are reported in  Appendices C, D, and E.
%for the linear model.

\begin{figure}
\centering
\centerline{\includegraphics[width=\textwidth]{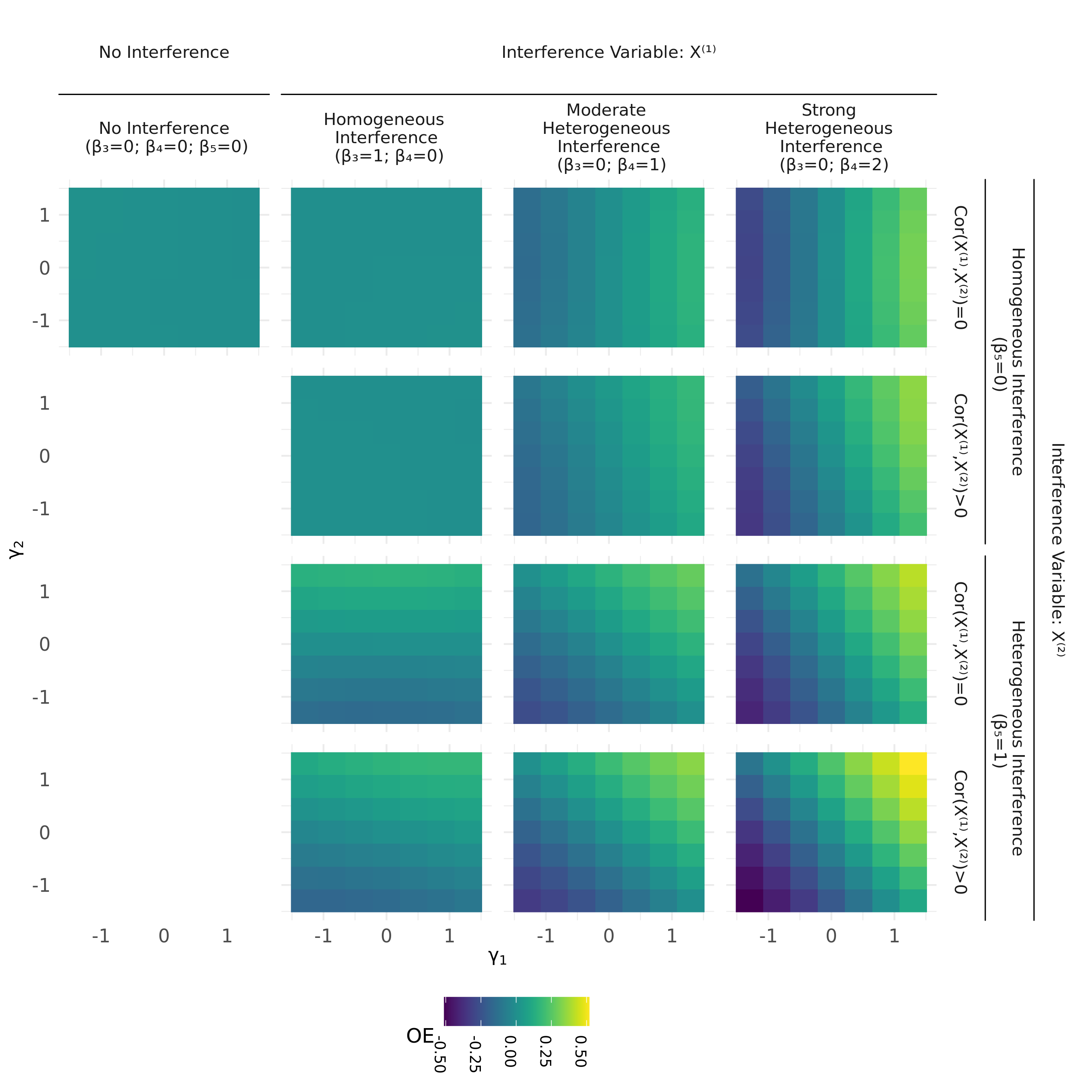}}
\caption{Scenario 1: Estimated overall effect (OE) under bivariate intervention strategies. The columns differentiate between scenarios where $X^{(1)}$ affects interference in different ways, either no interference, homogeneous interference, or heterogeneous interference (moderate or strong). The rows differentiate between scenarios where $X^{(2)}$ affects interference in different ways, either homogeneous interference or strong heterogeneous interference, with or without correlation with $X^{(1)}$. Within each heatmap, the x- and y-axes determine how units are preferentially treated based on their covariates $X^{(1)}$ and $X^{(2)}$, respectively. Colors represent the estimated causal effect. }
\label{fig:bivariate_OE_est}
\end{figure}

%
%To evaluate counterfactual interventions in the linear model setting, we generate a two dimensional grid of intervention strategies ranging from preferentially treating units where $X^{(1)}_{ij}$ and $X^{(2)}_{ij}$ are both 0 to preferentially treating people where both variables are 1. In any particular counterfactual intervention, the effect of $\bm{X}^{(1)}$ on treatment propensity is determined by $\gamma_1$, and the effect of $\bm{X}^{(2)}$ on treatment propensity is determined by $\gamma_2$. 
%
In Figure \ref{fig:bivariate_OE_est}, we report the average point estimates across data sets for the overall effects targeting $ X^{(1)}$ ($\gamma_1\neq 0$), $X^{(2)}$ ($\gamma_2\neq 0$), or both $ X^{(1)}$ and $X^{(2)}$ ($\gamma_1, \gamma_2\neq 0$) in Scenario 1. We show the cases when there is no interference ($\beta_3=0$, $\beta_4=0$, $\beta_5=0$), homogeneous interference with respect to $X^{(1)}$ ($\beta_4=0$), $X^{(2)}$ ( $\beta_5=0$), or both ($\beta_4=0$, $\beta_5=0$), or heterogeneous interference with respect to $X^{(1)}$ ($\beta_4=1$ for moderate, $\beta_4=2$ for strong), $X^{(2)}$ ($\beta_5=1$%for moderate, $\beta_5=2$ for strong
), or both ( $\beta_4=1,2$, $\beta_5=1%2
$), and $X^{(1)}, X^{(2)}$ are correlated or not.

When there is no heterogeneous interference due to ${X}^{(1)}$ and ${X}^{(2)}$, and interference is not present or only homogeneous (the intersection of the first two rows and first two columns in Figure~\ref{fig:bivariate_OE_est}), the overall effect is constant and equal to zero across all values of $\gamma_1$ and $\gamma_2$. %Any distribution of covariates in the treated group results in the same average potential outcome. 

%When there is no interference or only homogeneous interference through $X^{(1)}$ and $X^{(2)}$, then OE is constant across different intervention strategies which assign treatment using these two variables. When there is interference through $X^{(1)}$ but not $X^{(2)}$, OE depends on whether $X^{(1)}$ was targeted but is constant across how strategies use $X^{(2)}$, and vice versa. When interference is heterogeneous through both variables, OE depends on how interventions target treatment based on both variables. When a single variable drives spillover heterogeneity, assigning higher treatment probability based on that variable can increase overall effect. Intervening on a correlated variable can also drive higher overall effect. When two variables drive spillover heterogeneity, the largest overall effect is reached when designing an intervention that uses both variables to assign treatment probabilities, especially if the variables are correlated.

When there is heterogeneous interference in $ {X}^{(1)}$ ($\beta_4 = 1,2$) and not ${X}^{(2)}$ ($\beta_5 = 0$), and the two covariates are uncorrelated (the intersection of the first row and the last two columns of Figure~\ref{fig:bivariate_OE_est}), OE depends on $ \gamma_1$ but is constant across values of $ \gamma_2$ for a fixed $ \gamma_1$. In these cases, treatment strategies that give units with ${X}^{(1)}_{ij}=1$ a higher treatment propensity have higher OE, regardless of how they prioritize units with different values of ${X}^{(2)}_{ij}$. In the same setting, but when the two covariates are correlated (the intersection of the second row and the last two columns of Figure~\ref{fig:bivariate_OE_est}), OE depends on both $\gamma_1$ and $\gamma_2$.  

When there is heterogeneous interference in $ X^{(2)}$ ($\beta_5 = 1$) but not $ X^{(1)}$ ($\beta_4 = 0$) (the intersection of the second column and last two rows of Figure~\ref{fig:bivariate_OE_est}), OE depends on $\gamma_2$ but is constant across values of $\gamma_1$ for a fixed $\gamma_2$, unless the covariates are correlated. In these cases, treatment strategies that give units with ${X}^{(2)}_{ij}=1$ a higher treatment propensity have a higher OE. When the covariates are correlated, treatment strategies that give units with ${X}^{(2)}_{ij}=1$ and ${X}^{(1)}_{ij}=1$ a higher treatment propensity have a higher OE.

When there is heterogeneous interference in both $X^{(1)}$ and $ X^{(2)}$ ($\beta_4 = 1,2; \beta_5 = 1$) (the intersection of the last two columns and last two rows of Figure~\ref{fig:bivariate_OE_est}), the OE depends on both $\gamma_1$ and $ \gamma_2$. The largest OE is seen when $\gamma_1$ and $ \gamma_2$ are both large and positive, i.e.,  when we hypothesize assigning treatment to units with $X_{ij}^{(1)}=X_{ij}^{(2)}=1$ 
%$X_{ij}^{(1)}=1$ and $X_{ij}^{(2)}=1$ 
with a higher probability. %When both variables are used for determining counterfactual treatment propensity, we evaluate overall effects with $\gamma_1, \gamma_2\neq 0$, in the setting where there is heterogeneous interference with respect to both variables, i.e., $\beta_4, \beta_5\neq 0$ (the intersection of the last two columns and last two rows of Figure~\ref{fig:bivariate_OE_est}), we can see a larger OE relative to the setting where there is heterogeneous interference only with respect to one variable  (the intersection of the first two columns and last two rows of Figure~\ref{fig:bivariate_OE_est}; the intersection of the last two columns and first two rows of Figure~\ref{fig:bivariate_OE_est}). 

If interference is heterogeneous only with respect to $X^{(1)}$ ($\beta_4\neq 0$ and $\beta_5 = 0$, we can capture some of the changes in OE by designing a treatment strategy around ${X}^{(2)}$ ($\gamma_1 = 0$ and $\gamma_2\neq 0$), as long as $ {X}^{(1)}$ and $ {X}^{(2)}$ are correlated ($\rho=\mathbb{P}(X^{(1)}_{ij} = X^{(2)}_{ij})>0.5$) (second and fourth rows of Figure \ref{fig:bivariate_OE_est}). Furthermore, in the case where ${X}^{(1)}$ and ${X}^{(2)}$ are correlated, it could be advantageous to design treatment strategies around both variables, even if the underlying spillover mechanism only depends on a single variable. 

%
%\section{Diffusion Scenario}
%One mechanism through which interference, or partial interference, can occur is when an outcome diffuses through a social network along social ties, or interference by contagion \citep{ogburn2014contagion}. Consider a scenario where each cluster has a self contained social network, where there are ties between individuals within the same network and no ties between individuals in separate clusters. Like in the previous set of simulations, the population is a set of clustered individuals that make up a social network with partial interference. Let us use a binary treatment and a binary outcome. We can observe $Y_{ij} = 1$ under two possible scenarios. In the first scenario, unit $ij$ is assigned to the treatment and its outcome is fixed at 1. In the second scenario, unit $ij$ is not assigned to treatment, but one of unit $ij$'s network neighbors is assigned to treatment. Then, if diffusion occurs down the network edge connecting the two units with some Bernoulli probability, unit $ij$ will also experience outcome $Y_{ij} = 1$.

\begin{figure}
\centering
\centerline{\includegraphics[width=\textwidth]{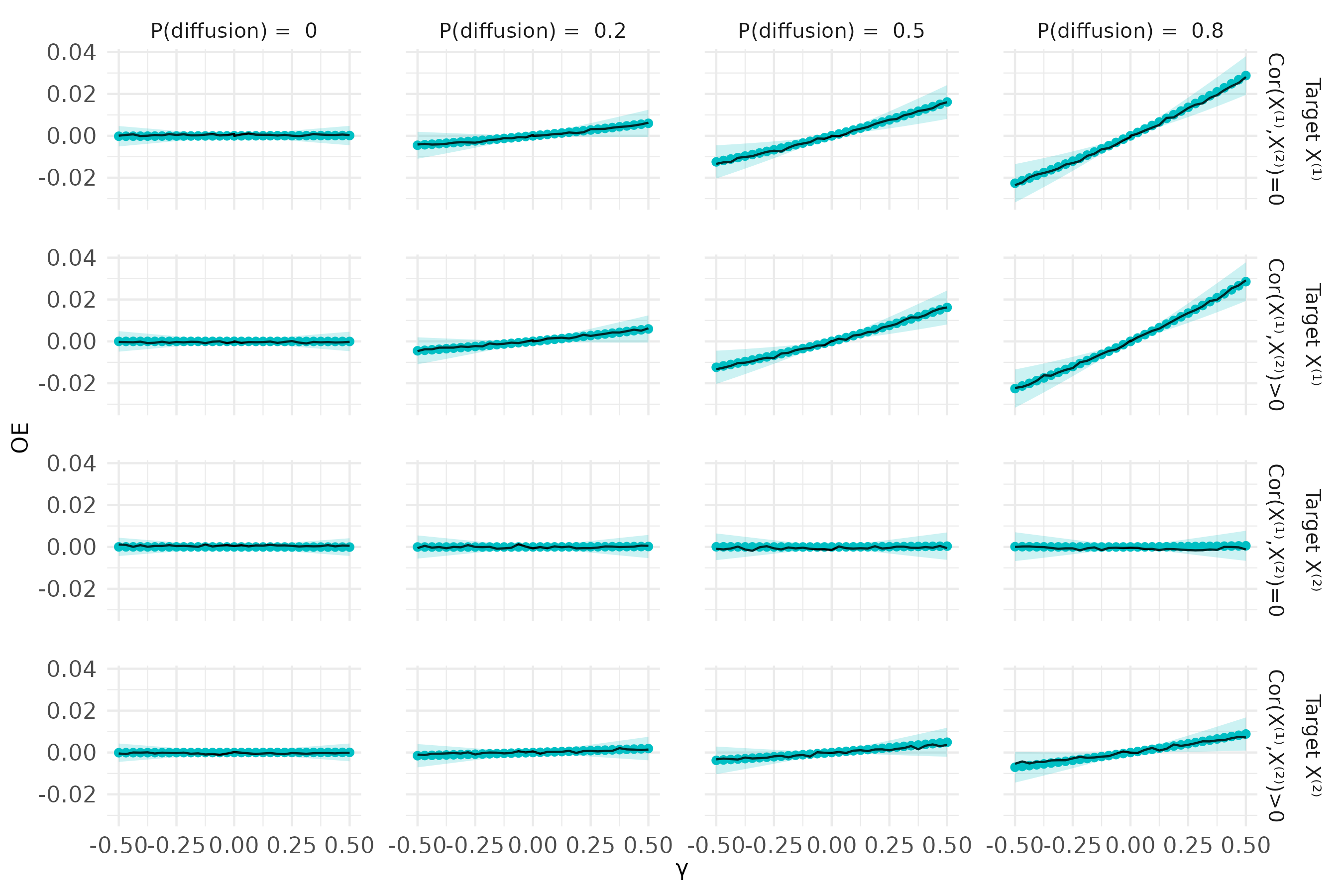}}
\caption{Scenario 2: Estimated overall effect and confidence intervals in a diffusion scenario. The first column shows scenarios where there is no diffusion of treatment, and therefore no interference. The latter three columns show scenarios with increasing amounts of diffusion, and therefore increasing interference. 
%In this example, $\bm{X}^{(1)}$ is a binary variable identifying if a unit is a central unit or an alter unit. The heterogeneity in interference relies on the fact that central units have more social ties and therefore are responsible for more spillover. 
The first two rows show interventions where individual treatment propensity depends on ${X}^{(1)}$ ($\gamma_1\neq 0$), and the bottom two rows show interventions where individual treatment propensity depends on  ${X}^{(2)}$ ($\gamma_2\neq 0$), with or without correlation with $X^{(1)}$. The $\gamma$ on the x-axis represents either $\gamma_1$ or $\gamma_2$ depending on the variable used for targeting. Black lines represent the true overall effect. Blue points represent OE point estimates, and shaded regions represent 95\% confidence intervals.}
\label{fig:diff_oe_est}
\end{figure}

For Scenario 2, with star networks and a diffusion process, 
we present the point estimates and confidence intervals of the overall effects in Figure~\ref{fig:diff_oe_est}.
% we present results for $\gamma_1 \in (-0.5, 0.5)$ and with $\gamma_2$ fixed at zero. Other parameters include the concordance between covariates $\bm{X}^{(1)}$ (network degree) and $\bm{X}^{(2)}$, $\rho$, and the probability that a unit's treatment diffuses to each neighbor, $p_d$. When $p_d = 0$, there is no interference. Individuals' outcomes are affected only by their own treatment. When $p_d > 0$, there is heterogeneous interference, depending on a unit's network degree. 
% %The treatment of high-degree individuals has more opportunities to diffuse than the treatment of low-degree individuals. In other words, a central node with $\bm{X}^{(1)}_{ij} = 1$ has more influence on the cluster than a peripheral node. 
% We considered scenarios where $p_d = (0,0.2, 0.5, 0.8)$ and where $\rho = (0.5, 0.65)$.
% 
% % In the network diffusion scenario simulation, 
% an overall effect is the contrast between the average population outcome under a counterfactual that assigns a higher treatment propensity to individuals with a certain degree and a counterfactual that assigns treatment independent of degree. 
We consider overall effects for two univariate hypothetical allocation strategies: one that assigns treatment depending on the value of $X^{(1)}$, that is, centrality  ($\gamma_1\neq 0$), and one where the individual treatment propensity depends on  ${X}^{(2)}$ ($\gamma_2\neq 0$), a variable that may or may not be correlated with centrality.
In the absence of interference ($p_d = 0$, column 1 of Figure~\ref{fig:diff_oe_est}), the overall effect is estimated to be zero regardless of $\gamma_1$ or $\gamma_2$, that is, it does not depend on which units are treated. In the presence of interference ($p_d>0$, columns 2-4 of Figure~\ref{fig:diff_oe_est}), the overall effect is higher for larger values of $\gamma_1$ reflecting preference for treating central units. The higher the probability of diffusion, the higher the overall effect when preferentially treating central nodes. Furthermore, 
%the overall effect is non-zero 
if the treatment probability depends on $ X^{(2)}$ ($\gamma_2\neq 0$), when this variable is correlated with centrality the overall effect is non-zero, that is, preferentially treating those with $ X^{(2)}=1$ is still beneficial.
%there are still gains in overall effect under treatment strategies that depend on the non-network characteristic $X^{(2)}_{ij}$ if it is associated with network degree 
%$\bm{X}^{(2)}$ in the simulation 
(row 4 of Figure~\ref{fig:diff_oe_est}). 

Bias is consistently minimal and coverage levels of the 95\% confidence intervals are close to the nominal level. Detailed bias and coverage figures are in Web Appendices C, D, and E, with further supplemental results.

In Web Appendix I, we also report a simulation study 
on the performance of the hypothesis test for overall, direct, and indirect effects. Simulation results show that the level of the test is close to nominal, with good power under the alternative hypotheses. 

%Analyses where the probability of treatment depends on a single variable and the outcome model is linear are in Web Appendix A, along with figures comparing bootstrapped and analytical variances. Analyses where the probability of treatment depends on two variables and the outcome model is linear are included below, with supplemental results for indirect and direct effects in Web Appendix B. Diffusion scenario analyses are included below, with indirect and direct effects relegated to Web Appendix C. 

\section{Application to insurance uptake intervention}
\label{s:cai}

We apply the proposed estimators to study the covariates' importance for treatment assignment in a randomized trial designed to increase insurance uptake among farmers in rural China \citep{cai2015insurance}. 
The original trial assigns individuals to information sessions according to a factorial design with multiple factors including the intensity (intensive vs simple) and the time (early vs delayed) of the session. 
%The outcome of interest is whether an individual purchases weather insurance. 
%The treatment is derived from two variables: whether an individual received a simple versus intense information session encouraging the purchase of insurance and whether an individual received an information session at an earlier or later date. 
Here, we considered individuals to be treated if they were assigned to an early, intensive information session. 
The outcome of interest is whether an individual purchases weather insurance. 
Friendship network information and covariates including area of rice cultivation and anticipated probability of future disaster were collected for each individual.

In this setting, interference may occur if individuals who are assigned to and attend the intensive information sessions in round 1 encourage (or discourage) others in their village to purchase the weather insurance. %Interference would be heterogeneous 
The overall effect on the population average insurance uptake may depend on allocation strategies that 
 give the intensive information session to different individuals 
 %may change the average insurance uptake in the population due 
 if there is a heterogeneous response to the information session or if different individuals have a different influence over others. The friendship network structure may play a role in determining this process. 
%if some people have more friends or are more influential to their friends. 
%
The data includes 4586 individuals in 47 villages (clusters) of varying sizes. We make the partial interference assumption, which here means that we assume a farmer's insurance purchase decisions can only be influenced by individuals in the same village. This assumption is reasonable -- the friendship network was collected for the entire data set, and on average 98\% of an individual's friends reside in the same village.

We estimate the direct, indirect, and overall effects of intensive information sessions on purchase of weather insurance under realistic counterfactual scenarios where intervention strategies depend on covariates. We first focus on covariates describing network characteristics: degree (the number of friends) and betweenness (the number of pathways between other units that pass through a fixed unit) \citep{rawlings2023networkcentrality}. 
%However, it is not a requirement for covariates to be related to network characteristics. 
%
%We assume consistency, positivity, and conditional ignorability. Treatment was randomly assigned and distributed through group information sessions where individuals were exposed to the same information.
%
For each covariate, we consider a range of $\gamma$s reported in \cref{tab:app_gammas} and chosen by the procedure described in \cref{sec: targets}.
%
% For analyses including all clusters, we considered 
% $\gamma_{\text{degree}} \in (-0.29, 0.091)$, 
% $\gamma_{\text{betweenness}} \in (-0.0037, 0.0021)$, 
% $\gamma_{\text{rice area}} \in (-0.70  1.00)$, and 
% $\gamma_{\text{future disaster}} \in (-0.42  0.37)$. 
% %
% For analyses including only clusters of 80 or fewer units, we considered 
% $\gamma_{\text{degree}} \in (-0.30, 0.16)$, 
% $\gamma_{\text{betweenness}} \in (-0.014, 0.0029)$, 
% $\gamma_{\text{rice area}} \in (-0.94, 1.23)$, and 
% $\gamma_{\text{future disaster}} \in (-0.50, 0.47)$. 
% %
% For analyses including only cluster of greater than 80 units, we considered $\gamma_{\text{degree}} \in (-0.17, 0.079)$, $\gamma_{\text{betweenness}} \in (-0.00077, 0.0012)$, $\gamma_{\text{rice area}} \in (-0.36, 0.69)$, and $\gamma_{\text{future disaster}} \in (-0.33, 0.26)$.
%
In scenarios where $\gamma_{\text{degree}}<0$ ($\gamma_{\text{degree}}>0$), the hypothetical allocation strategy prioritizes individuals with lower (higher) degree for receiving the treatment, whereas when $\gamma_{\text{degree}}=0$, the probability of treatment is the same for all individuals regardless of degree. %, and when $\gamma_{\text{degree}}>0$, the hypothetical intervention prioritizes individuals with higher degree for receiving the treatment. 
The remaining $\gamma$ values can be interpreted similarly.
For each covariate, we perform the hypothesis test outlined in \cref{sec:test}.

Intuition from the network analysis literature and from the simulations under Scenario 2 suggest that, with a simple diffusion process in a simple random graph, strategies treating high-degree and high betweenness individuals would increase overall and indirect effects because these individuals have larger numbers of neighbors to potentially influence or are placed in paths of information flow \citep{lee2023influencers}.

\begin{table}
\caption{Range of values for $\gamma$ for each covariate considered in our analysis, in the analysis of all clusters, clusters with up to 80 units, and clusters larger than 80 units.}
\label{tab:app_gammas}
    \begin{tabular}{lccc}
    Covariate & All clusters & $\leq 80$ units & $> 80$ units \\ \hline
    Degree & (-0.29, 0.091) & (-0.30, 0.16) & (-0.17, 0.079) \\
    Betweenness & (-0.0037, 0.0021) & (-0.014, 0.0029) & (-0.00077, 0.0012) \\
    Rice Area & (-0.70  1.00) & (-0.94, 1.23) & (-0.36, 0.69) \\
    Future disaster & (-0.42  0.37) & (-0.50, 0.47) & (-0.33, 0.26) \\ \hline
    \end{tabular}
\end{table}

\begin{table}
\caption{P-values for the hypothesis test on the overall effect for interventions based on individual characteristics. Results are separated by the clusters used, for all clusters, clusters with at most 80 units, and clusters with more than 80 units. Bold values correspond to p-values below 0.05.}
\label{tab:app_pvals}
    \begin{tabular}{lccc}
    \hline
    Covariate & All clusters & $\leq 80$ units & $> 80$ units \\ 
    \hline \\[-10pt]
    \multicolumn{4}{l}{{\it Allocation strategies based on one characteristic, compared to random treatment}} \\ \hline \hline
    Degree & $\bm{<0.001}$ & $\bm{0.041}$ & $\bm{<0.001}$ \\
    Betweenness & 0.311 & 0.845 & 0.406 \\
    Rice area & $\bm{0.028}$ & 0.232 & 0.201 \\
    Future disaster & 0.165 & 0.615 & 0.193 \\
    \hline \\[-10pt]
    
    \multicolumn{4}{l}{{\it Allocation strategies based on two characteristics, compared to random treatment}} \\ \hline \hline
    Degree \& Betweenness & 0.064 &0.1 & 0.063\\
    Degree \& Rice area & $\bm{<0.001}$ & 0.056 & $\bm{0.001}$ \\
    \hline \\[-10pt]
    
    \multicolumn{4}{l}{{\it Allocation strategies based on degree and a second characteristic, compared to only degree}} \\ \hline \hline
    Degree \& Betweenness & 0.131 & $\bm{0.005}$ & 0.118\\
    Degree \& Rice area & 0.052 & 0.327 & 0.604 \\
    \hline \\

    \end{tabular}
\end{table}

\begin{figure}
\begin{subfigure}[t]{\linewidth}
\centerline{\includegraphics[width=\textwidth]{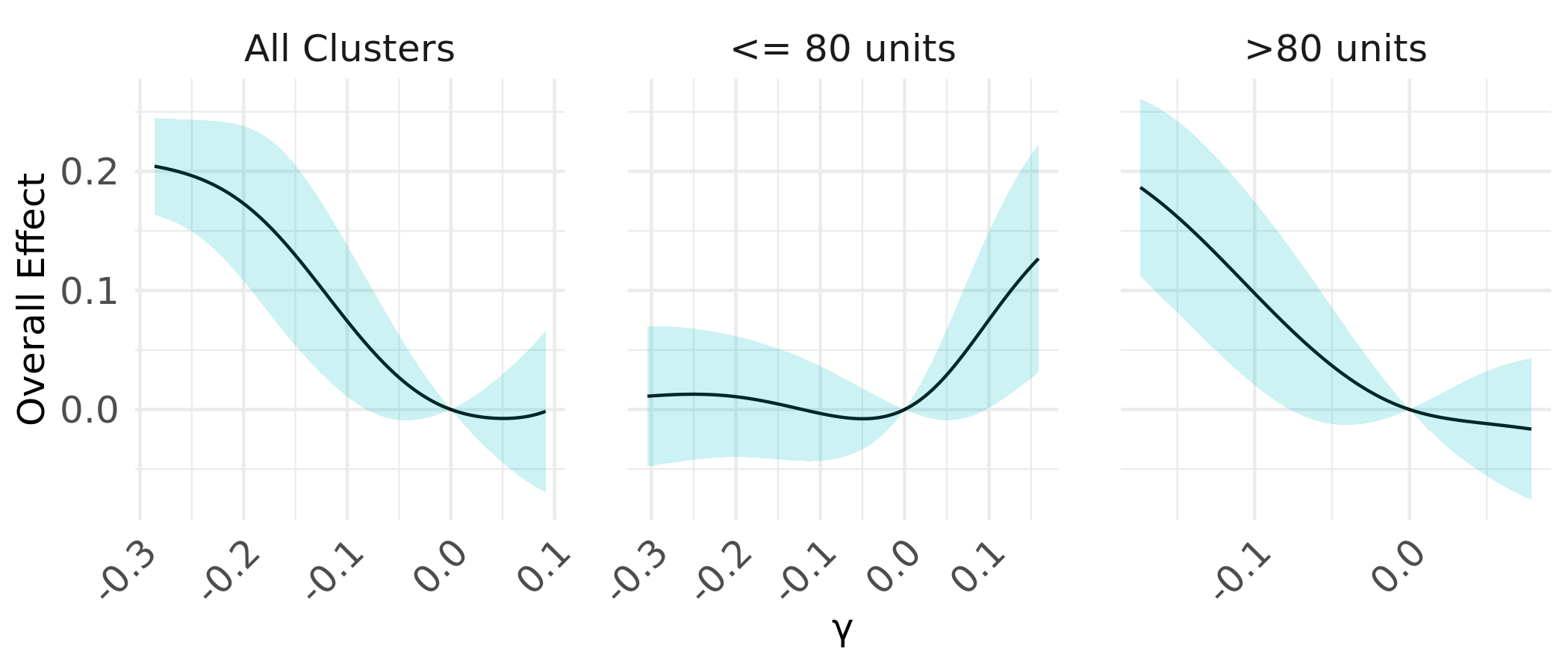}}
\caption{Estimated overall effect when targeting treatment based on degree. Larger values of $\gamma$ correspond to strategies that assign a larger probability of treatment to individuals with a larger degree. %In the all- analysis and that for the larger villages, targeting smaller degree individuals leads to a larger overall effect. In smaller villages, targeting larger degree individuals leads to a larger overall effect. 
}
\label{fig:cai_oe_degree}
\end{subfigure}
\begin{subfigure}{\linewidth}
\centerline{\includegraphics[width=\textwidth]{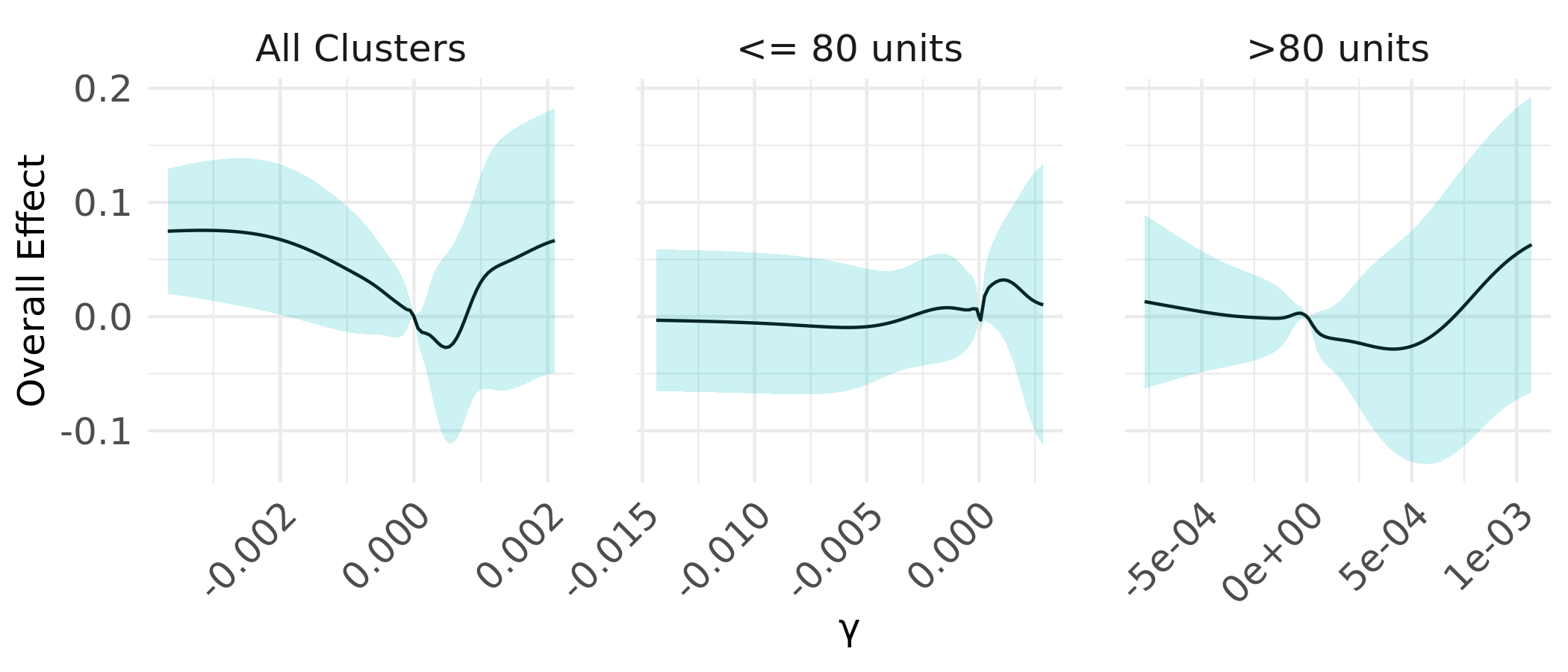}}
\caption{Estimated overall effect when targeting treatment based on betweenness. Larger values of $\gamma$ correspond to strategies that assign a larger probability of treatment to individuals with a larger betweenness.
}
\label{fig:cai_oe_btwn}
\end{subfigure}
\caption{Estimated overall effects for insurance uptake when targeting network characteristics. Estimates show overall effects and confidence intervals when targeting treatment based on (a) degree, and (b) betweenness. Black lines show point estimates and blue shaded regions show 95\% confidence intervals. Results are presented for all villages and stratified into villages with less or greater than 80 individuals in the study.}
\label{fig:cai_oe}
\end{figure}

The results for the overall effects are shown in \cref{fig:cai_oe} and p-values are reported in \cref{tab:app_pvals}. We observe that a significantly greater overall effect is attained when targeting treatment to low-degree individuals when considering villages of all sizes. Analyzing the data for villages of $\le80$ and $>80$ individuals, using the median cluster size as the threshold, reveals a potential explanation for this counterintuitive result. In the smaller villages, it is beneficial to assign the intensive session with higher probability to those with higher degree, while in larger villages it is beneficial to target individuals with lower degree. This might be due to the underlying network structure within clusters of different sizes: larger villages have a much larger concentration of units and many peripheral individuals with few connections (Web Appendix F). Therefore, in large clusters, central individuals will be reached easily irrespective of who is treated, while peripheral individuals will be hard to reach when targeting treatment to high-degree individuals. 
We observe no significant changes in overall effect when targeting treatment based on individual's betweenness.

\begin{figure}
\centerline{\includegraphics[width=\textwidth]{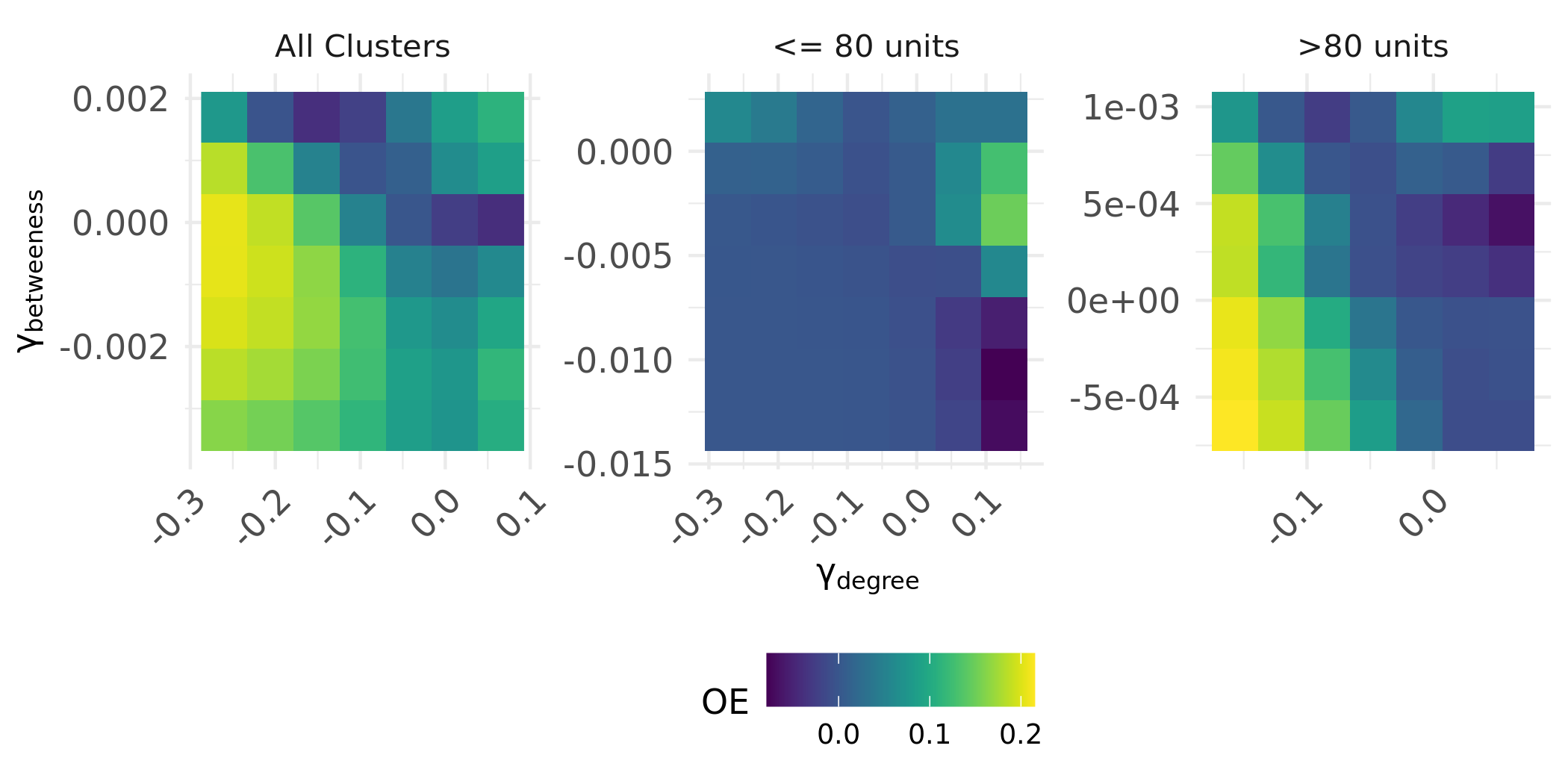}}
\caption{The overall effect when targeting treatment based on both degree and betweenness. Larger values of $\gamma_{\text{degree}}$ correspond to strategies that assign a larger probability of treatment to individuals with a larger degree. Larger values of $\gamma_{\text{betweenness}}$ correspond to strategies that assign a larger probability of treatment to individuals with larger betweenness. When either $\gamma$ is 0, that corresponds to a strategy agnostic to the corresponding variable. Results are stratified by cluster size.}
\label{fig:cai_bivar_dgrbtwn}
\end{figure}

We also examined counterfactual scenarios where the probability of treatment depends on both degree and betweenness through a vector of $\bm \gamma = (\gamma_{\text{degree}}, \gamma_{\text{betweenness}})$. The results are shown in \cref{fig:cai_bivar_dgrbtwn}.
In scenarios where we inform treatment strategy using both degree and betweenness, changes in OE across all clusters appear to be driven by changes in OE for larger clusters of $>80$ units, while smaller clusters with $\le80$ units do not show great heterogeneity in overall effect. 
As in the hypothetical allocations where treatment probability depends only on betweenness or degree, degree appears to drive more heterogeneity in OE, such that in larger clusters assigning a higher treatment probability to low-degree individuals leads to a relatively larger OE, while in smaller clusters assigning a higher treatment probability to high-degree individuals leads to a relatively larger OE. However, there can be additional gain in OE when targeting based on both variables. If we consider a targeting strategy where we assign a higher treatment probability to individuals with the lowest degree values, there is a significant change in OE when we also target based on betweenness in smaller clusters. %(p-value = 0.16 for all, 0.01 for $\le80$, and 0.12 for $>80$)

%thetical interventions where treatment probability depends only on betweenness, this variable does not appear to drive much heterogeneity in overall effects. However, like in the univariate intervention on degree, it remains evident that in larger clusters assigning a higher treatment probability to low-degree individuals leads to a relatively larger OE, while in smaller clusters assigning a higher treatment probability to high-degree individuals leads to a relatively larger OE.  

\begin{figure}
\begin{subfigure}{\linewidth}
\centerline{\includegraphics[width=\textwidth]{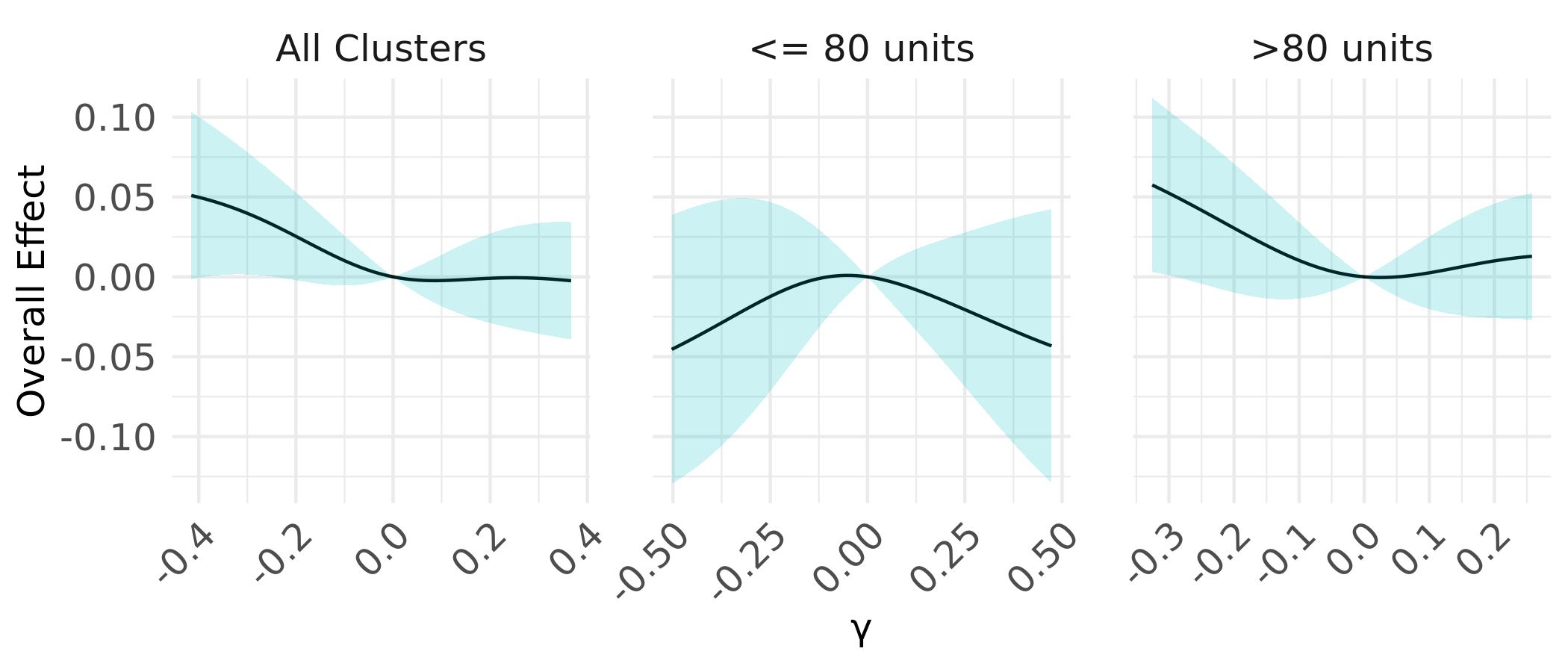}}
\caption{Estimated overall effect when targeting treatment based on an individual's perceived probability of a future disaster. Larger values of $\gamma$ correspond to strategies that assign a larger probability of treatment to individuals with a larger perceived probability of a future disaster. }
\label{fig:cai_disaster_oe}
\end{subfigure}
\begin{subfigure}{\linewidth}
\centerline{\includegraphics[width=\textwidth]{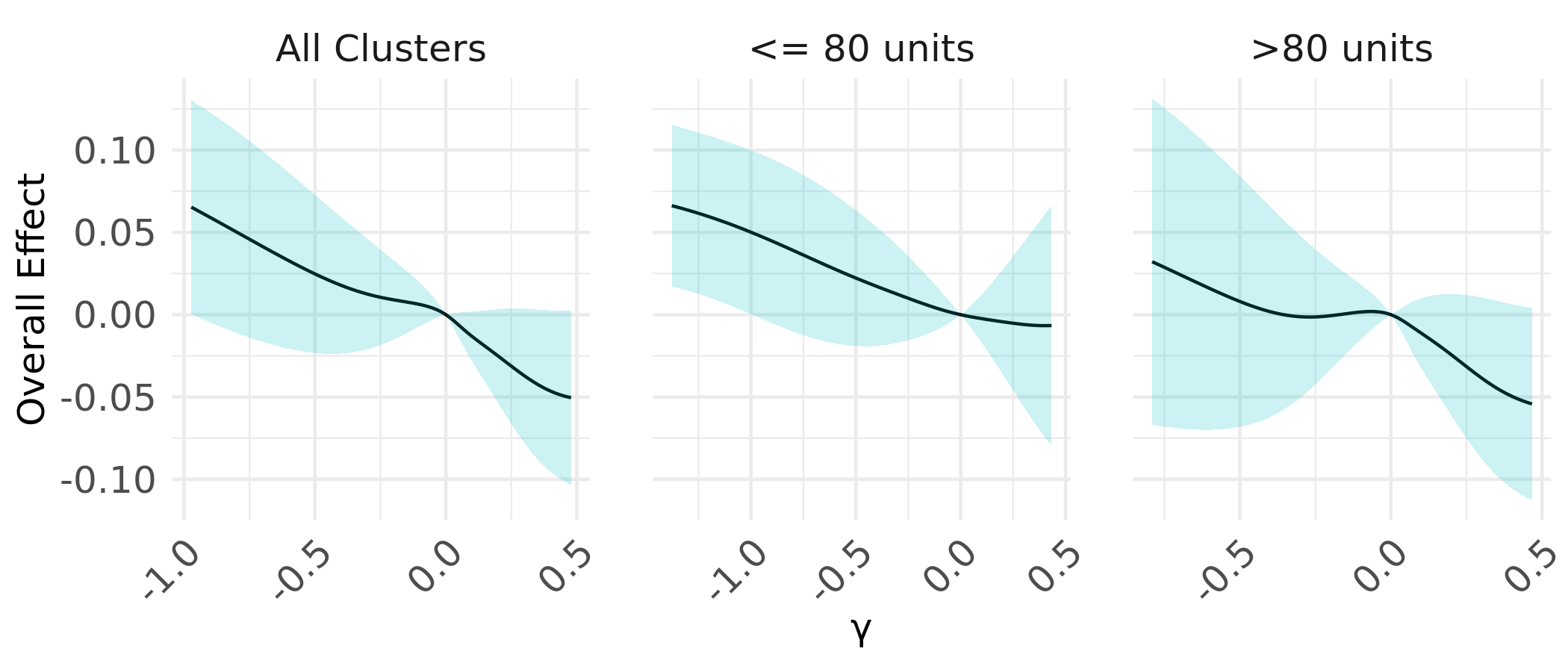}}
\caption{Estimated overall effect when targeting treatment based on the area of rice cultivation. Larger values of $\gamma$ correspond to strategies that assign a larger probability of treatment to individuals with a larger area of rice cultivation.}
\label{fig:cai_rice}
\end{subfigure}
\caption{Estimated overall effects for insurance uptake when targeting individual characteristics. Estimates show overall effects and confidence intervals when targeting treatment based on (a) their perceived probability of a future disaster, and (b) the area of rice cultivation. Black lines show point estimates and blue shaded regions show 95\% confidence intervals. Results are presented for all villages and stratified into villages with less or greater than 80 individuals in the study. }
\label{fig:cai_oe_nonetwork}
\end{figure}

Beyond measures of network centrality such as degree and betweenness, we examined non-network characteristics including an individual's perceived probability of future disaster and their area of rice production. 
The results are shown in \cref{fig:cai_oe_nonetwork}.
When assigning 
the information session intervention based on an individual's perceived probability of future disaster, there is not a statistically significant difference between overall effects under targeting strategies with different $\gamma$ for that covariate
in either the stratified or pooled analyses (\cref{fig:cai_disaster_oe}).
%
%, we see the highest DE when preferentially treating those with a high perceived probability (Figure~\ref{fig:cai_disaster_de}). However, targeting those same individuals corresponds to a lower OE (Figure~\ref{fig:cai_disaster_oe}). A possible explanation is that while these risk-aware individuals are very sensitive to being treated themselves, they are also very likely to be influenced by their neighbors being treated so it is a more efficient use of resources to treat their neighbors, who may be more hesitant to purchase insurance and need the extra push of direct treatment. Even so, there is not a statistically significant difference between overall effects of different targeting strategies for this variable in either the stratified or pooled analyses (p$>$0.05).
%
%
When assigning the information session intervention based on an individual's area of rice production, OE is larger when preferentially treating individuals with small areas of production 
%compared to when preferentially treating individuals with large areas of production 
(Figure~\ref{fig:cai_rice}), with results being statistically significant for all clusters. A possible explanation is if people with higher rice area are clustered together socially, treating people with lower rice area might better leverage spillover to reach the most people. Further, people with a larger area of rice production maybe be more likely to buy insurance regardless of the intensive information session.

%This could be explained if we consider that the "no treatment" group here includes individuals who received either a less intense information session or an intense session with a time delay. 
%The direct effect may appear smaller when we target people who cultivate large areas because these individuals would be very likely to purchase insurance regardless of their treatment assignment.

\begin{figure}
\centerline{\includegraphics[width=\textwidth]{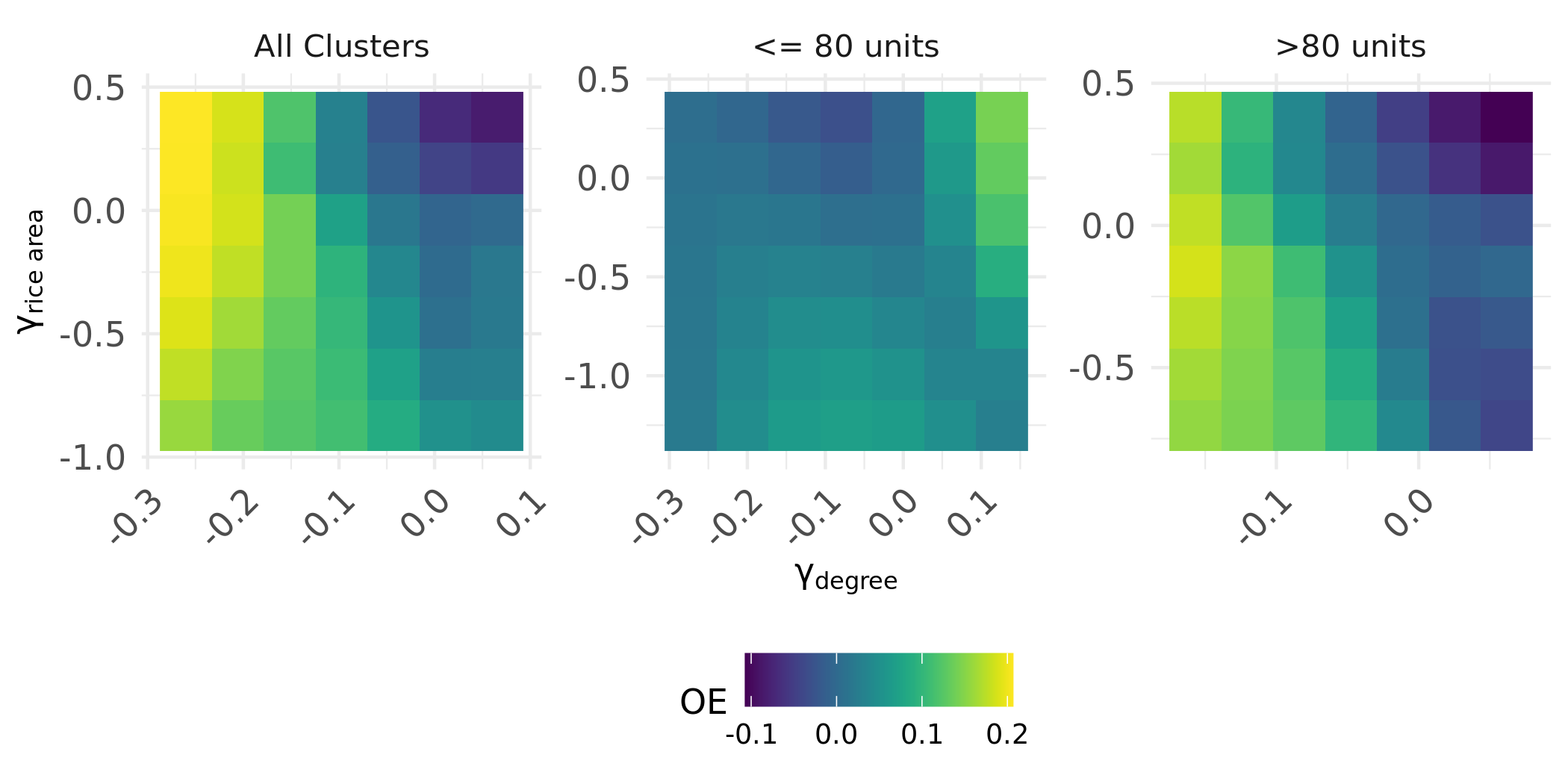}}
\caption{Estimated overall effect when targeting treatment based on the important network, and non-network characteristics of degree and the area of rice cultivation. Larger values of $\gamma_{\text{rice area}}$ correspond to strategies that assign a larger probability of treatment to individuals with a larger area of rice cultivation. Larger values of $\gamma_{\text{degree}}$ correspond to strategies that assign a larger probability of treatment to individuals with a larger network degree. Results are stratified by cluster size.}
\label{fig:cai_dgrrice}
\end{figure}

We also considered the comparative OEs of bivariate intervention strategies that assign treatment probability based on both degree and the area of rice production. The results are shown in \cref{fig:cai_dgrrice}.
The largest OE is observed when assigning higher treatment probabilities to individuals with high rice area and low degree. However, the smallest OE is observed when assigning higher treatment probabilities to individuals with high-degree and high rice area, suggesting an interaction between these two variables when determining the interference heterogeneity in the individuals' response to the intensive session and in their influence on their peers. Further, this differs from the results of a univariate targeting strategy on rice area, where assigning a larger treatment probability to an individual with a low rice area was associated with a larger OE. The heterogeneity in OE is statistically significant at the 5\% level. However, when we fix the targeting strategy to assign higher treatment probabilities to individuals with low degree, we do not see significant changes in OE depending on the extent to which the treatment is assigned also based on rice area.

Results for direct and indirect effects under the same set of hypothetical interventions are included in Web Appendix F.

\section{Discussion}
This paper proposes methods for estimating causal effects under partial interference with realistic counterfactual scenarios where the marginal treatment probability is fixed but the treatment probability conditional on covariates of interest varies. 
%Thus, individual treatment propensities depend on observed characteristics such as network centrality or sex. 
Because of different mechanisms of heterogeneity, including heterogeneous interference, treating one group of individuals could have a different impact on population average outcomes compared to treating a different group of individuals.
Therefore, by comparing the OE of different treatment strategies we evaluate the importance of targeting individuals with different covariates for treatment. 
In a limited resource setting, our proposed methods would allow providers to compare the population level effects of different treatment allocation strategies.
%, with the goal of identifying the  strategy which would lead to a higher OE of an intervention. 
However, it is worth noting that this work is not focused on identifying the optimal treatment allocation strategy, but rather on estimating overall effects under different potential stochastic allocation strategies based on different sets of covariates, and testing whether targeting each of these sets would change the overall effect in the population. In this sense, this paper can be seen as complementing existing work on optimal targeting under interference \citep[e.g.,][]{viviano2024policy, zhang2024policylearning}.

In our simulation study, we examined what would happen if we designed a treatment strategy around a variable that is correlated with the variable that truly affects interference. We find that we are still able to take advantage of interference to increase overall effects, although the overall effects are higher under direct interventions on the variable affecting interference. However, in cases where the true variable of interest is difficult to measure, it is valuable to know that designing treatment strategies around a correlated variable still has a payoff in terms of increasing OEs. For example, instead of intervening on network centrality, whose measure requires significant resources to collect detailed network data, one could target more accessible variables that are correlated with centrality.

The application to the insurance uptake dataset showed that multiple variables were driving heterogeneity in spillover, and univariate or bivariate strategies could be designed to leverage this heterogeneity into a larger overall effect. We also found that overall effects can look different when stratified by cluster size, as we saw in the case of degree and perceived probability of a future disaster.

This work could be extended to allow for more complex treatment assignment mechanisms. One possible extension would be to non-binary treatments, where rather than treatment propensities, estimators depend on treatment densities. Another possible extension could allow for longitudinal treatments with multiple, dependent time points, or investigate how average potential outcomes vary by individual covariates. 
%Further extensions could address designing optimal targeting strategies under interference. 

%  The \backmatter command formats the subsequent headings so that they
%  are in the journal style.  Please keep this command in your document
%  in this position, right after the final section of the main part of 
%  the paper and right before the Acknowledgements, Supporting Information (Supplementary %  Materials),   and References sections. 

%\backmatter

%  This section is optional.  Here is where you will want to cite
%  grants, people who helped with the paper, etc.  But keep it short!

\section*{Acknowledgements}

We thank Forrest Crawford for helpful comments. This work was supported by NSF grants DGE-2139841 and DMS-2124124 and the NIH grant R01MH134715. 
%\vspace*{-8pt}

%  Here, we create the bibliographic entries manually, following the
%  journal style.  If you use this method or use natbib, PLEASE PAY
%  CAREFUL ATTENTION TO THE BIBLIOGRAPHIC STYLE IN A RECENT ISSUE OF
%  THE JOURNAL AND FOLLOW IT!  Failure to follow stylistic conventions
%  just lengthens the time spend copyediting your paper and hence its
%  position in the publication queue should it be accepted.

%  We greatly prefer that you incorporate the references for your
%  article into the body of the article as we have done here 
%  (you can use natbib or not as you choose) than use BiBTeX,
%  so that your article is self-contained in one file.
%  If you do use BiBTeX, please use the .bst file that comes with 
%  the distribution.  In this case, replace the thebibliography
%  environment below by 
%
% \bibliographystyle{biom} 
% \bibliography{mybibilo.bib}

\bibliographystyle{plainnat}
\bibliography{formatted_manuscript}

%\begin{thebibliography}{}

%\bibitem{ } Cox, D. R. (1972). Regression models and life tables (with
%discussion).  \textit{Journal of the Royal Statistical Society, Series B}
%\textbf{34,} 187--200.

%\bibitem{ }  Hastie, T., Tibshirani, R., and Friedman, J. (2001). \textit{The 
%Elements of Statistical Learning: Data Mining, Inference, and Prediction}.
%New York: Springer.

%\end{thebibliography}

%  If your paper refers to supporting web material, then you MUST
%  include this section!!  See Instructions for Authors at the journal
%  website http://www.biometrics.tibs.org

%\appendix
%  To get the journal style of heading for an appendix, mimic the following.
%\section{}
%Put your short appendix here.  Remember, longer appendices are
%possible when presented as Supplementary Web Material.  Please 
%review and follow the journal policy for this material, available
%under Instructions for Authors at \texttt{http://www.biometrics.tibs.org}.

%\label{lastpage}
\newpage
%  
%\documentclass[12pt]{article}

%\usepackage{graphicx}
%\usepackage{float}
%\usepackage{setspace}
%\usepackage{geometry}
%\usepackage{amsmath}
%\usepackage{amssymb}
%\usepackage{caption}
%\usepackage{bm}
%\usepackage[capitalize,noabbrev]{cleveref}
%\geometry{margin=1in}
%\doublespacing

%\begin{document}

\section*{Web Appendix}
This appendix contains details on estimator asymptotics and additional figures for bias and coverage of proposed estimators and their variance measures. It also contains the procedures we followed to generate true values against which to validate the proposed estimators.  

\subsection*{Web appendix A - details of estimator asymptotics}
We assume that potential outcomes are bounded. Estimating functions $\psi$, to be defined shortly, are finite, integrable, and twice differentiable. 

The vector of estimating functions, $\bm \psi$ is defined: 
    \begin{equation*}
    \begin{aligned}
            %\bmath{\psi}(\bmath{Y}_i, \text{X}_i, \bmath{A}_i; \bmath{\mu}, \bmath{\gamma}) = (\psi_{1}(\bmath{Y}_i, \text{X}_i, \bmath{A}_i; \mu_1, \bm \gamma_1), \psi_{2}(\bmath{Y}_i, \text{X}_i, \bmath{A}_i; \mu_2, \bm \gamma_2), \ldots \psi_{R}(\bmath{Y}_i, \text{X}_i, \bmath{A}_i; \mu_R, \bm \gamma_R))
            %
            \bm {\psi}(\bm {Y}_i, \bm{X}_i, \bm A_i; \bm \mu_{\alpha, \bm\Gamma^*}) = 
            % & ( \psi_{0, \alpha, \bm \gamma_1}(\bmath{Y}_i, \text{X}_i, \bmath{A}_i; \mu_{0, \alpha, \bm \gamma_1} ), \ldots, \psi_{0, \alpha, \bm \gamma_R}(\bmath{Y}_i, \text{X}_i, \bmath{A}_i; \mu_{0, \alpha, \bm \gamma_R}),\\
            % &\psi_{1, \alpha, \bm \gamma_1}(\bmath{Y}_i, \text{X}_i, \bmath{A}_i; \mu_{1, \alpha, \bm \gamma_1}),
            % \ldots, \psi_{1, \alpha, \bm \gamma_R}(\bmath{Y}_i, \text{X}_i, \bmath{A}_i; \mu_{1, \alpha, \bm \gamma_R}),\\
            % &\psi_{\alpha, \bm \gamma_1}(\bmath{Y}_i, \text{X}_i, \bmath{A}_i; \mu_{\alpha, \bm \gamma_1}), 
            % \ldots, \psi_{\alpha, \bm \gamma_R}(\bmath{Y}_i, \text{X}_i, \bmath{A}_i; \mu_{\alpha, \bm \gamma_R} )
            \bigg[&\big\{\psi_{0, \alpha, \bm \gamma}(\bm Y_i, \bm{X}_i, \bm A_i; \mu_{0, \alpha, \bm \gamma} )\big\}_{\bm \gamma \in \bm\Gamma^*}, \\
            &\big\{\psi_{1, \alpha, \bm \gamma}(\bm Y_i, \bm{X}_i, \bm A_i; \mu_{1, \alpha, \bm \gamma} )\big\}_{\bm \gamma \in \bm\Gamma^*}, \\&\big\{\psi_{\alpha, \bm \gamma}(\bm Y_i, \bm{X}_i, \bm A_i; \mu_{ \alpha, \bm \gamma} )\big\}_{\bm \gamma \in \bm\Gamma^*} \bigg]
    \end{aligned}  
    \end{equation*}
    where
    \begin{equation*}
    \begin{aligned}
        \psi_{a, \alpha, \bm \gamma}& (\bm Y_i, \bm{X}_i, \bm A_i; \mu_{a, \alpha, \bm \gamma}) = \\ 
        &\frac{1}{n_i} \sum_{j=1}^{n_i} \left[\frac{I(A_{ij} = a) P_{\alpha, \bm \gamma, \text{X}}(\bm A_{i, -j}  \mid  \bm{X}_i)}{f(\bm A_i  \mid  \bm{X}_i)}Y_{ij} - \mu_{a, \alpha, \bm \gamma} \frac{I(A_{ij} = a)P_{\alpha, \bm \gamma, \text{X}}(\bm A_{i}  \mid \bm{X}_i)}{f(A_{ij}  \mid \bm{X}_i)} \right],
    \end{aligned}
    \end{equation*}
    and
    \begin{equation*}
        \psi_{\alpha, \gamma}(\bm Y_i, \bm{X}_i, \bm A_i; \mu_{\alpha, \gamma}) = \frac{1}{n_i} \sum_{j=1}^{n_i} \left[\frac{P_{\alpha, \bm \gamma, \text{X}}(\bm A_{i}  \mid \bm{X}_i)}{f(\bm A_i  \mid  \text{X}_i)}Y_{ij} - \mu_{\alpha, \gamma} \frac{P_{\alpha, \bm \gamma, \text{X}}(\bm A_{i}  \mid \bm{X}_i)}{f(A_{ij}  \mid \bm{X}_i)} \right]
    \end{equation*}

\subsection*{Web appendix B - simulation scenarios}
Our main results include simulation studies for two scenarios in which we apply our novel estimators. Here, we enumerate the parameters we considered and varied for each simulation scenario (\cref{tab:sim_list}).

\begin{table}[H]
    \caption{This table shows the range of values for parameters used in the simulation studies.}
    %\documentclass[11pt]{article}
%\documentclass[a4paper,landscape]{article}
%\usepackage{geometry}
%\usepackage{amsmath}
%\geometry{margin=1in}

%\begin{document}

% \sf
\begin{center}
\begin{tabular}{ |c|c|c| }
 \hline
Category & Parameter & Values Simulated\\ 
 \hline
Scenario 1: & $\rho = P(X^{(1)}_{ij} = X^{(2)}_{ij})$ & 0.5, .65\\
 Outcome Linear Model Parameters 
& $\beta_0$ & 0.1 \\ 
&$\beta_1$ & 3\\
& $\boldsymbol{\beta}_2$ & $\mathbf{0}$ \\
&$\beta_3$ & 0 \\ 
&$\beta_4$ & 0, 0.5, 1 \\
&$\beta_5$ & 0, 1 \\
&$\sigma^2$ & 1\\

\hline
Scenario 2: & $p_\text{diff}$ & 0, 0.2, 0.5, 0.8 \\
Diffusion Process Parameters & $\rho = P(X^{(1)}_{ij} = X^{(2)}_{ij})$ & 0.5, 0.8 \\

 \hline
\end{tabular}
\end{center}

%\end{document}
    \label{tab:sim_list}
\end{table}

\subsection*{Web appendix C - univariate intervention plots}
\begin{figure}[H] 
\centerline{\includegraphics[width=\textwidth]{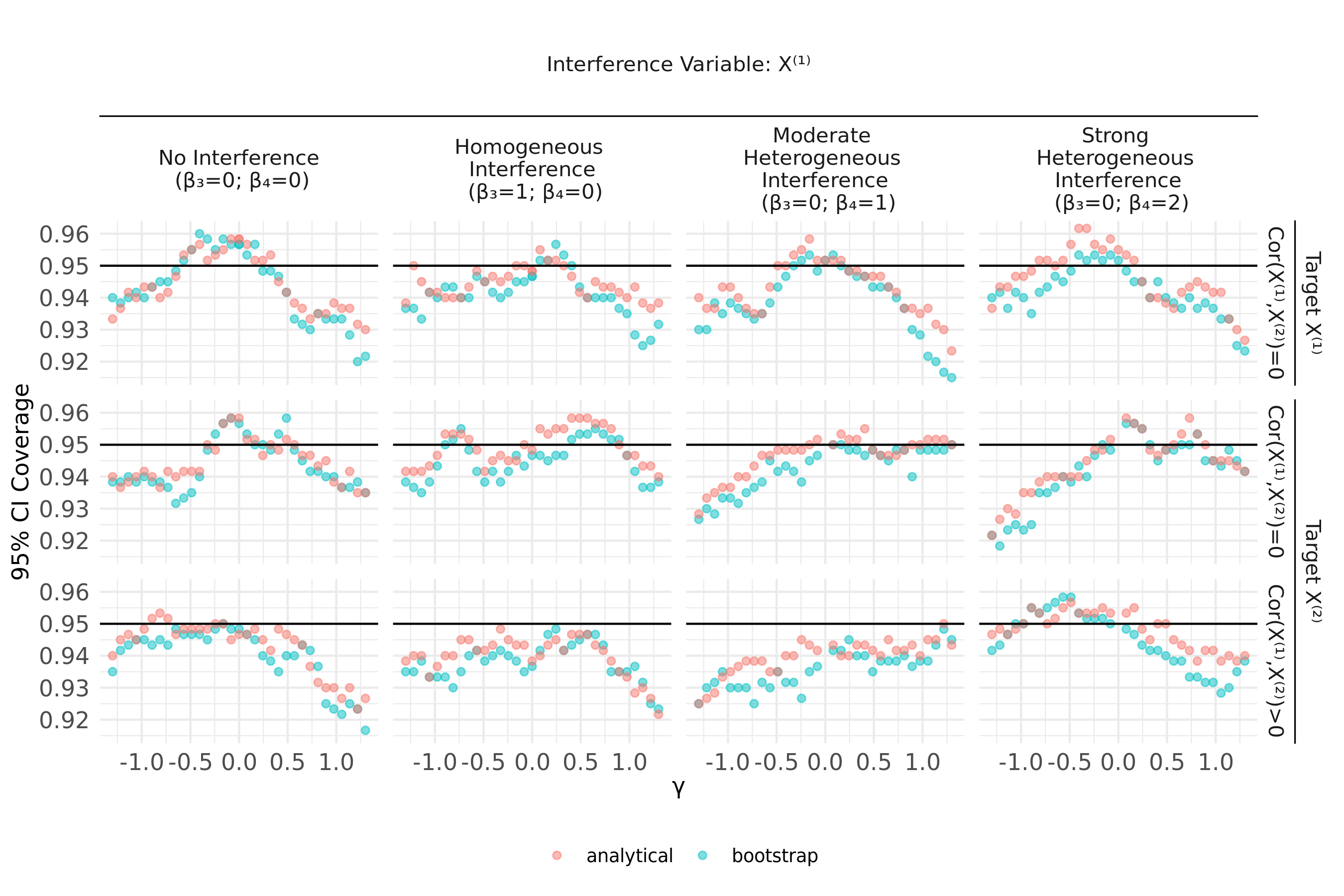}}
\caption{Coverage of confidence intervals for estimated direct effect under various intervention strategies.  Coverage represents the proportion of simulation trials in which the simulated true direct effect is contained by the confidence intervals. Coverage is shown both for confidence intervals calculated using the analytical, m-estimation variance and using a bootstrapped variance. The black line indicated the expected 95\% level of coverage. Columns represent level of interference present, including no interference, only homogeneous interference, moderate heterogeneous interference through $\mathbf{X}^{(1)}$, or strong heterogeneous interference through $\mathbf{X}^{(1)}$. Rows represent whether counterfactual treatment propensities depend on $\mathbf{X}^{(1)}$ (row 1) or $\mathbf{X}^{(2)}$ (row 3 and 4), and whether the two are correlated. Larger values of $\gamma$ denote counterfactual treatment scenarios where individuals with larger values of the targeted covariate have a higher treatment propensity. 
%The observed coverages are close to the expected 95\%. 
}
\label{fig:univariate_DE_cov}
\end{figure}

\begin{figure}[H] 
\centerline{\includegraphics[width=\textwidth]{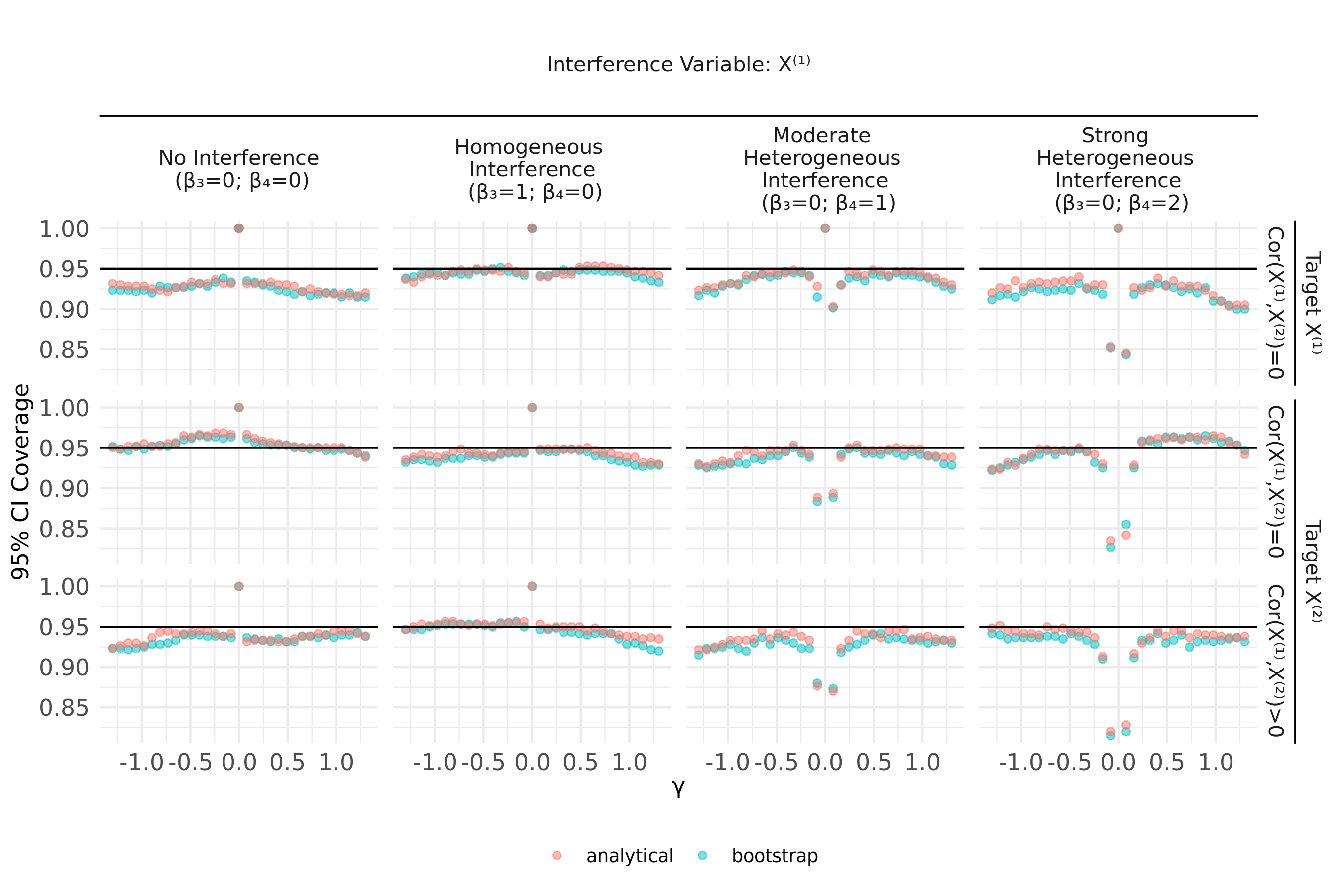}}
\caption{Coverage of confidence intervals for estimated indirect effect on untreated under various intervention strategies. See Figure~\ref{fig:univariate_DE_cov} for a detailed description of scenarios. The observed coverages are close to the expected 95\%.}
\label{fig:univariate_IE0_cov}
\end{figure}

\begin{figure}[H] 
\centerline{\includegraphics[width=\textwidth]{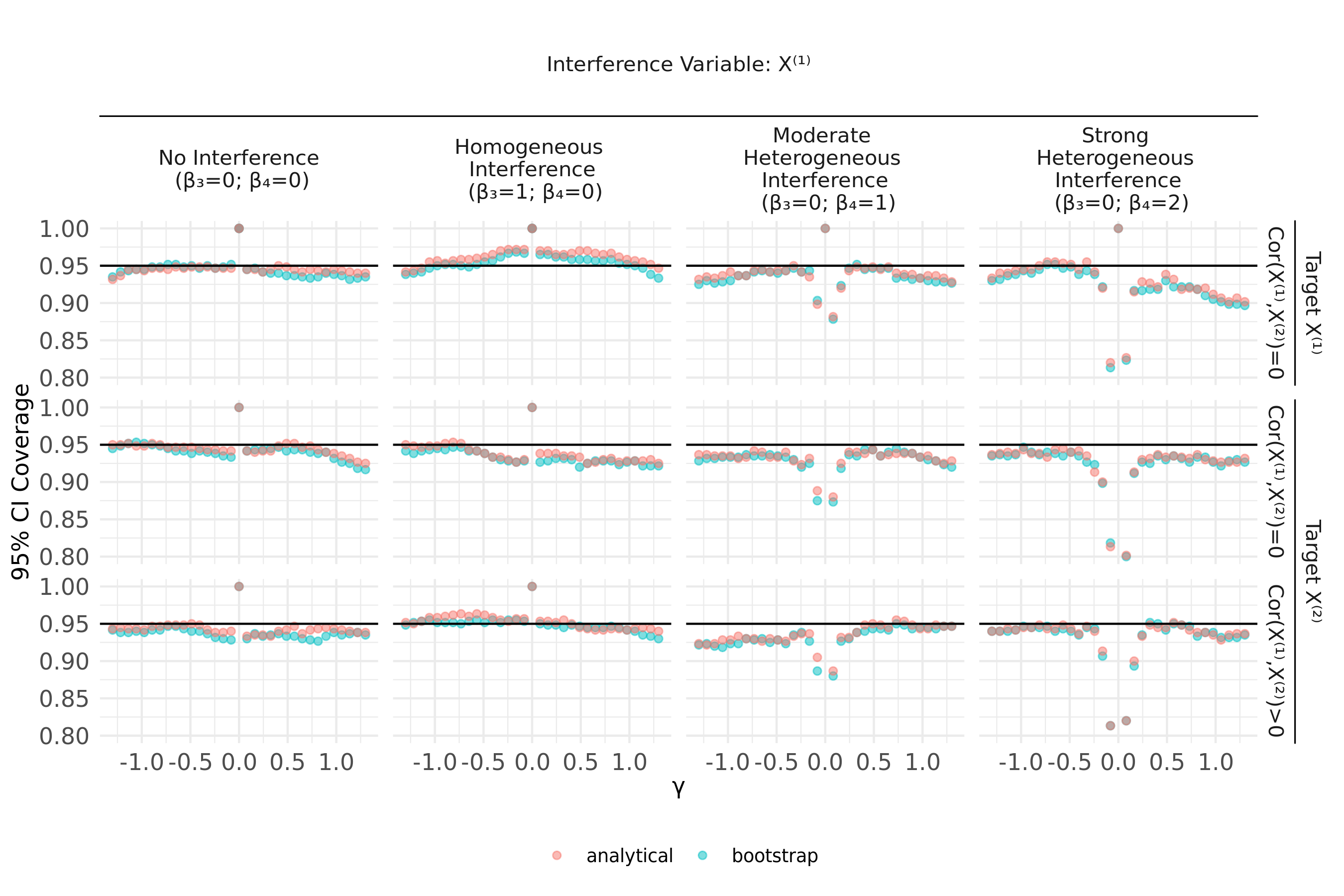}}
\caption{Coverage of confidence intervals for estimated indirect effect on treated under various intervention strategies. See Figure~\ref{fig:univariate_DE_cov} for a detailed description of scenarios. 
%The observed coverages are close to the expected 95\%.
}
\label{fig:univariate_IE1_cov}
\end{figure}

\begin{figure}[H]
\centerline{\includegraphics[width=\textwidth]{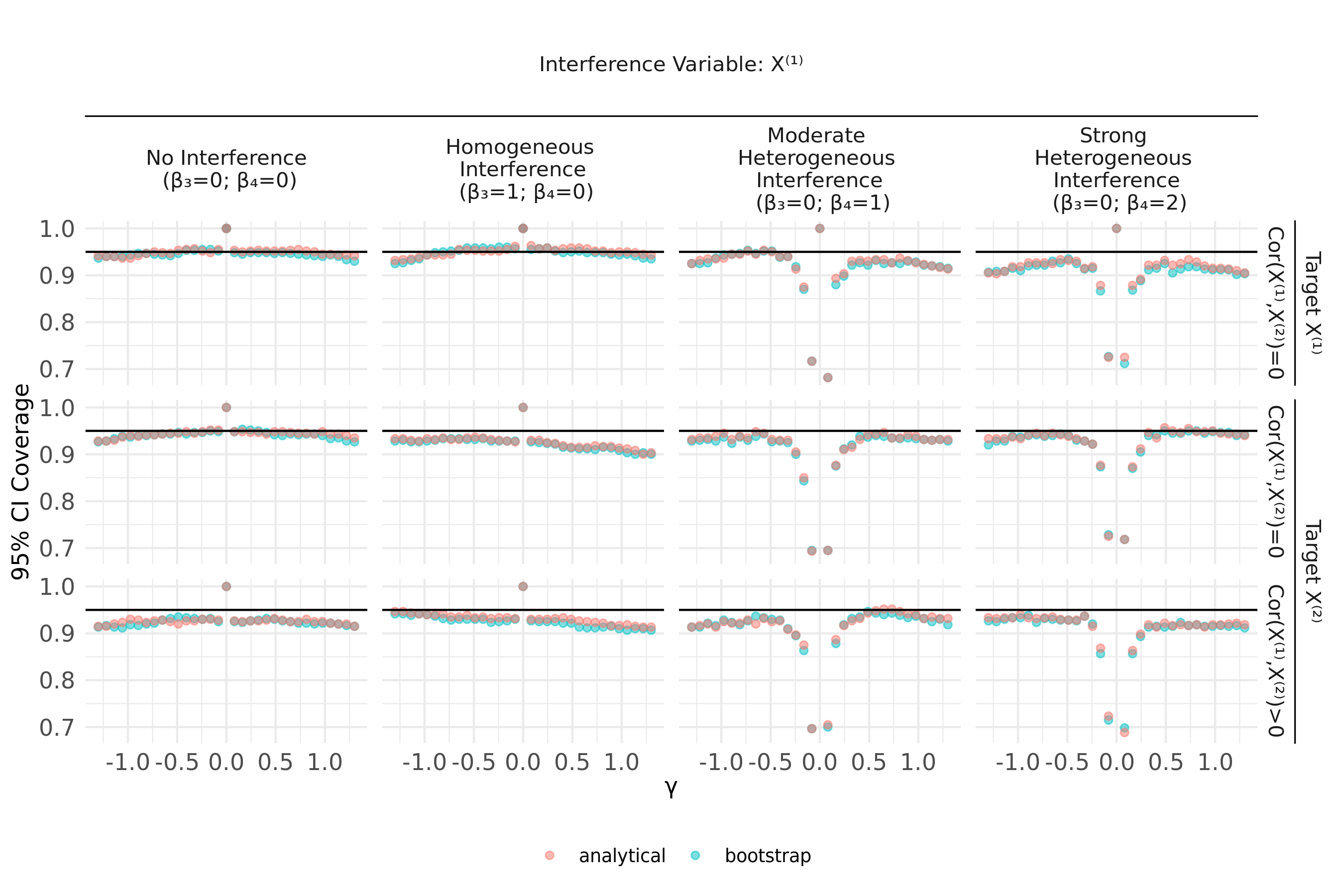}}
\caption{Coverage of confidence intervals for estimated overall effect under univariate intervention strategies. Columns represent different interference structures, including no interference, homogeneous interference, and heterogeneous interference. Rows represent whether treatment propensity depends on $\mathbf{X}^{(1)}$ or $\mathbf{X}^{(2)}$, and whether the two are correlated. 
%When calculated analytically or bootstrapped, 95\% confidence intervals for the proposed overall effect estimator contain the true effect approximately 95\% of the time. 
}
\label{fig:univariate_OE_cov}
\end{figure}

\begin{figure}[H]
\centerline{\includegraphics[width=\textwidth]{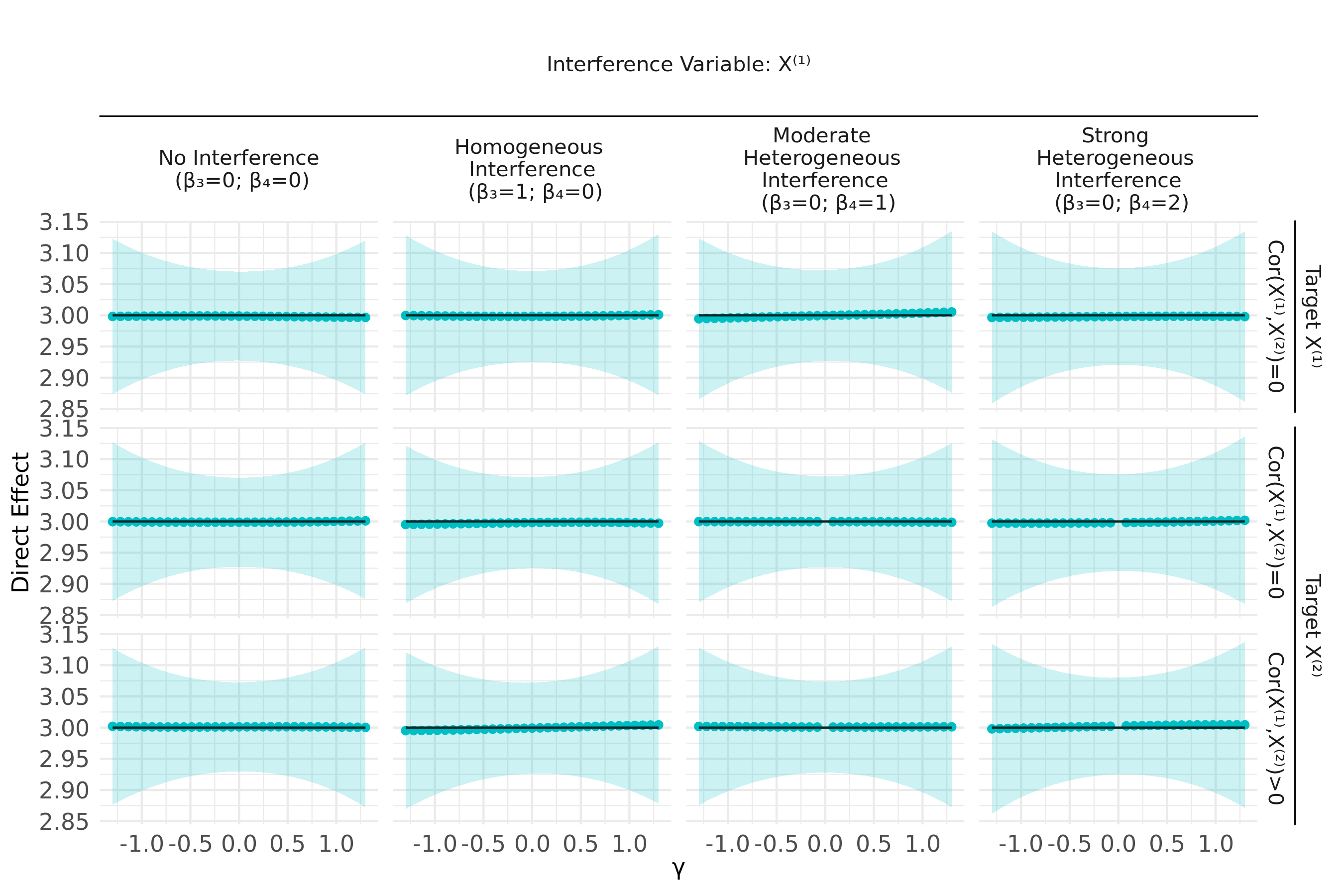}}
\caption{Estimated direct effect under various intervention strategies. See Figure~\ref{fig:univariate_DE_cov} for a detailed description of scenarios. Direct effect is constant in the scenarios we consider.}
\label{fig:univariate_DE_est}
\end{figure}

\begin{figure}[H] 
\centerline{\includegraphics[width=\textwidth]{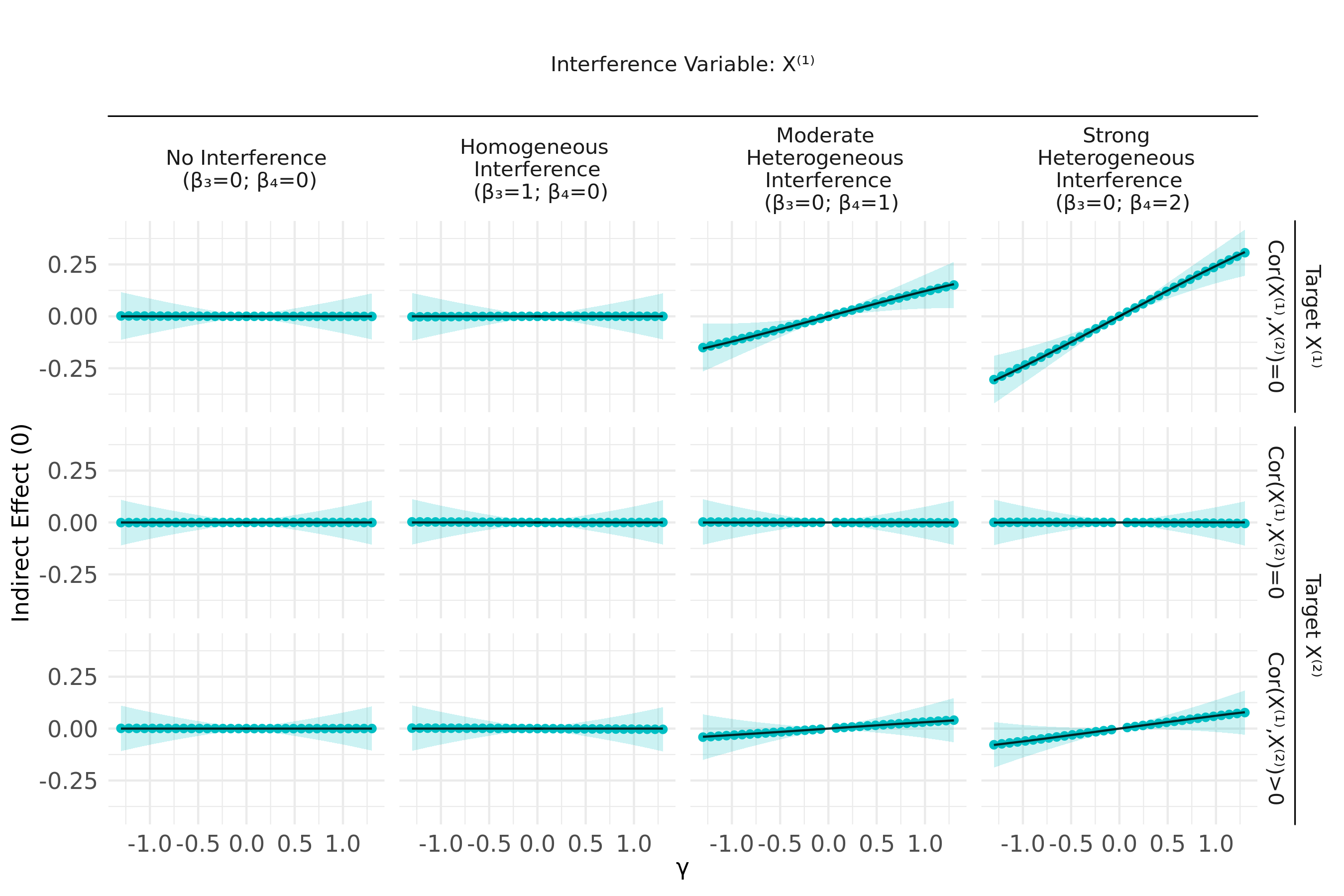}}
\caption{Estimated indirect effect on untreated IE(0) under various intervention strategies. See Figure~\ref{fig:univariate_DE_cov} for a detailed description of scenarios. When there is no interference or only homogeneous interference, IE(0) is constant across counterfactual treatment strategies (columns 1 and 2). When there is heterogeneous interference, counterfactual treatment strategies that assign higher treatment propensity to individuals with certain values of $\mathbf{X}^{(1)}$ (or $\mathbf{X}^{(2)}$ in the case that the two are correlated) have a higher IE(0) (columns 3 and 4).}
\label{fig:univariate_IE0_est}
\end{figure}

\begin{figure}[H]
\centerline{\includegraphics[width=\textwidth]{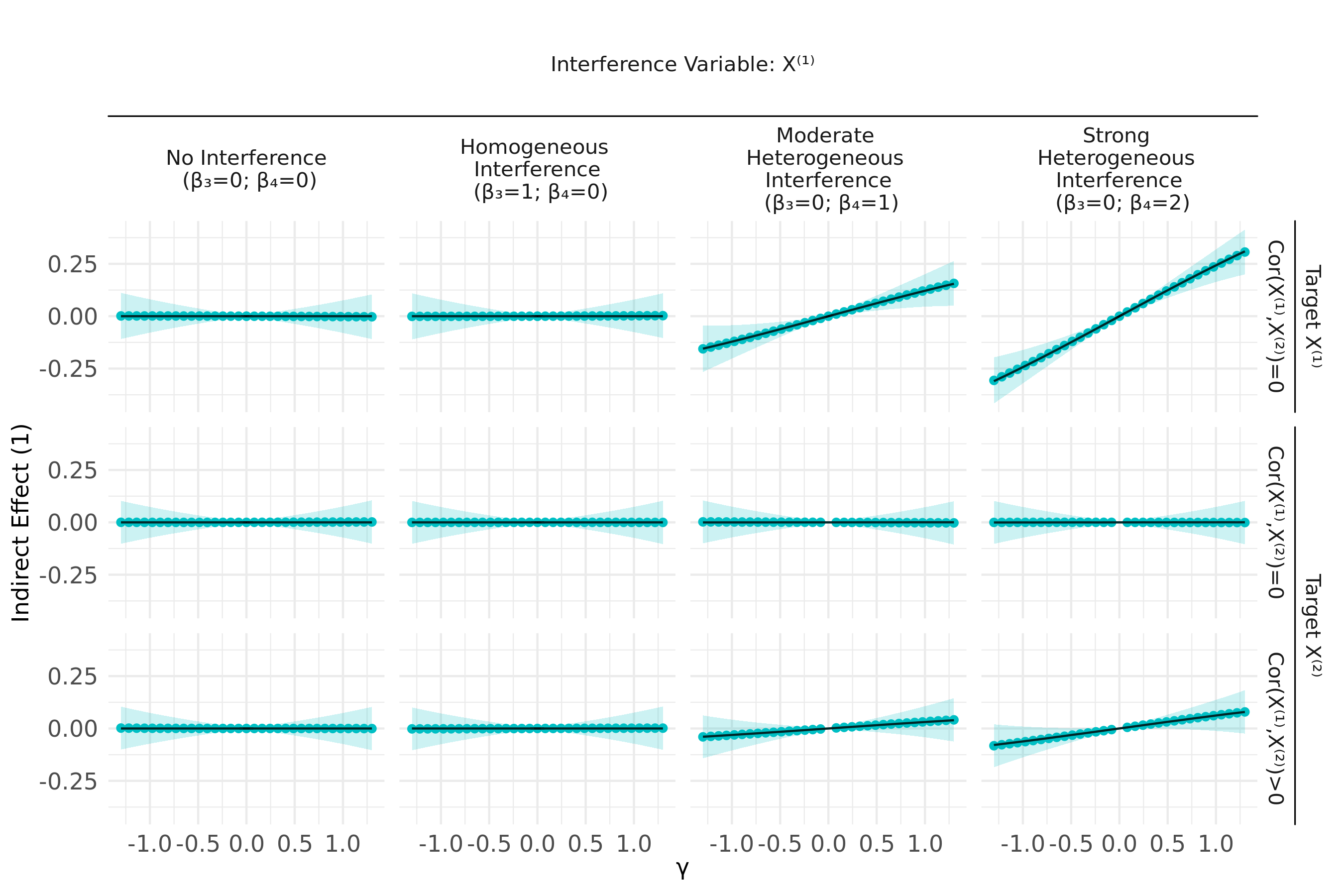}}
\caption{Estimated indirect effect on treated (IE(1)) under various intervention strategies. See Figure~\ref{fig:univariate_DE_cov} for a detailed description of scenarios. When there is no interference or only homogeneous interference, IE(1) is constant across counterfactual treatment strategies (columns 1 and 2). When there is heterogeneous interference, counterfactual treatment strategies that assign higher treatment propensity to individuals with certain values of $\mathbf{X}^{(1)}$ (or $\mathbf{X}^{(2)}$ in the case that the two are correlated) have a higher IE(1) (columns 3 and 4).}
\label{fig:univariate_IE1_est}
\end{figure}

In this appendix, we include supplemental tables and figures for the simulation study where we allow for hypothetical treatment allocation strategies that depend on a single covariate. We present a table of the average bias for direct effect, indirect effect, and overall effect estimators \cref{tab:biastableuni}. We then present figures showing coverage of the analytically calculated and bootstrapped 95\% confidence intervals for each  causal effect(\cref{fig:univariate_DE_cov}, \cref{fig:univariate_IE0_cov}, \cref{fig:univariate_IE1_cov}, \cref{fig:univariate_OE_cov}). Finally, we present figures of the estimated direct effect and indirect effects (\cref{fig:univariate_DE_est}, \cref{fig:univariate_IE0_est}, \cref{fig:univariate_IE1_est}). 
%In \cref{fig:univariate_OE_cov}, the high coverage is evident for no or homogeneous interference settings. When there is heterogeneous interference, this is less obvious. However, overall effects are a contrast of $Y(\gamma) - Y(\gamma_0 = 0)$. When $\gamma$ is very close to 0 the variance shrinks to become infinitesimally small, which makes coverage appear low. When simulation sizes increase towards infinity, these coverages converge to 95\%. In short, there is a numerical explanation for the apparently low coverage when $\gamma$ is close to 0.

\begin{table}[H]
\centering
\caption{This table shows the bias, rounded to the 6th digit, of estimators for direct effect (DE), indirect effect on the treated (IE(1)) and untreated (IE(0)), and overall effect (OE) from the linear model simulation study with univariate interventions where the target variable shows whether an allocation strategy depends on $X^{(1)}$ or $X^{(2)}$. The parameter $\beta_3$ determines homogeneous interference. The parameter $\beta_4$ determines heterogeneous interference related to $\mathbf{X}^{(1)}$ and the parameter $\beta_5$ determines heterogeneous interference related to $\mathbf{X}^{(2)}$ and is fixed at 0. The bias presented is averaged across values of $\gamma$ and $\rho$, and is minimal across tested parameters.}
\label{tab:biastableuni}

\begin{tabular}{|c|c|c|c|c|c|c|c|}
  \hline
$\beta_3$ & $\beta_4$ & $\beta_5$ & Targeted variable & DE bias & IE(0) bias & IE(1) bias & OE bias \\ 
  \hline
0 & 0 & 0 & 0 & -0.001768 & 0.000257 & -0.000252 & -0.000005 \\ 
  0 & 0 & 0 & 0.65 & 0.000711 & -0.000125 & -0.000589 & -0.000258 \\ 
  0 & 1 & 0 & 0 & -0.000315 & 0.000013 & 0.000068 & 0.000251 \\ 
  0 & 1 & 0 & 0.65 & 0.000674 & 0.000351 & 0.000207 & -0.000180 \\ 
  0 & 2 & 0 & 0 & -0.002192 & 0.000391 & 0.000146 & 0.000166 \\ 
  0 & 2 & 0 & 0.65 & 0.003174 & -0.000537 & 0.000186 & -0.000000 \\ 
  1 & 0 & 0 & 0 & -0.001041 & -0.000426 & 0.000327 & -0.000387 \\ 
  1 & 0 & 0 & 0.65 & -0.000333 & -0.000211 & 0.000257 & -0.000240 \\ 
  1 & 1 & 0 & 0 & -0.003137 & 0.000108 & 0.000019 & -0.000447 \\ 
  1 & 1 & 0 & 0.65 & 0.000523 & -0.000184 & -0.000621 & -0.000704 \\ 
  1 & 2 & 0 & 0 & 0.001860 & 0.000130 & 0.000897 & 0.000384 \\ 
  1 & 2 & 0 & 0.65 & -0.001080 & 0.000713 & 0.000209 & 0.001389 \\ 
  \hline
\end{tabular}

\end{table} 

%
%We present a table of the bias of our proposed estimators for direct effect (DE), indirect effect on the untreated (IE(0)), indirect effect on the treated (IE(1)), and overall effect (OE) 
In Table~\ref{tab:biastableuni},  bias is minimal across parameters and estimators. 
%We present figures of the estimated coverage of the 95\% confidence intervals for the same estimators 
In Figures~\ref{fig:univariate_DE_cov}, \ref{fig:univariate_IE0_cov}, \ref{fig:univariate_IE1_cov}, \ref{fig:univariate_OE_cov}, 95\% confidence intervals, calculated analytically or bootstrapped, contain the true effect approximately 95\% of the time, i.e., coverage is approximately 95\%. For overall effects (\cref{fig:univariate_OE_cov}), this is evident for no or homogeneous interference settings. When there is heterogeneous interference, this is less obvious. However, overall effects are a contrast of $\overline{Y}(\alpha, \bm \gamma) - Y(\alpha, \bm  0)$. When $\gamma$ is very close to 0 the variance shrinks to become infinitesimally small, which makes coverage appear low. When simulation sizes increase towards infinity, these coverages converge to 95\%. In short, there is a numerical explanation for the apparently low coverage when $\gamma$ is close to 0.

In the simulated setting, direct effect measures the average effect of treatment on an individual when the intervention strategy for the rest of the cluster is fixed. Because of this, direct effect does not depend on the magnitude of interference or which covariate(s) the intervention targets. As expected in the simple scenarios we simulated, which do not include an interaction between treatment and covariates, the novel estimators return a flat direct effect across the tested intervention strategies (Figure~\ref{fig:univariate_DE_est}).

In the univariate simulation, estimates of indirect effect hold the individual treatment effect constant and contrast average potential outcomes under two different counterfactual treatment strategies for surrounding units. We fix one of these strategies as on that is agnostic to covariates, and index the second strategy by $\gamma$. Because the simulations introduce spillover through variable $\mathbf{X}^{(1)}$ in the scenarios with heterogeneous interference, IE(0) and IE(1) are larger when intervention strategies depend on  $\mathbf{X}^{(1)}$ or a variable correlated with $\mathbf{X}^{(1)}$ (Figures~\ref{fig:univariate_IE0_est}, \ref{fig:univariate_IE1_est}). 

\subsection*{Web appendix D - bivariate intervention plots}
\begin{figure}[H] 
\centerline{\includegraphics[width=\textwidth]{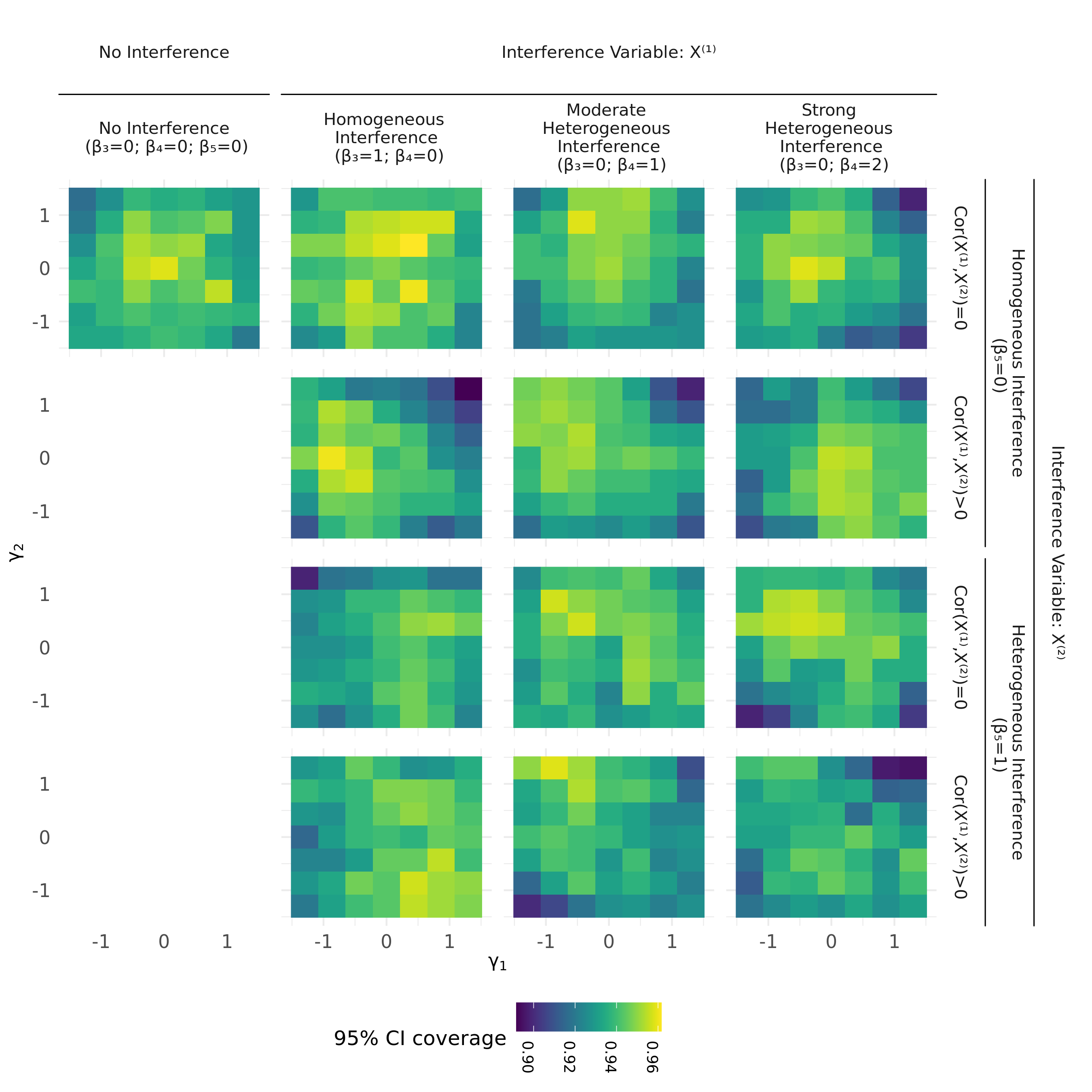}}
\caption{Coverage of estimated direct effects under various bivariate intervention strategies.The columns differentiate between scenarios where $\mathbf{X}^{(1)}$ affects interference in different ways, either no interference, homogeneous interference, moderate heterogeneous interference, or strong heterogeneous interference. The rows differentiate between scenarios where $\mathbf{X}^{(2)}$ affects interference in different ways, either no interference or heterogeneous interference. Within each heatmap, the x- and y-axes determine how units are preferentially treated based on their covariates $X^{(1)}$ and $X^{(2)}$, respectively. A larger value of $\gamma_1$ corresponds to assigning a higher probability of treatment to individuals with larger values of $\mathbf{X}^{(1)}$, and likewise for $\gamma_2$ and $\mathbf{X}^{(2)}$. Coverage is close to the expected 95\% level.}
\label{fig:bivar_DE_cov}
\end{figure}

\begin{figure}[H] 
\centerline{\includegraphics[width=\textwidth]{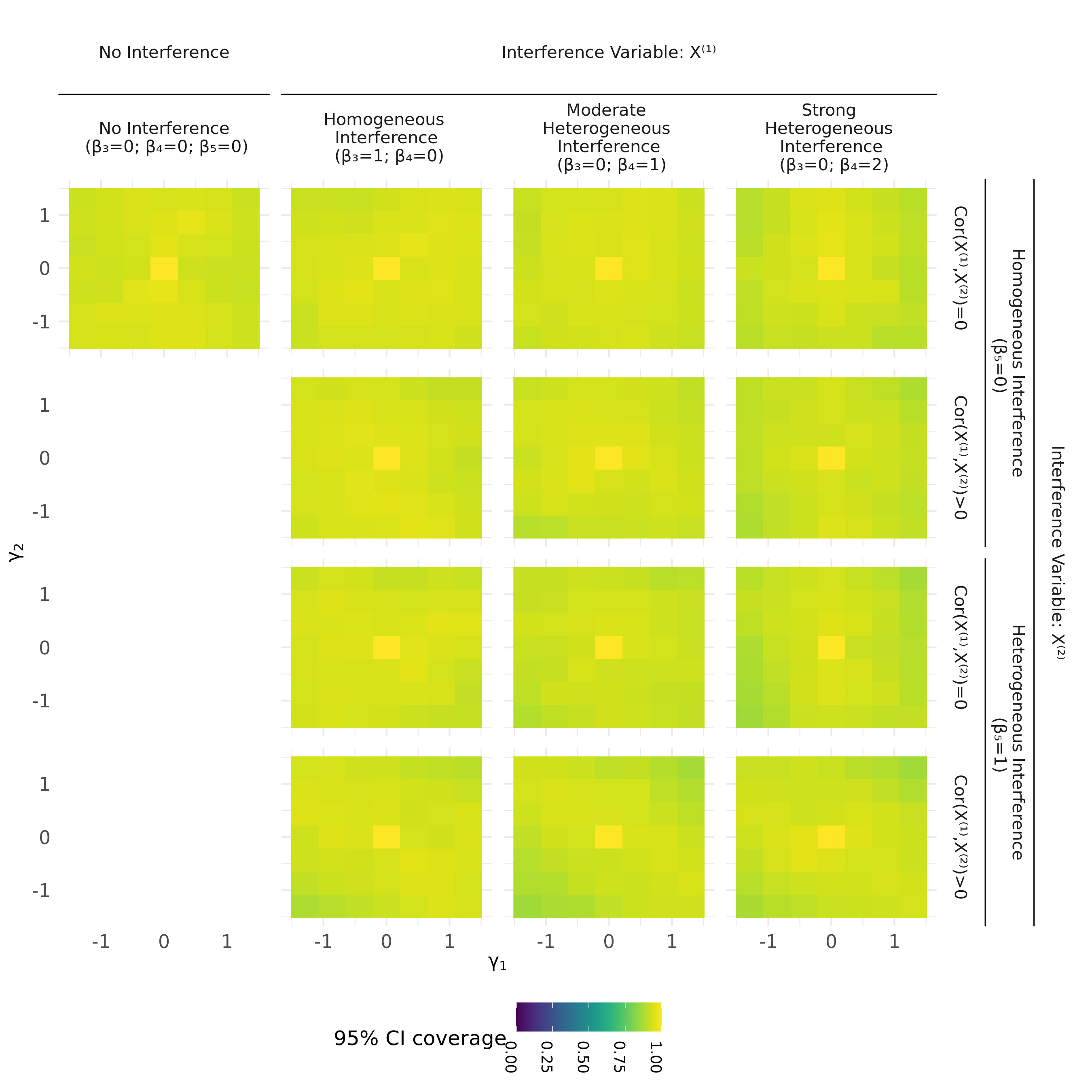}}
\caption{Coverage of estimated indirect effects on untreated under various bivariate intervention strategies. See Figure~\ref{fig:bivar_DE_cov} for a detailed description of scenarios. Coverage is close to the expected 95\% level. }
\label{fig:bivar_IE_cov}
\end{figure}

\begin{figure}[H] 
\centerline{\includegraphics[width=\textwidth]{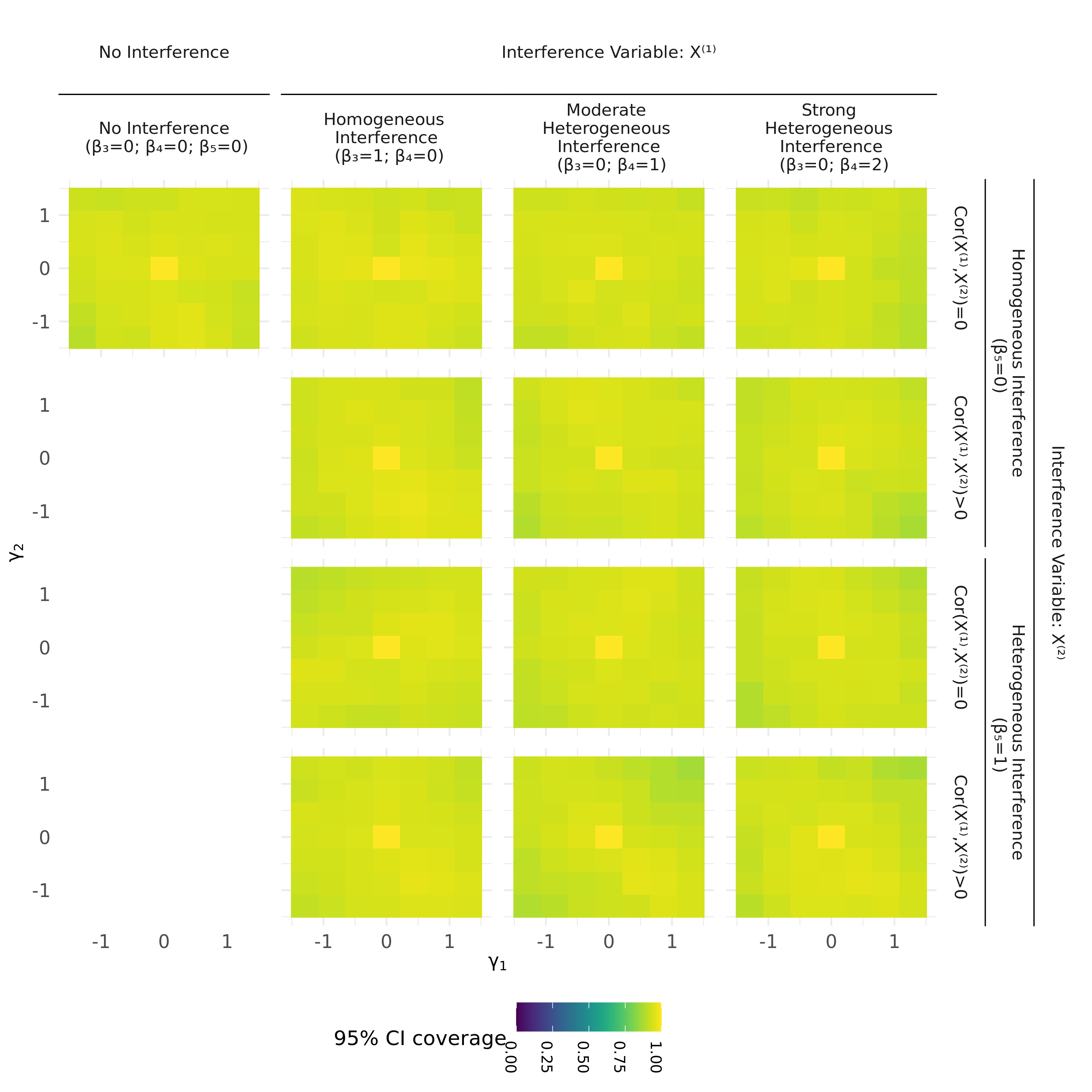}}
\caption{Coverage of estimated indirect effects on treated under various bivariate intervention strategies. See Figure~\ref{fig:bivar_DE_cov} for a detailed description of scenarios. Coverage is close to the expected 95\% level. }
\label{fig:bivar_IE1_cov}
\end{figure}

\begin{figure}[H]
\centerline{\includegraphics[width=\textwidth]{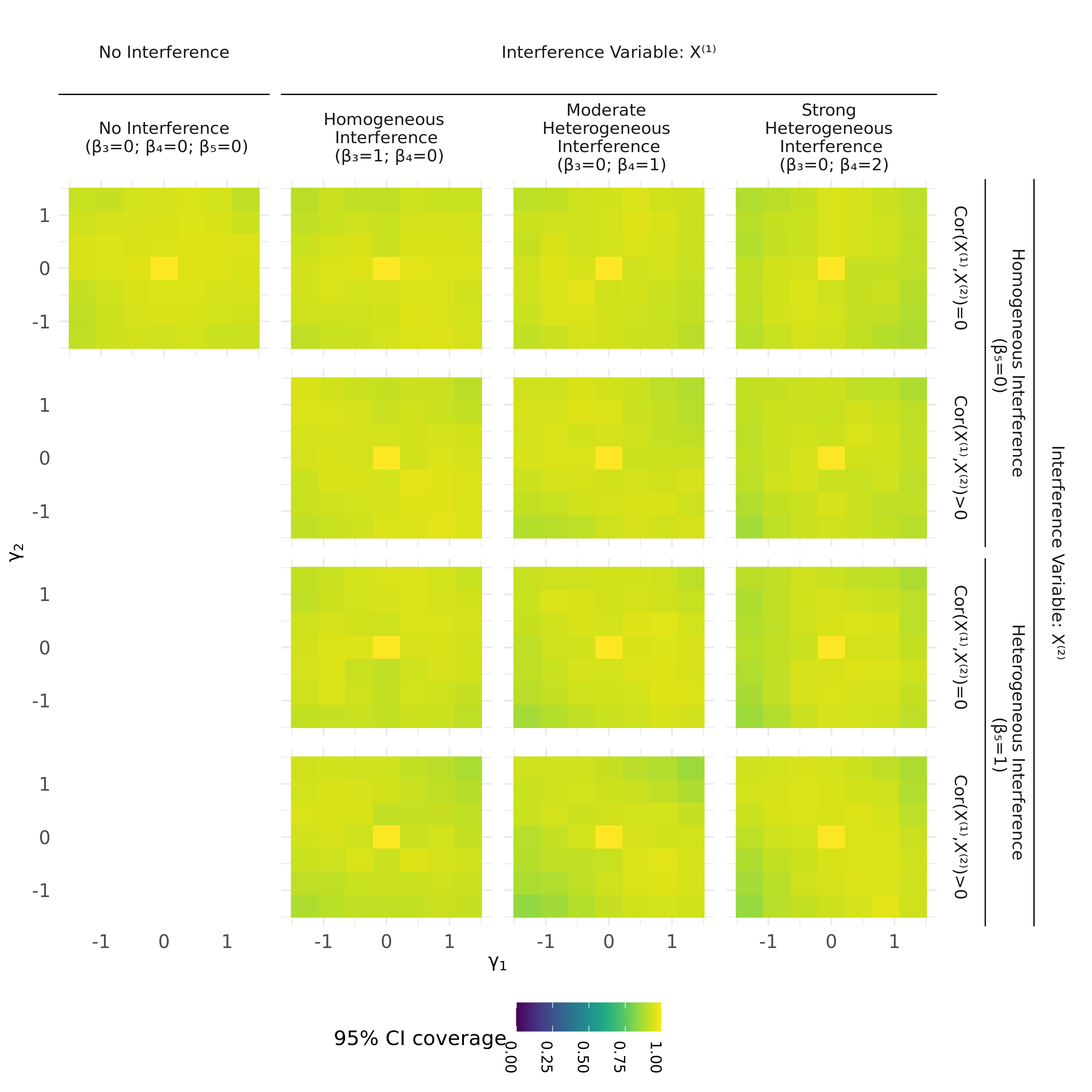}}
\caption{Coverage of confidence intervals for estimated overall effect under bivariate intervention strategies. See Figure~\ref{fig:bivar_DE_cov} for a detailed description of scenarios. Coverage is close to the expected 95\% level. }
\label{fig:bivariate_OE_cov}
\end{figure}

\begin{figure}[H] 
\centerline{\includegraphics[width=\textwidth]{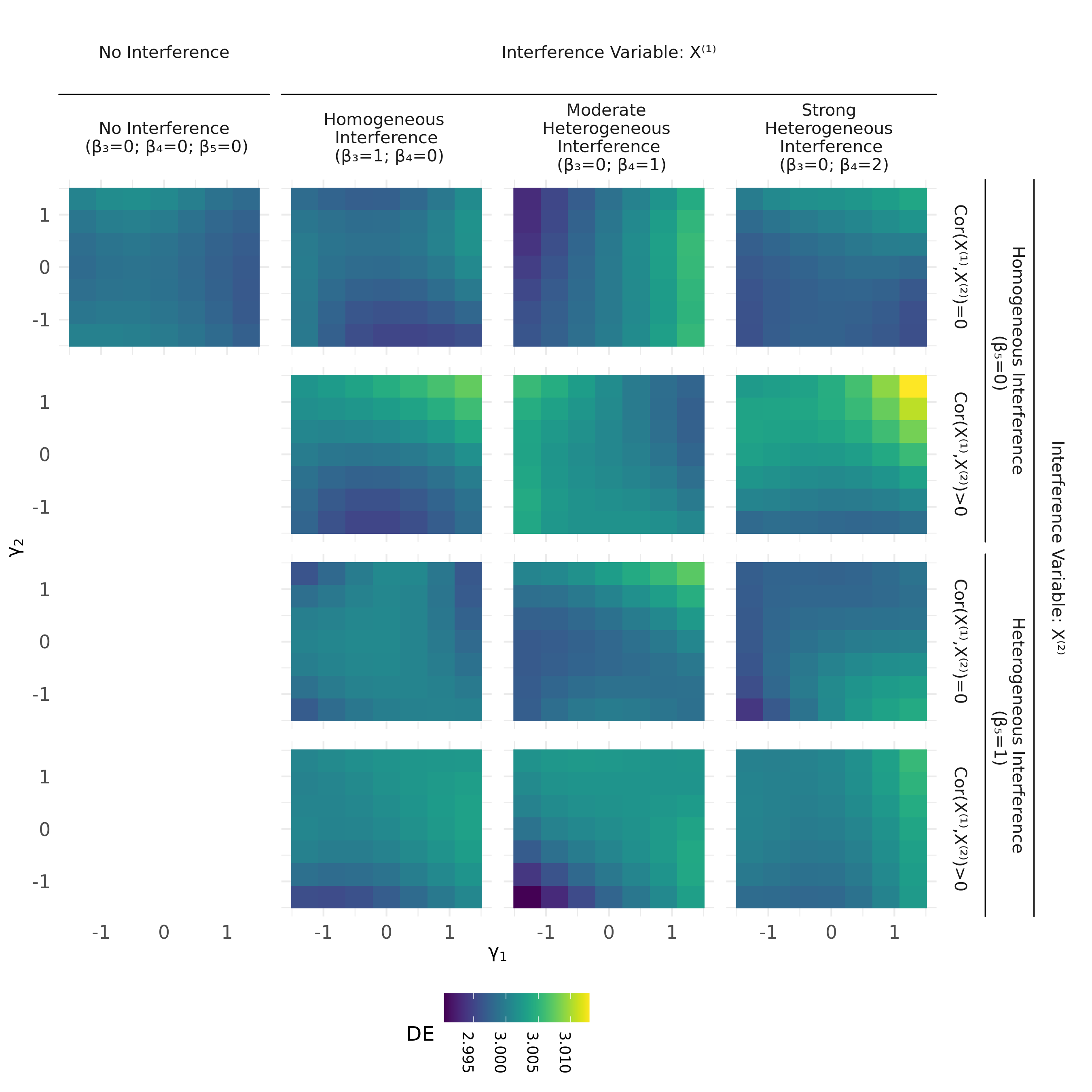}}
\caption{Estimated direct effects under various bivariate intervention strategies. See Figure~\ref{fig:bivar_DE_cov} for a detailed description of scenarios. Direct effect is constant in the scenarios we consider.}
\label{fig:bivar_DE_results}
\end{figure}

\begin{figure}[H]
\centerline{\includegraphics[width=\textwidth]{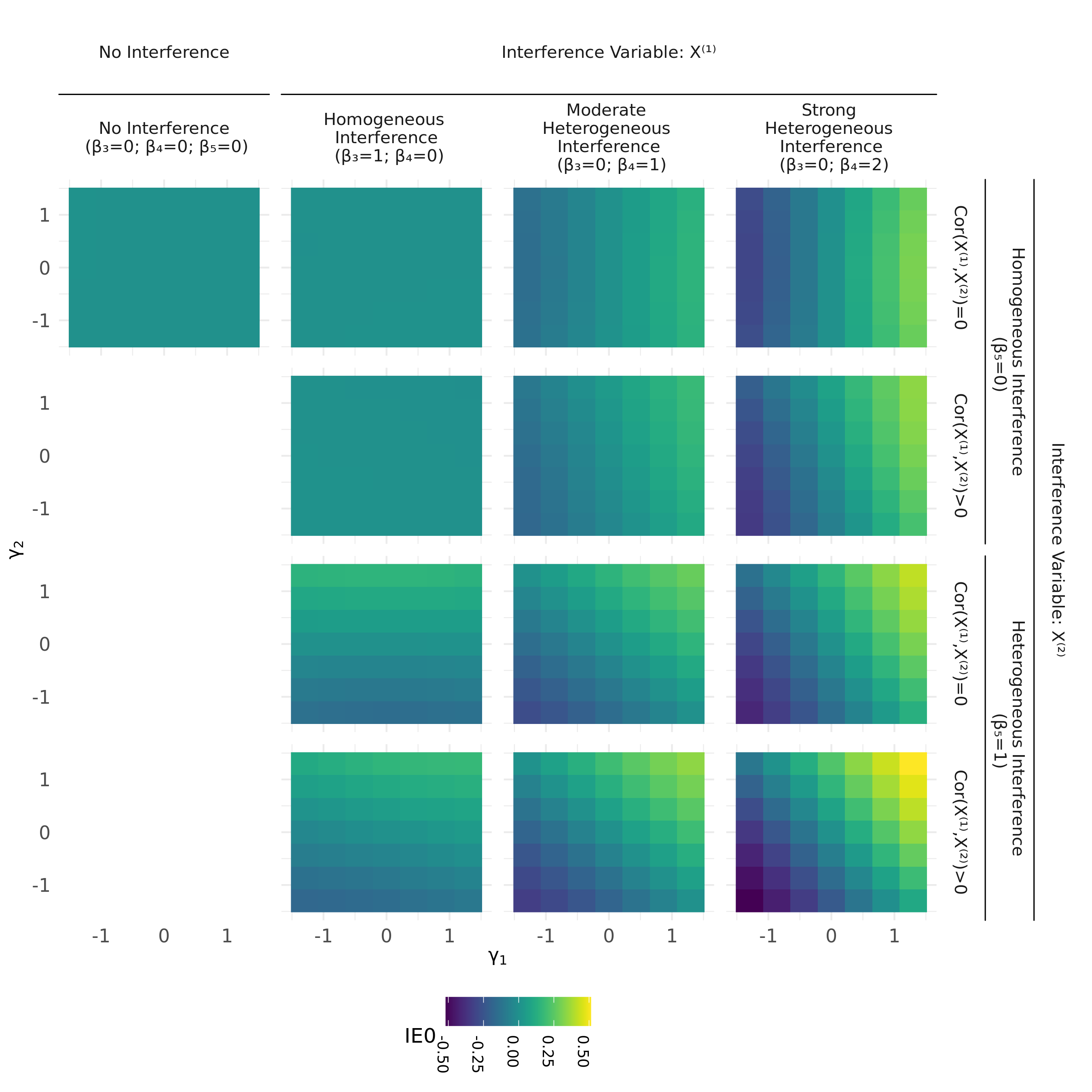}}
\caption{Estimated indirect effects on untreated under various intervention strategies. See Figure~\ref{fig:bivar_DE_cov} for a detailed description of scenarios. When there is no interference or only homogeneous interference in both $\mathbf{X}^{(1)}$ and $\mathbf{X}^{(2)}$, then IE(0) is constant. When there is only heterogeneous interference through $\mathbf{X}^{(1)}$ or $\mathbf{X}^{(2)}$, IE(0) only depends on the $\gamma$ corresponding to the variable related to heterogeneous interference. When there is heterogeneous interference through both $\mathbf{X}^{(1)}$ and $\mathbf{X}^{(2)}$, IE(0) depends on both $\gamma_1$ and $\gamma_2$, such that the largest IE(0) is observed when the counterfactual treatment strategy assigns a high treatment probability to individuals with large values of both covariates.  }
\label{fig:bivar_IE_results}
\end{figure}

\begin{figure}[H] 
\centerline{\includegraphics[width=\textwidth]{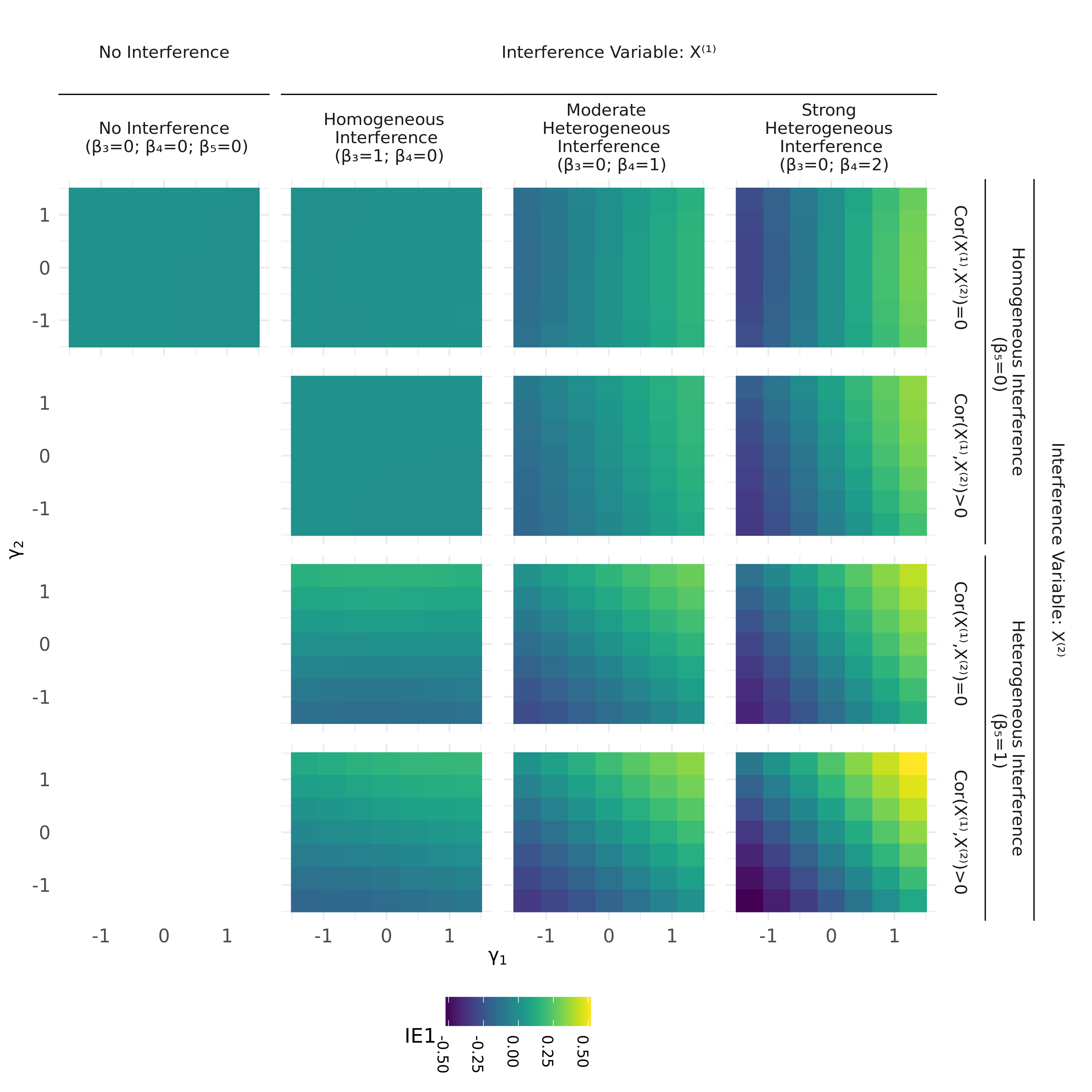}}
\caption{Estimated indirect effects on treated (IE(1)) under various intervention strategies. See Figure~\ref{fig:bivar_DE_cov} for a detailed description of scenarios. When there is no interference or only homogeneous interference in both $\mathbf{X}^{(1)}$ and $\mathbf{X}^{(2)}$, then IE(1) is constant. When there is only heterogeneous interference through $\mathbf{X}^{(1)}$ or $\mathbf{X}^{(2)}$, IE(1) only depends on the $\gamma$ corresponding to the variable related to heterogeneous interference. When there is heterogeneous interference through both $\mathbf{X}^{(1)}$ and $\mathbf{X}^{(2)}$, IE(1) depends on both $\gamma_1$ and $\gamma_2$, such that the largest IE(1) is observed when the counterfactual treatment strategy assigns a high treatment probability to individuals with large values of both covariates. }
\label{fig:bivar_IE1_results}
\end{figure}

In this appendix, we include supplemental tables and figures for the simulation study where we allow for interventions that depend on two covariates. We present a table of the average bias for direct effect, indirect effect, and overall effect estimators \cref{tab:biastable}. We then present figures showing coverage of the analytically calculated 95\% confidence intervals for each of the same causal effects (\cref{fig:bivar_DE_cov}, \cref{fig:bivar_IE_cov}, \cref{fig:bivar_IE1_cov}, \cref{fig:bivariate_OE_cov}). Finally, we present figures of the estimated direct effect and indirect effects (\cref{fig:bivar_DE_results}, \cref{fig:bivar_IE_results}, \cref{fig:bivar_IE1_results}). 

\begin{table}[H]
\centering
\caption{This table shows the bias, rounded to the 6th digit, of estimators for direct effect (DE), indirect effect on the treated (IE(1)) and untreated (IE(0)), and overall effect (OE) from the linear model simulation study with bivariate interventions. The parameter $\beta_3$ determines homogeneous interference. The parameter $\beta_4$ determines heterogeneous interference related to $\mathbf{X}^{(1)}$ and the parameter $\beta_5$ determines heterogeneous interference related to $\mathbf{X}^{(2)}$. The bias presented is averaged across values of $\gamma_1$ and $\gamma_2$, and is small for all estimators and sets of parameters.}
\label{tab:biastable}
\begin{tabular}{|c|c|c|c|c|c|c|c|}
  \hline
$\beta_3$ & $\beta_4$ & $\beta_5$ & $P(X^{(1)}_{ij} = X^{(2)}_{ij})$ & DE bias & IE(0) bias & IE(1) bias & OE bias \\ 
  \hline
0 & 0 & 0 & 0 & -0.001101 & -0.000256 & -0.000097 & 0.000126 \\ 
  0 & 0 & 0 & 0.65 & 0.001075 & 0.000108 & 0.000007 & -0.001813 \\ 
  0 & 0 & 1 & 0 & -0.002667 & 0.003694 & 0.001770 & 0.003447 \\ 
  0 & 0 & 1 & 0.65 & -0.003134 & 0.003050 & 0.001210 & 0.002843 \\ 
  0 & 1 & 0 & 0 & -0.000400 & 0.002159 & 0.002129 & 0.002127 \\ 
  0 & 1 & 0 & 0.65 & 0.001770 & -0.000802 & 0.000150 & -0.000814 \\ 
  0 & 1 & 1 & 0 & -0.001646 & 0.000685 & 0.001085 & 0.002460 \\ 
  0 & 1 & 1 & 0.65 & -0.001370 & 0.005661 & 0.002915 & 0.003270 \\ 
  0 & 2 & 0 & 0 & -0.003049 & 0.001406 & 0.000303 & -0.000584 \\ 
  0 & 2 & 0 & 0.65 & -0.000203 & 0.001071 & -0.001584 & -0.000827 \\ 
  0 & 2 & 1 & 0 & -0.000029 & 0.002219 & 0.002828 & 0.001438 \\ 
  0 & 2 & 1 & 0.65 & -0.000462 & 0.001831 & 0.001461 & 0.003210 \\ 
  1 & 0 & 0 & 0 & -0.003031 & 0.001650 & 0.000413 & 0.001013 \\ 
  1 & 0 & 0 & 0.65 & -0.002984 & 0.001339 & -0.000844 & 0.001057 \\ 
  1 & 0 & 1 & 0 & -0.000044 & 0.002105 & 0.001091 & 0.001108 \\ 
  1 & 0 & 1 & 0.65 & -0.000991 & 0.001191 & -0.000832 & 0.001242 \\ 
  1 & 1 & 0 & 0 & -0.005335 & 0.000665 & -0.001622 & 0.001245 \\ 
  1 & 1 & 0 & 0.65 & -0.001601 & 0.001130 & -0.001431 & -0.001559 \\ 
  1 & 1 & 1 & 0 & -0.000021 & 0.000915 & 0.000603 & 0.000208 \\ 
  1 & 1 & 1 & 0.65 & -0.002448 & 0.002004 & 0.001365 & 0.003805 \\ 
  1 & 2 & 0 & 0 & 0.001282 & 0.000504 & 0.000692 & 0.001231 \\ 
  1 & 2 & 0 & 0.65 & -0.002611 & 0.005719 & 0.003683 & 0.005333 \\ 
  1 & 2 & 1 & 0 & -0.000840 & 0.001860 & 0.002996 & 0.002862 \\ 
  1 & 2 & 1 & 0.65 & -0.001080 & 0.002897 & 0.002467 & 0.003259 \\ 
   \hline
\end{tabular}

\end{table} 

We present a table of the bias of our proposed estimators for direct effect (DE), indirect effect on the untreated (IE(0)), indirect effect on the treated (IE(1)), and overall effect (OE) (Table~\ref{tab:biastableuni}). Bias is minimal across parameters and estimators. We present figures of the estimated coverage of the 95\% confidence intervals for the same estimators (Figures~\ref{fig:univariate_DE_cov}, \ref{fig:univariate_IE0_cov}, \ref{fig:univariate_IE1_cov}, \ref{fig:univariate_OE_cov}). Coverage is approximately 95\%. 

In the bivariate simulation, estimates of indirect effect hold the individual treatment effect constant and contrast average potential outcomes under two different counterfactual treatment strategies for surrounding units. These counterfactual treatment strategies assign individual treatment propensity dependent on an individual's values of covariates $\mathbf{X}^{(1)}$ and $\mathbf{X}^{(2)}$.  We fix one of the contrasted strategies to be agnostic to covariates, and index the second strategy by $\gamma_1$ and $\gamma_2$. Direct effect (DE) is approximately constant across all simulations because we do not vary parameters affecting direct effect (Figure~\ref{fig:bivar_DE_results}). 
Indirect effect on the untreated (IE(0)) and the treated (IE(1)) increase when there is heterogeneous interference through the variable targeted by the counterfactual intervention strategy, as in the univariate case (Figures~\ref{fig:bivar_IE_results}, \ref{fig:bivar_IE1_results}). 

\subsection*{Web appendix E - diffusion plots}
\begin{figure}[H]
\centerline{\includegraphics[width=.5\textwidth]{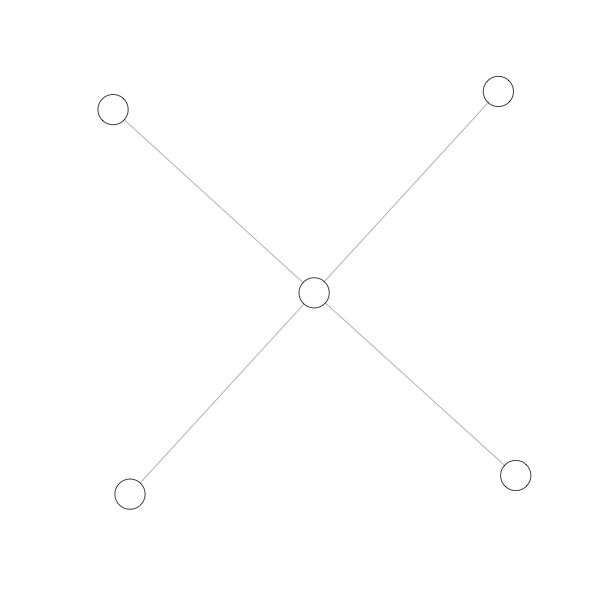}}
\caption{Cluster structure for Scenario 2. The structure is a simple star network with one central unit and four alters.}
\label{fig:star}
\end{figure}

\begin{figure}[H]
\centerline{\includegraphics[width=\textwidth]{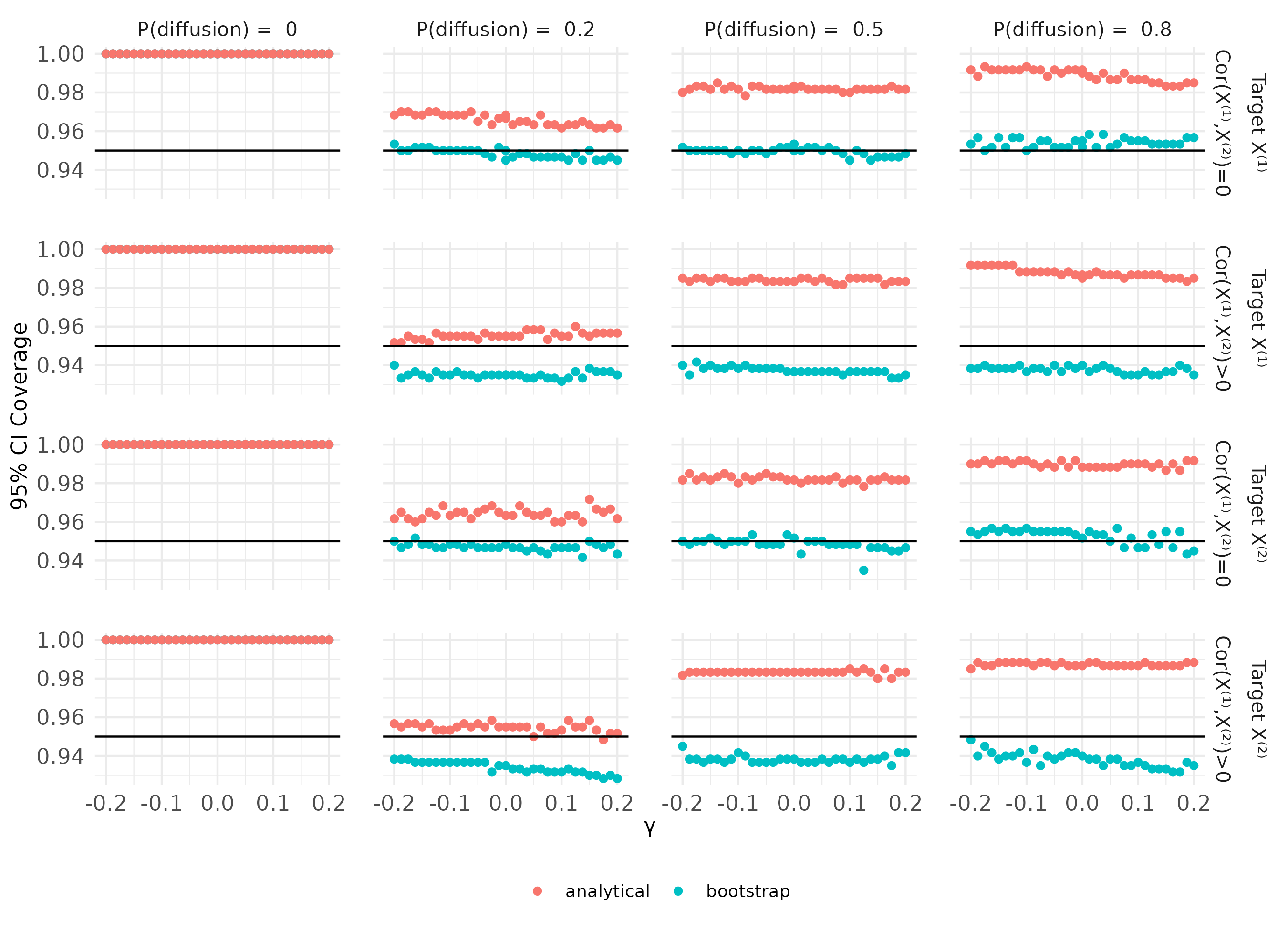}}
\caption{Coverage of 95\% confidence intervals of direct effect for diffusion scenario. Columns show different levels of diffusion. In this setting, no diffusion is equivalent to no heterogeneous interference, and increasing diffusion is equivalent to increasing heterogeneous interference. The top two rows show interventions where counterfactual treatment propensities depend on $\mathbf{X}^{(1)}$, here an indicator for network centrality, and the bottom two rows show interventions where counterfactual treatment propensities depend on $\mathbf{X}^{(2)}$. Across scenarios, coverage is close to or greater than the expected 95\%.}
\label{fig:diffusion_DE_coverage}
\end{figure}

\begin{figure}[H]
\centerline{\includegraphics[width=\textwidth]{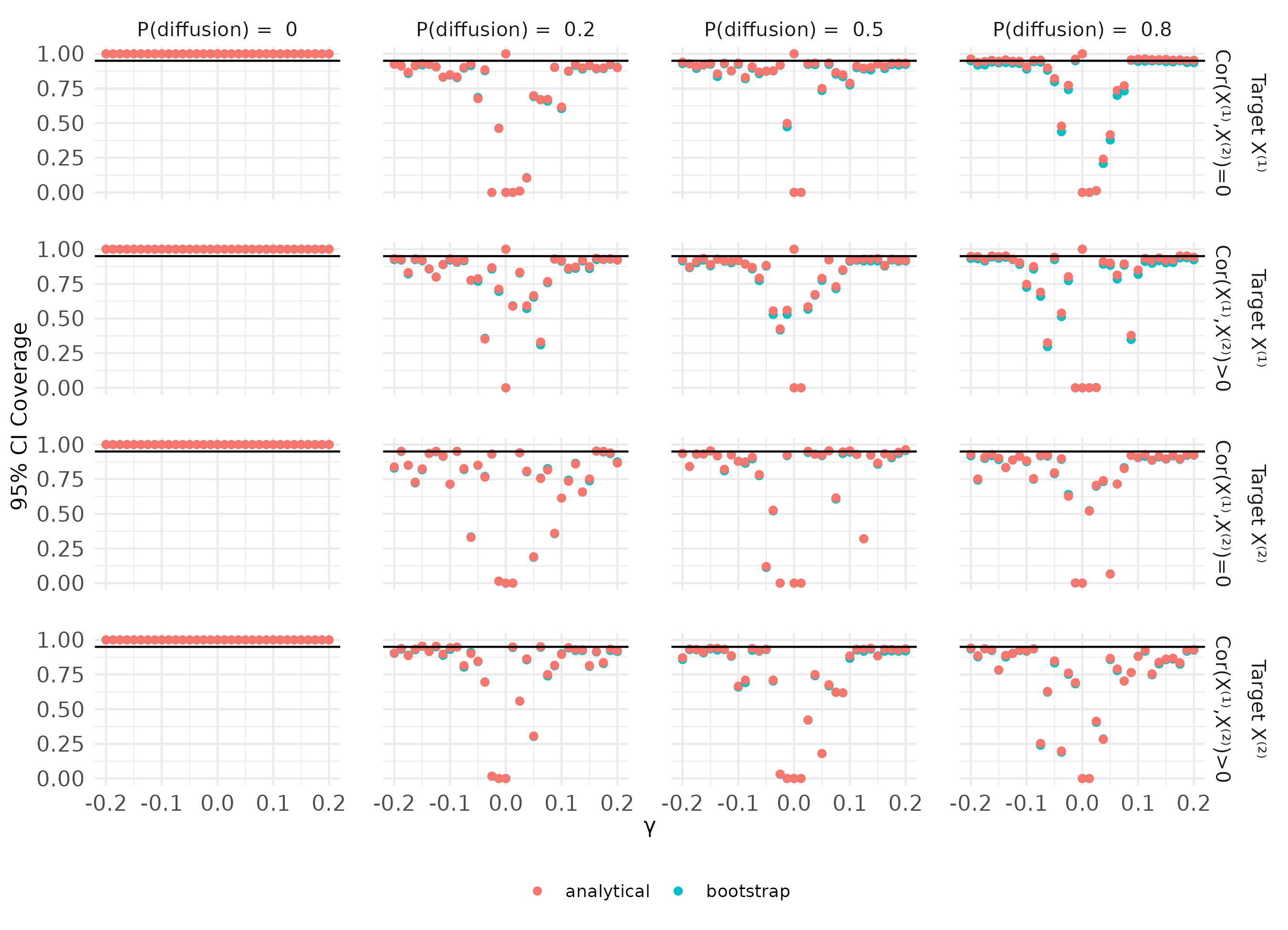}}
\caption{Coverage of 95\% confidence intervals of indirect effect on the untreated for diffusion scenario. See Figure~\ref{fig:diffusion_DE_coverage} for a detailed description of scenarios. Across scenarios, coverage is close to or greater than the expected 95\%. }
\label{fig:diffusion_IE0_coverage}
\end{figure}

\begin{figure}[H] 
\centerline{\includegraphics[width=\textwidth]{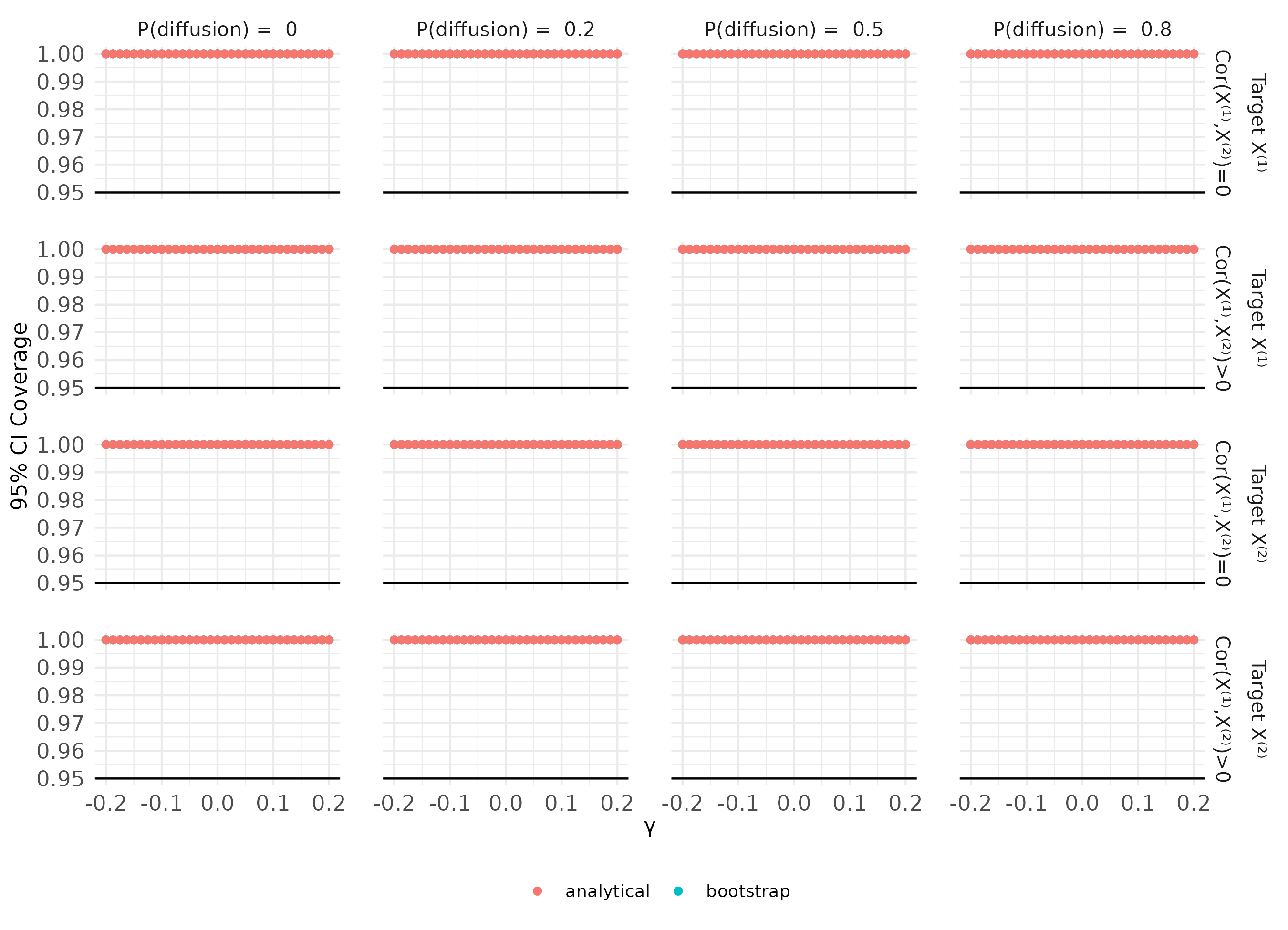}}
\caption{Coverage of 95\% confidence intervals of indirect effect on the treated for diffusion scenario. See Figure~\ref{fig:diffusion_DE_coverage} for a detailed description of scenarios. Coverage is 100\% because IE(1) is fixed at 0 in these scenarios.}
\label{fig:diffusion_IE1_coverage}
\end{figure}

\begin{figure}[H] 
\centerline{\includegraphics[width=\textwidth]{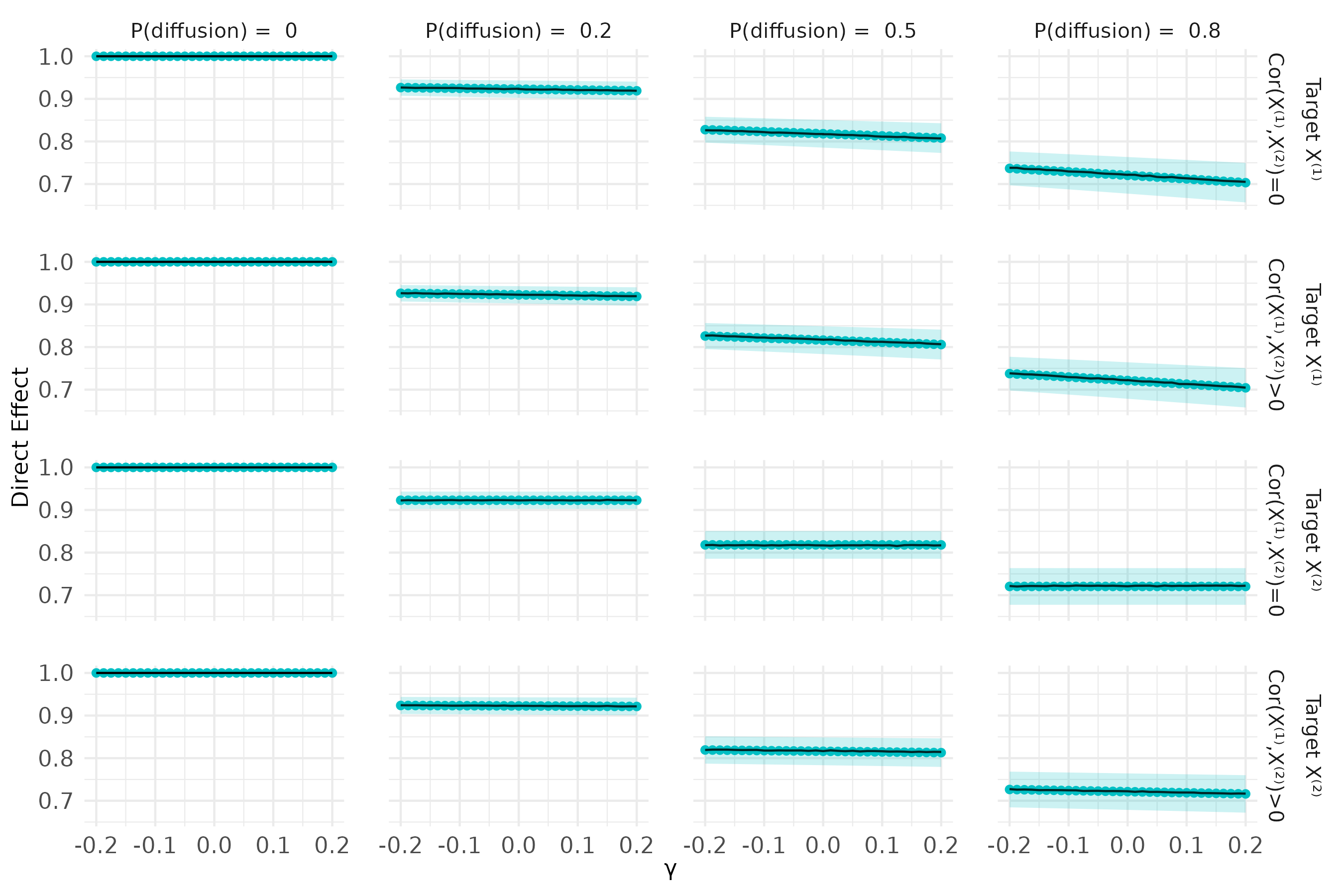}}
\caption{Estimated direct effect (DE) for diffusion scenario. See Figure~\ref{fig:diffusion_DE_coverage} for a detailed description of scenarios. The black lines shows the true value of DE. As diffusion increases, DE tends to decrease. In scenarios with diffusion, counterfactual treatment strategies with a larger treatment propensity for central units have larger DE than counterfactual treatment strategies that assign a smaller treatment propensity to central units.}
\label{fig:diffusion_DE_results}
\end{figure}

\begin{figure}[H] 
\centerline{\includegraphics[width=\textwidth]{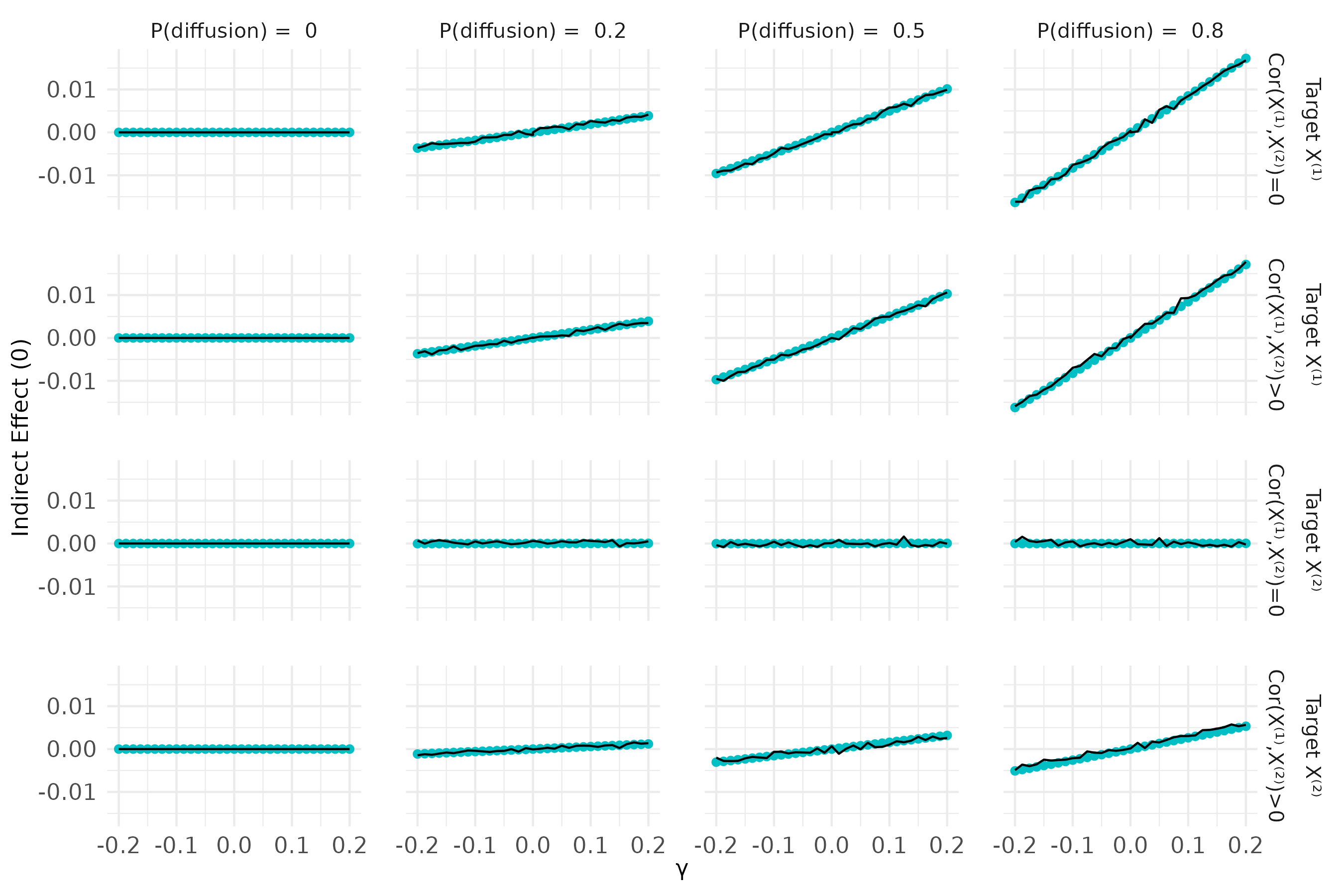}}
\caption{Estimated indirect effect on the untreated (IE(0)) for diffusion scenario. See Figure~\ref{fig:diffusion_DE_coverage} for a detailed description of scenarios. When there is no diffusion, IE(0) is constant regardless of how treatment propensities depend on a unit's centrality. When there is diffusion, there is a larger IE(0) for counterfactual treatment strategies that assign larger treatment propensity to central units. }
\label{fig:diffusion_IE0_results}
\end{figure}

\begin{figure}[H]
\centerline{\includegraphics[width=\textwidth]{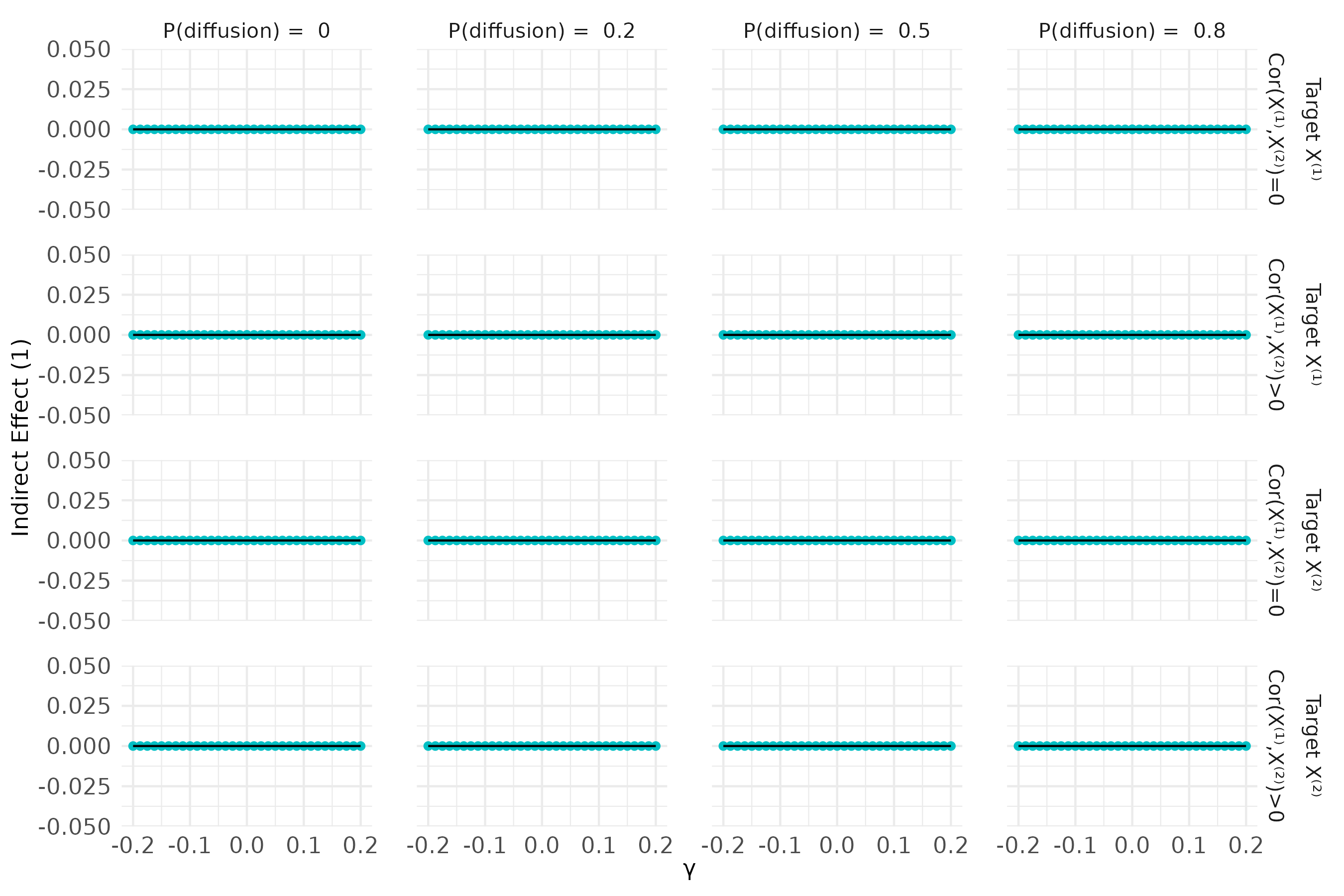}}
\caption{Estimated indirect effect on the treated (IE(1)) for diffusion scenario. See Figure~\ref{fig:diffusion_DE_coverage} for a detailed description of scenarios. When there is no diffusion, IE(1) is constant regardless of how treatment propensities depend on a unit's centrality. When there is diffusion, there is a larger IE(1) for counterfactual treatment strategies that assign larger treatment propensity to central units. }
\label{fig:diffusion_IE1_results}
\end{figure}

This appendix includes additional details about the set of simulations for Scenario 2, where interference occurs through a diffusion process. We show a figure of the network structure used in these simulations (\cref{fig:star}); a table of bias for causal estimators (\cref{tab:biastablediff}); figures for coverage of 95\% confidence intervals (\cref{fig:diffusion_DE_coverage}, \cref{fig:diffusion_IE0_coverage}, \cref{fig:diffusion_IE1_coverage}); and figures of estimated direct effect and indirect effects (\cref{fig:diffusion_DE_results}, \cref{fig:diffusion_IE0_results}, \cref{fig:diffusion_IE1_results}).

\begin{table}[H]
\centering
\caption{This table shows the bias, rounded to the 6th digit, of estimators for direct effect (DE), indirect effect on the treated (IE(1)) and untreated (IE(0)), and overall effect (OE) from the diffusion simulation study. The bias presented is averaged across values of $\gamma$. The variable $\mathbf{X}^{(1)}$ is an indicator of whether a unit is a central node or a peripheral node. Bias for IE(1) is fixed at 0 because IE(1) is fixed.}
% latex table generated in R 4.2.0 by xtable 1.8-4 package
% Tue Mar 26 11:11:27 2024
\begin{tabular}{|c|c|c|c|c|c|c|}
  \hline
P(diffusion) & $P(X^{(1)}_{ij} = X^{(2)}_{ij})$ & Targeted & DE bias & IE(0) bias& IE(1) bias& OE bias \\ 
&& variable &&&&\\
  \hline
0 & 0 &  $\mathbf{X}^{(1)}$ & 0.000000 & 0.000000 & 0.000000 & 0.000387 \\ 
  0 & 0 &  $\mathbf{X}^{(2)}$ & 0.000000 & 0.000000 & 0.000000 & 0.000468 \\ 
  0 & 0.65 &  $\mathbf{X}^{(1)}$ & 0.000000 & 0.000000 & 0.000000 & 0.000009 \\ 
  0 & 0.65 &  $\mathbf{X}^{(2)}$ & 0.000000 & 0.000000 & 0.000000 & -0.000103 \\ 
  0.2 & 0 &  $\mathbf{X}^{(1)}$ & -0.000096 & -0.000172 & 0.000000 & -0.000771 \\ 
  0.2 & 0 &  $\mathbf{X}^{(2)}$ & 0.000011 & -0.000279 & 0.000000 & -0.000721 \\ 
  0.2 & 0.65 &  $\mathbf{X}^{(1)}$ & -0.000384 & 0.000068 & 0.000000 & 0.000039 \\ 
  0.2 & 0.65 &  $\mathbf{X}^{(2)}$ & -0.000293 & -0.000023 & 0.000000 & 0.000079 \\ 
  0.5 & 0 &  $\mathbf{X}^{(1)}$ & 0.000984 & 0.000061 & 0.000000 & -0.000312 \\ 
  0.5 & 0 &  $\mathbf{X}^{(2)}$ & 0.000896 & 0.000149 & 0.000000 & -0.000501 \\ 
  0.5 & 0.65 &  $\mathbf{X}^{(1)}$ & -0.001184 & 0.000110 & 0.000000 & 0.000126 \\ 
  0.5 & 0.65 &  $\mathbf{X}^{(2)}$ & -0.001146 & 0.000072 & 0.000000 & 0.000005 \\ 
  0.8 & 0 &  $\mathbf{X}^{(1)}$ & -0.001387 & -0.000049 & 0.000000 & 0.000621 \\ 
  0.8 & 0 &  $\mathbf{X}^{(2)}$ & -0.001354 & -0.000082 & 0.000000 & 0.000869 \\ 
  0.8 & 0.65 &  $\mathbf{X}^{(1)}$ & -0.000528 & -0.000507 & 0.000000 & -0.000143 \\ 
  0.8 & 0.65 &  $\mathbf{X}^{(2)}$ & -0.000432 & -0.000603 & 0.000000 & -0.000033 \\ 
   \hline
\end{tabular}

\label{tab:biastablediff}
\end{table}
\begin{figure}[H] 
\centerline{\includegraphics[width=\textwidth]{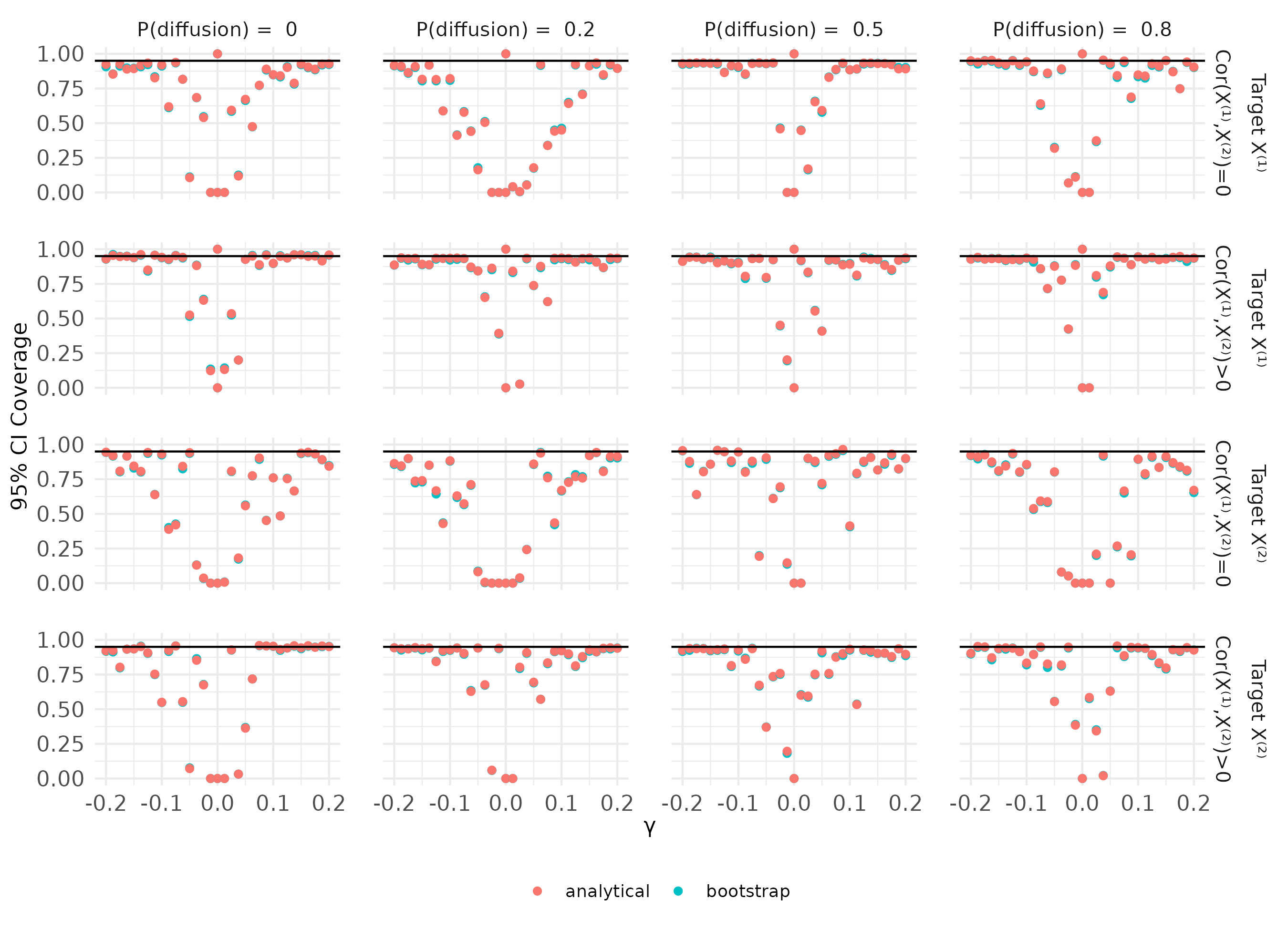}}
\caption{Coverage of estimated overall effect in a treatment diffusion scenario. See Figure~\ref{fig:diffusion_DE_coverage} for a detailed description of scenarios. Across scenarios, coverage is close to or greater than the expected 95\%.}
\label{fig:diff_oe_cov}
\end{figure}
We present a table of the bias of our proposed estimators for direct effect (DE), indirect effect on the untreated (IE(0)), indirect effect on the treated (IE(1)), and overall effect (OE) (Table~\ref{tab:biastablediff}). Bias is minimal across parameters and estimators. We present figures of the estimated coverage of the 95\% confidence intervals for the same estimators (Figures~\ref{fig:diffusion_DE_coverage}, \ref{fig:diffusion_IE0_coverage}, \ref{fig:diffusion_IE1_coverage}, \ref{fig:diff_oe_cov}). Coverage is approximately 95\%. 

In diffusion scenarios, the direct effect (DE) contrasts the average potential outcomes for a treated individual against an untreated individual. When there is no diffusion, a treated individual always has an outcome of 1 and an untreated individual always has an outcome of 0. When there is diffusion, a treated individual will still always have an outcome of 1, but an untreated individual can now have an outcome greater than 0 if the treatment of another unit diffuses. Therefore, the difference between these two average potential outcomes, and therefore DE, decreases and diffusion increases (Figure~\ref{fig:diffusion_DE_results}). When there is more diffusion there are larger indirect effects on the untreated when targeting treatment towards central individuals in each cluster (Figure~\ref{fig:bivar_IE_results}). Because all treated individuals have a fixed outcome of, there is no indirect effect on the treated individuals (Figure~\ref{fig:bivar_IE1_results}). 

\subsection*{Web appendix F - application plots}
\begin{figure}[t]
    \centering
    \includegraphics[width=0.8\textwidth]{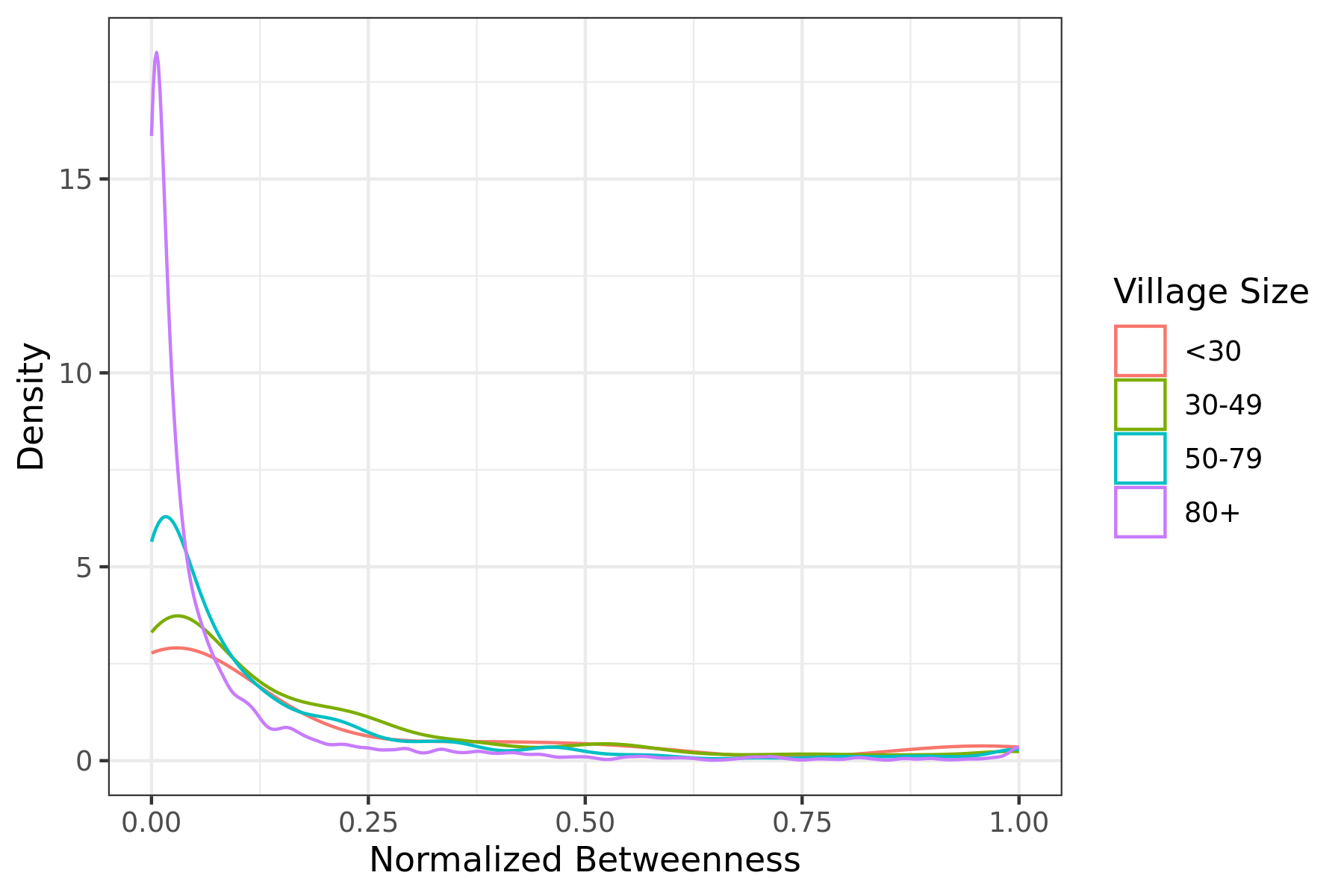}
    \caption{The distribution of normalized betweenness, stratified by village size.}
    \label{fig:betweenness}
\end{figure}

\begin{figure}[H]
    \centering
    \includegraphics[width=0.9\textwidth]{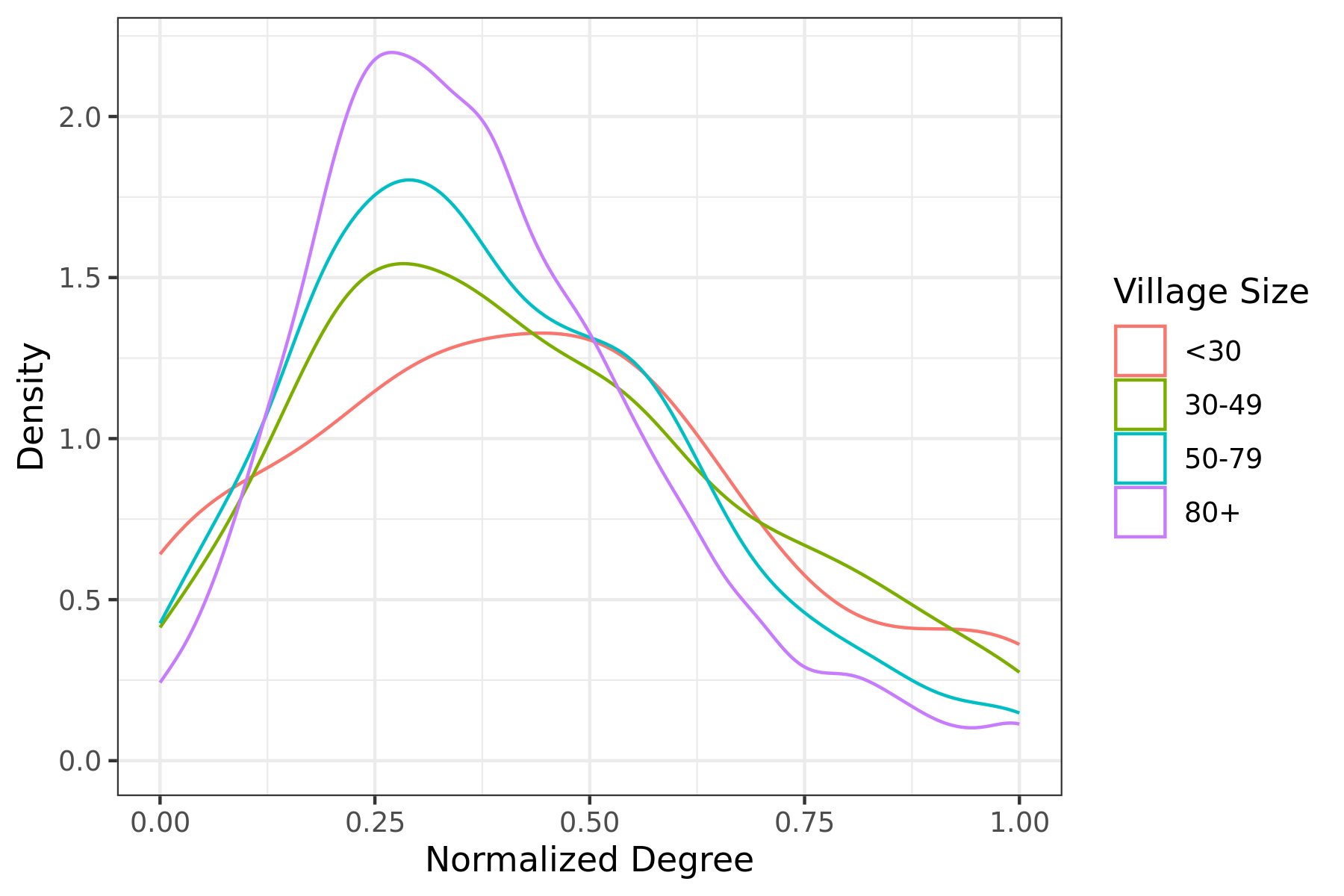}
    \caption{The distribution of normalized degree, stratified by village size.}
    \label{fig:degree}
\end{figure}

\begin{figure}[t]
\centerline{\includegraphics[width=0.9\textwidth]{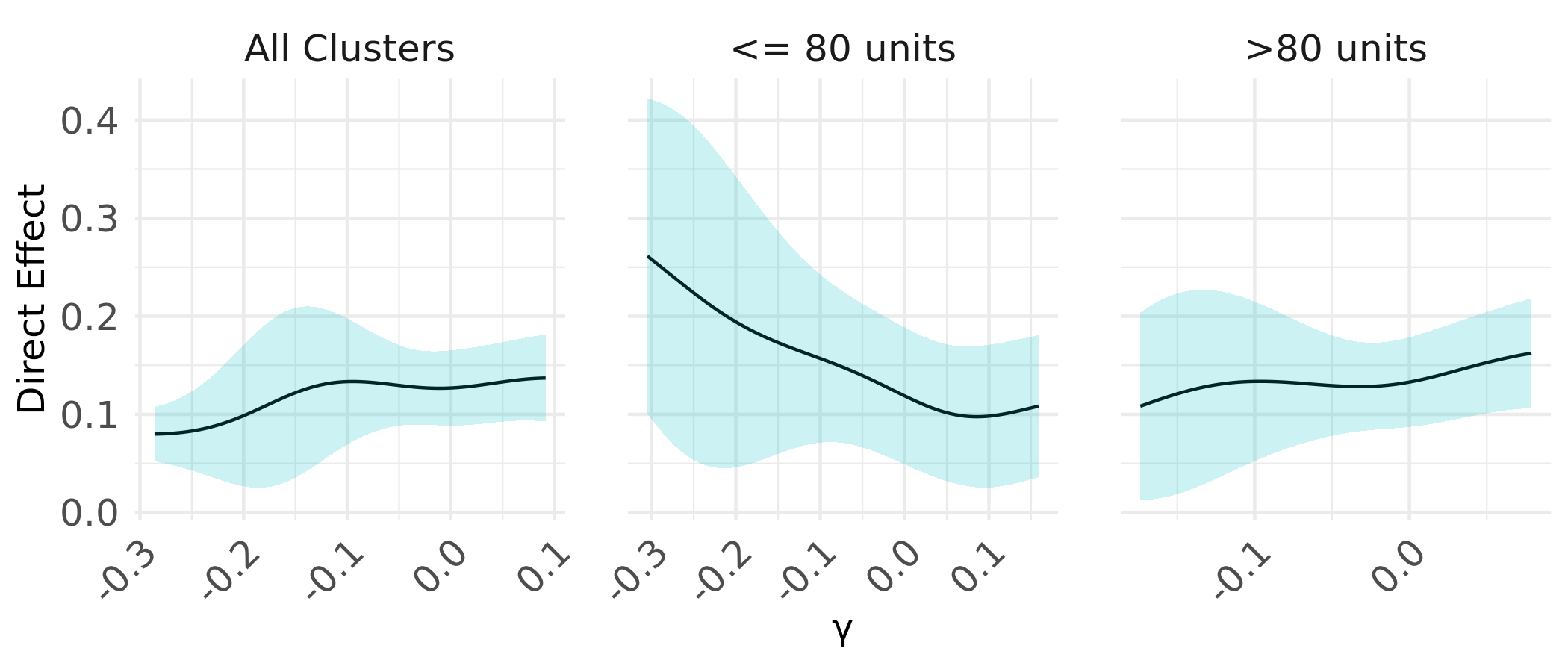}}
\caption{The direct effect when targeting treatment based on network degree. Larger values of $\gamma$ correspond to strategies that assign a larger probability of treatment to individuals with a larger network degree. Results are stratified by cluster size.}
\label{fig:cai_dgr_de}
\end{figure}

\begin{figure}[H]
\centerline{\includegraphics[width=0.9\textwidth]{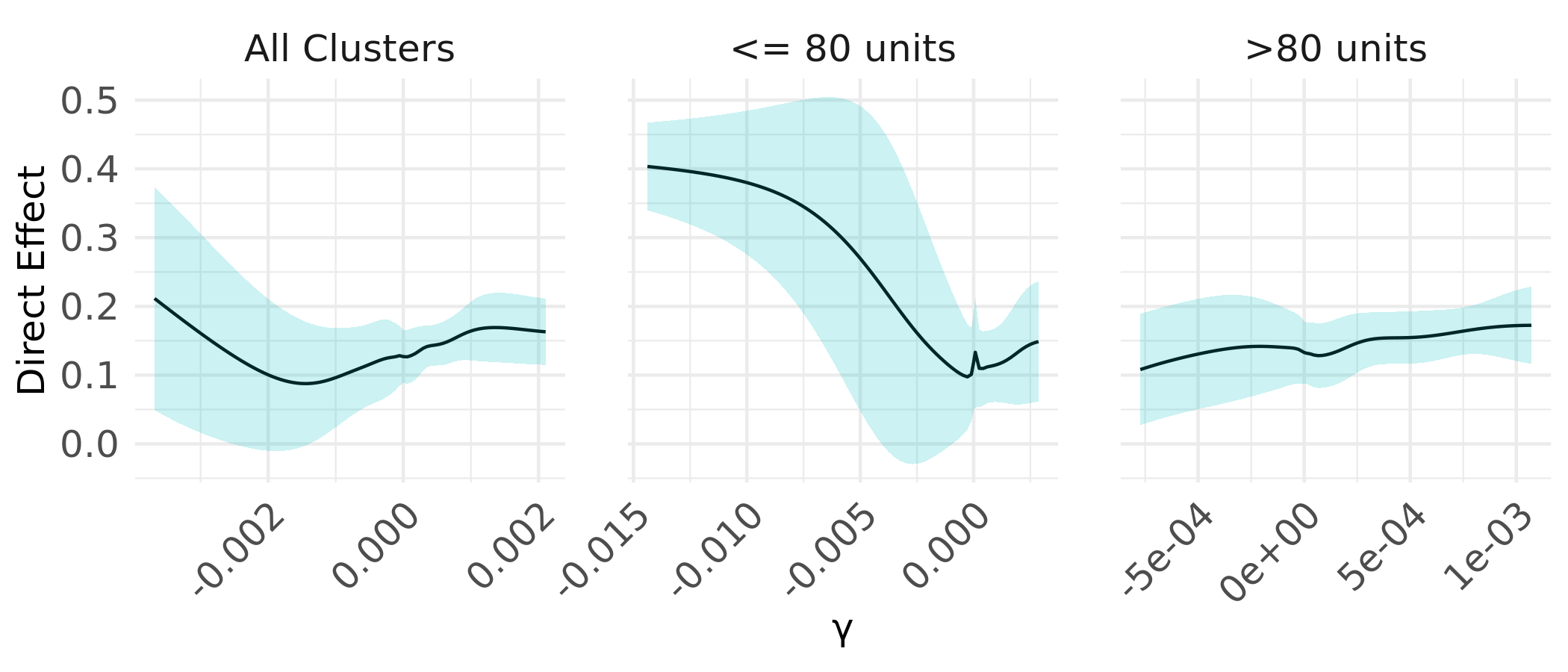}}
\caption{The direct effect when targeting treatment based on betweenness. Larger values of $\gamma$ correspond to strategies that assign a larger probability of treatment to individuals with a larger value of betweenness. Results are stratified by cluster size.}
\label{fig:cai_btwn_de}
\end{figure}

\begin{figure}[t]
\centerline{\includegraphics[width=0.9\textwidth]{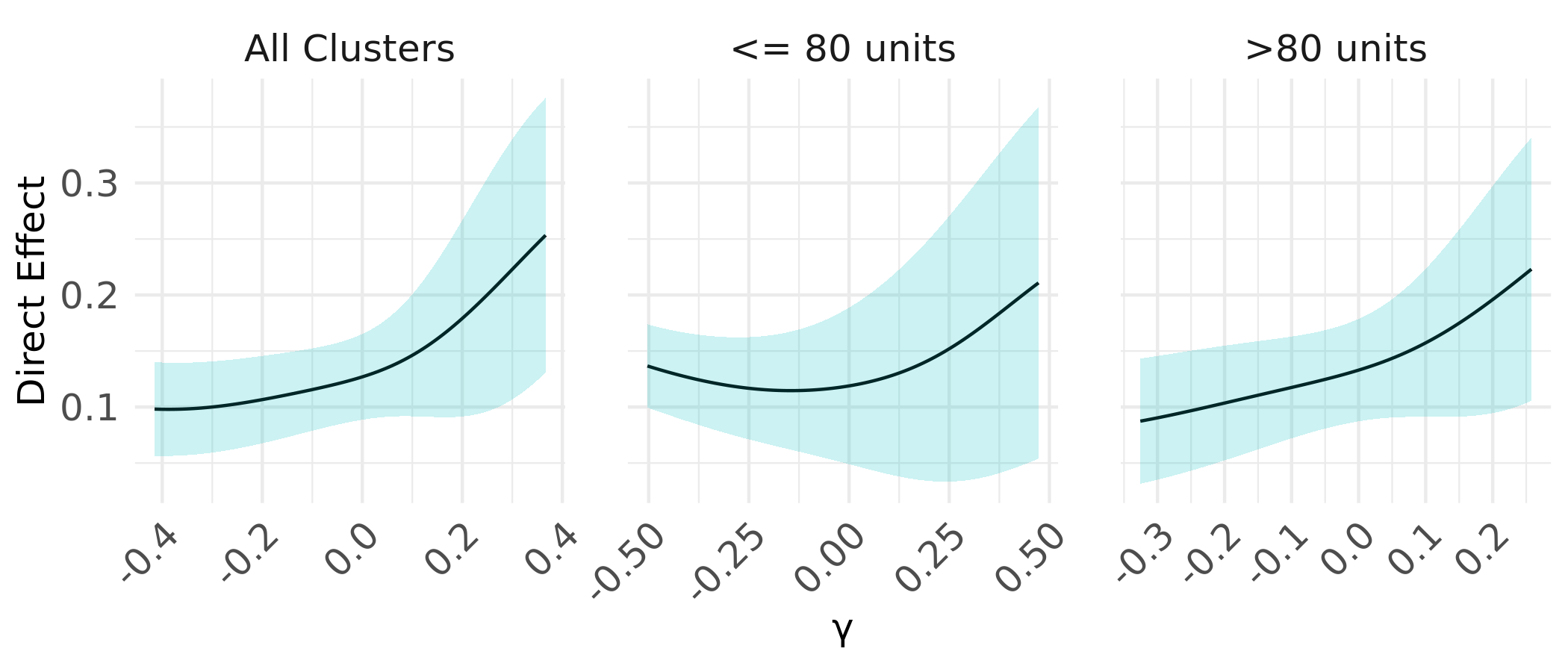}}
\caption{The direct effect when targeting treatment based on the perceived probability of a future disaster. Larger values of $\gamma$ correspond to strategies that assign a larger probability of treatment to individuals with a larger perceived probability of a future disaster. Results are stratified by cluster size.}
\label{fig:cai_disaster_de}
\end{figure}

\begin{figure}[H]
\centerline{\includegraphics[width=0.9\textwidth]{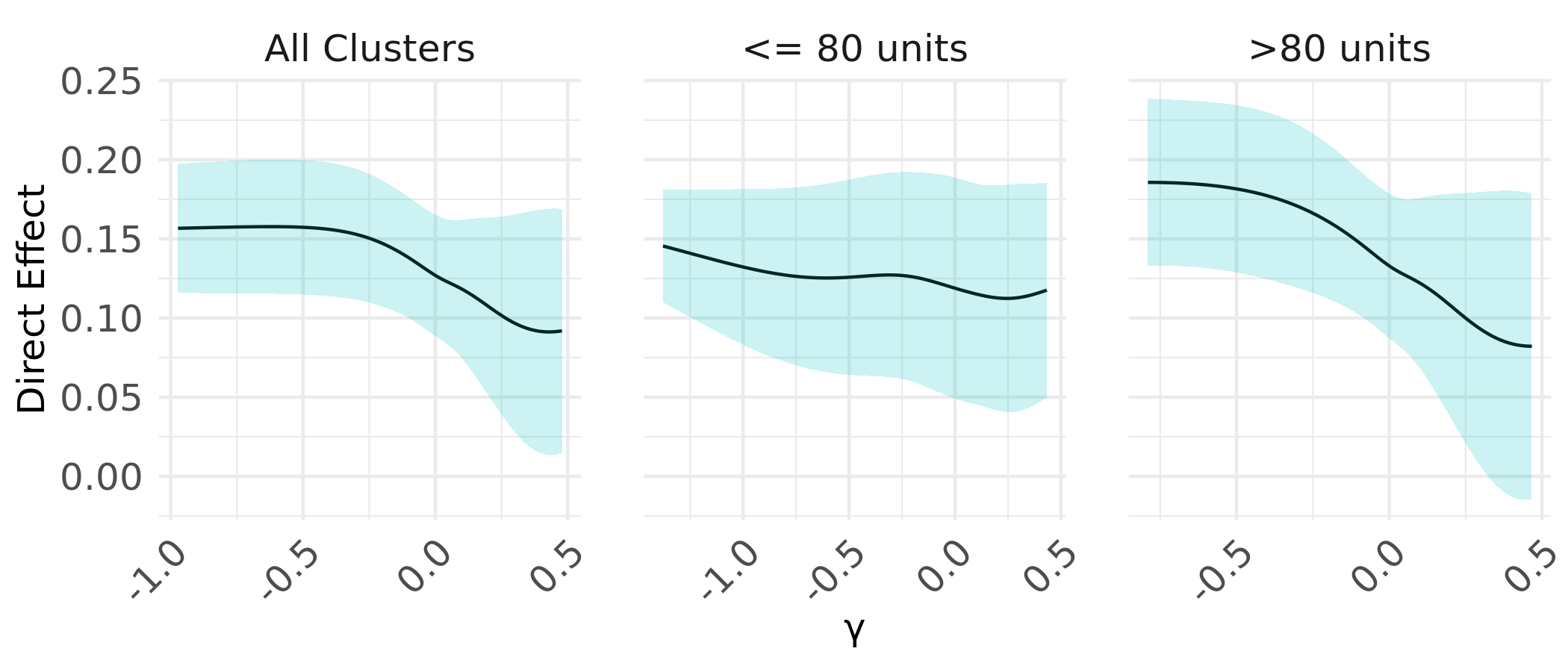}}
\caption{The direct effect when targeting treatment based on the area of rice cultivation. Larger values of $\gamma$ correspond to strategies that assign a larger probability of treatment to individuals with a larger area of rice cultivation. Results are stratified by cluster size.}
\label{fig:cai_rice_de}
\end{figure}

\begin{figure}[t]
\centerline{\includegraphics[width=0.9\textwidth]{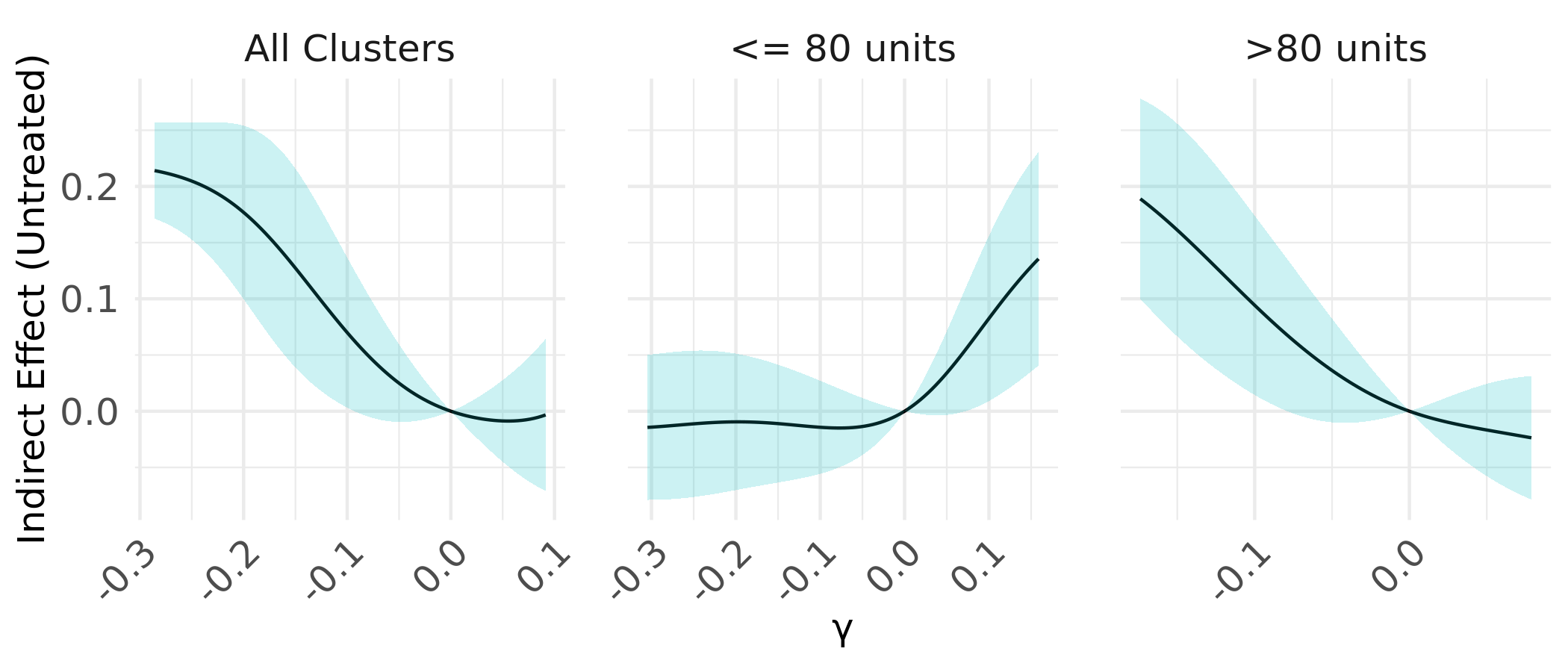}}
\caption{The indirect effect on untreated individuals when targeting treatment based on network degree. Larger values of $\gamma$ correspond to strategies that assign a larger probability of treatment to individuals with a larger network degree. Results are stratified by cluster size.}
\label{fig:cai_dgr_ie0}
\end{figure}

\begin{figure}[b!]
\centerline{\includegraphics[width=0.9\textwidth]{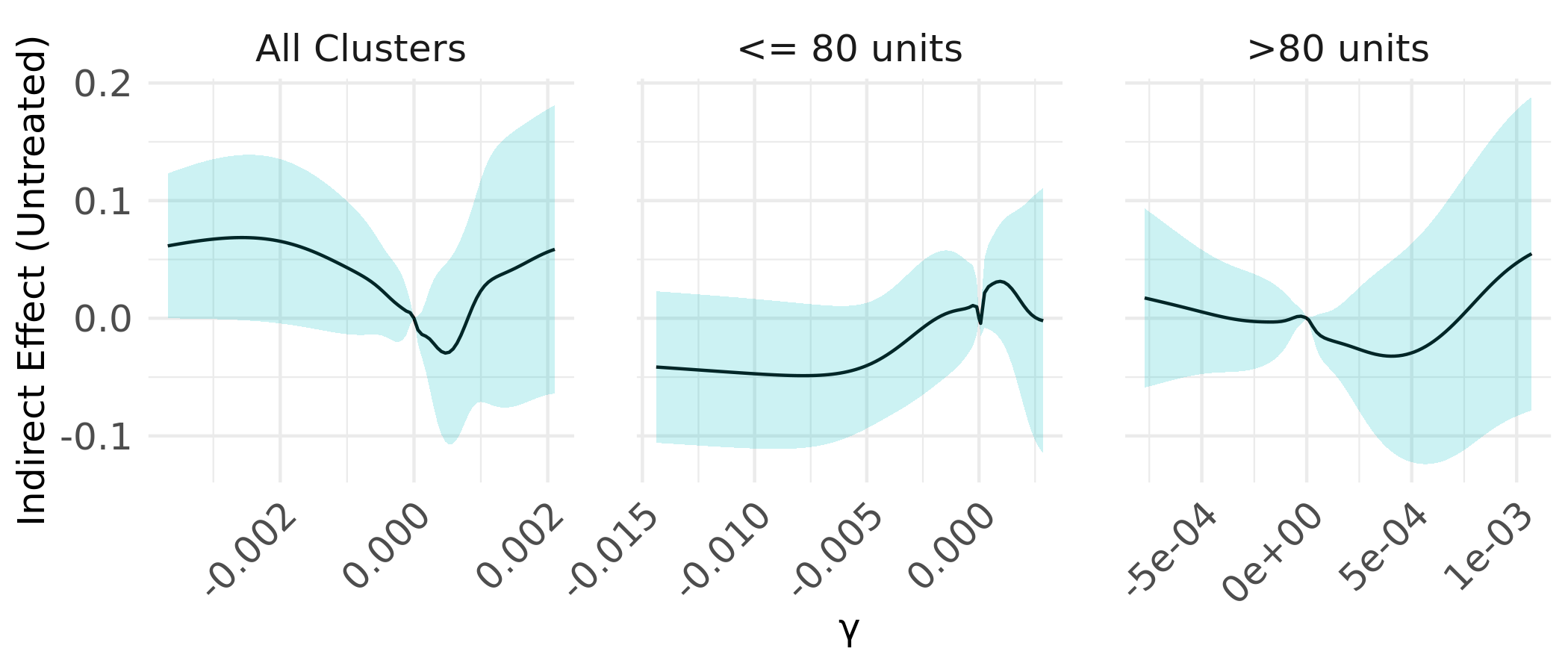}}
\caption{The indirect effect on untreated individuals when targeting treatment based on betweenness. Larger values of $\gamma$ correspond to strategies that assign a larger probability of treatment to individuals with a larger value of betweenness. Results are stratified by cluster size.}
\label{fig:cai_btwn_ie0}
\end{figure}

\begin{figure}[H]
\centerline{\includegraphics[width=0.9\textwidth]{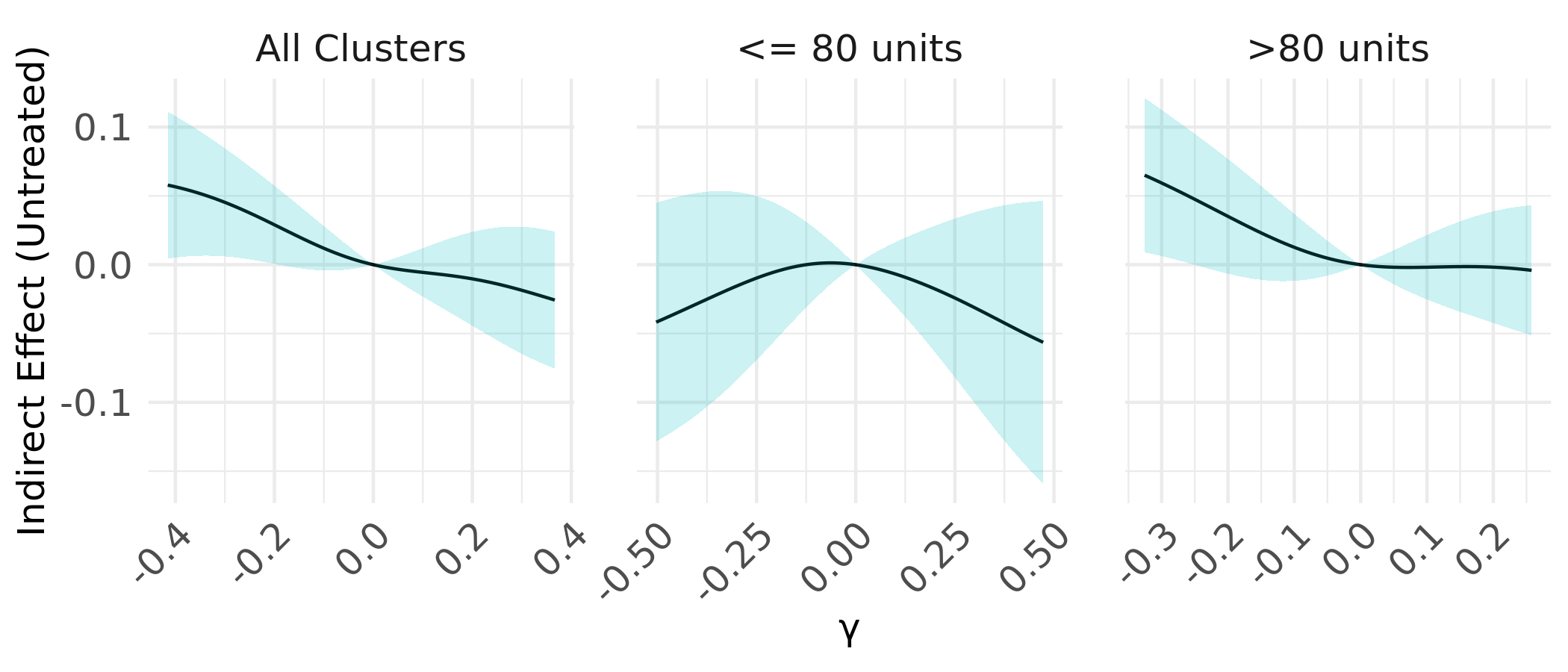}}
\caption{The indirect effect on untreated individuals when targeting treatment based on the perceived probability of a future disaster. Larger values of $\gamma$ correspond to strategies that assign a larger probability of treatment to individuals with a larger perceived probability of a future disaster. Results are stratified by cluster size.}
\label{fig:cai_disaster_ie0}
\end{figure}

\begin{figure}[H]
\centerline{\includegraphics[width=0.9\textwidth]{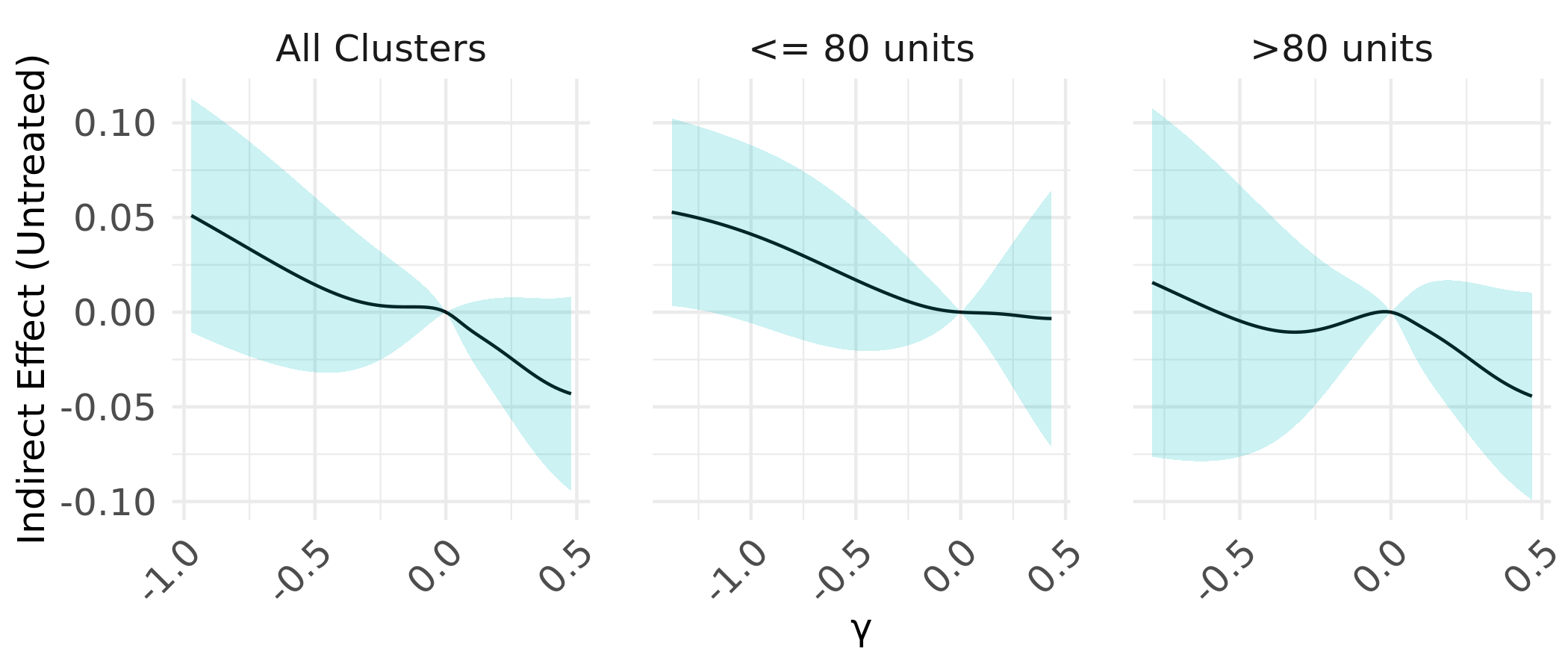}}
\caption{The indirect effect on untreated individuals when targeting treatment based on the area of rice cultivation. Larger values of $\gamma$ correspond to strategies that assign a larger probability of treatment to individuals with a larger area of rice cultivation. Results are stratified by cluster size.}
\label{fig:cai_rice_ie0}
\end{figure}

\begin{figure}[H]
\centerline{\includegraphics[width=0.9\textwidth]{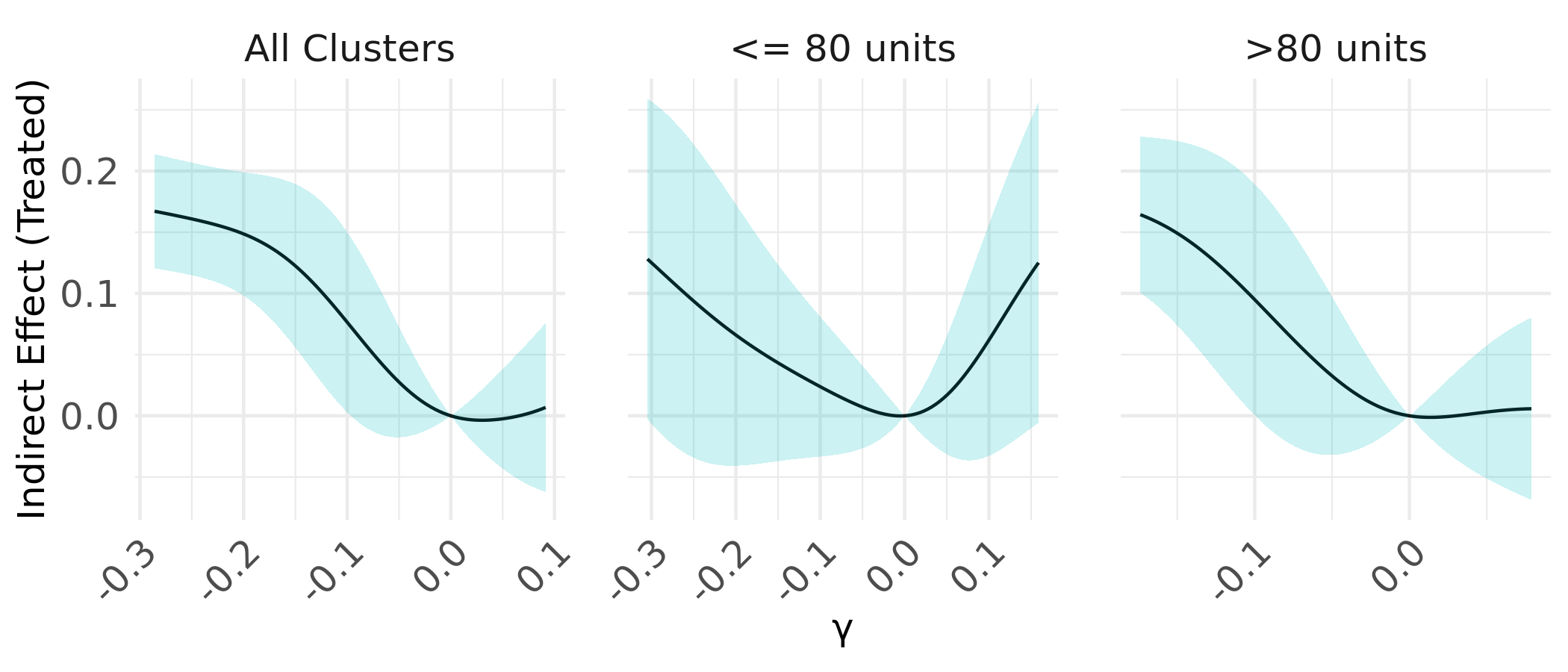}}
\caption{The indirect effect on treated individuals when targeting treatment based on network degree. Larger values of $\gamma$ correspond to strategies that assign a larger probability of treatment to individuals with a larger network degree. Results are stratified by cluster size.}
\label{fig:cai_dgr_ie1}
\end{figure}

\begin{figure}[H]
\centerline{\includegraphics[width=0.9\textwidth]{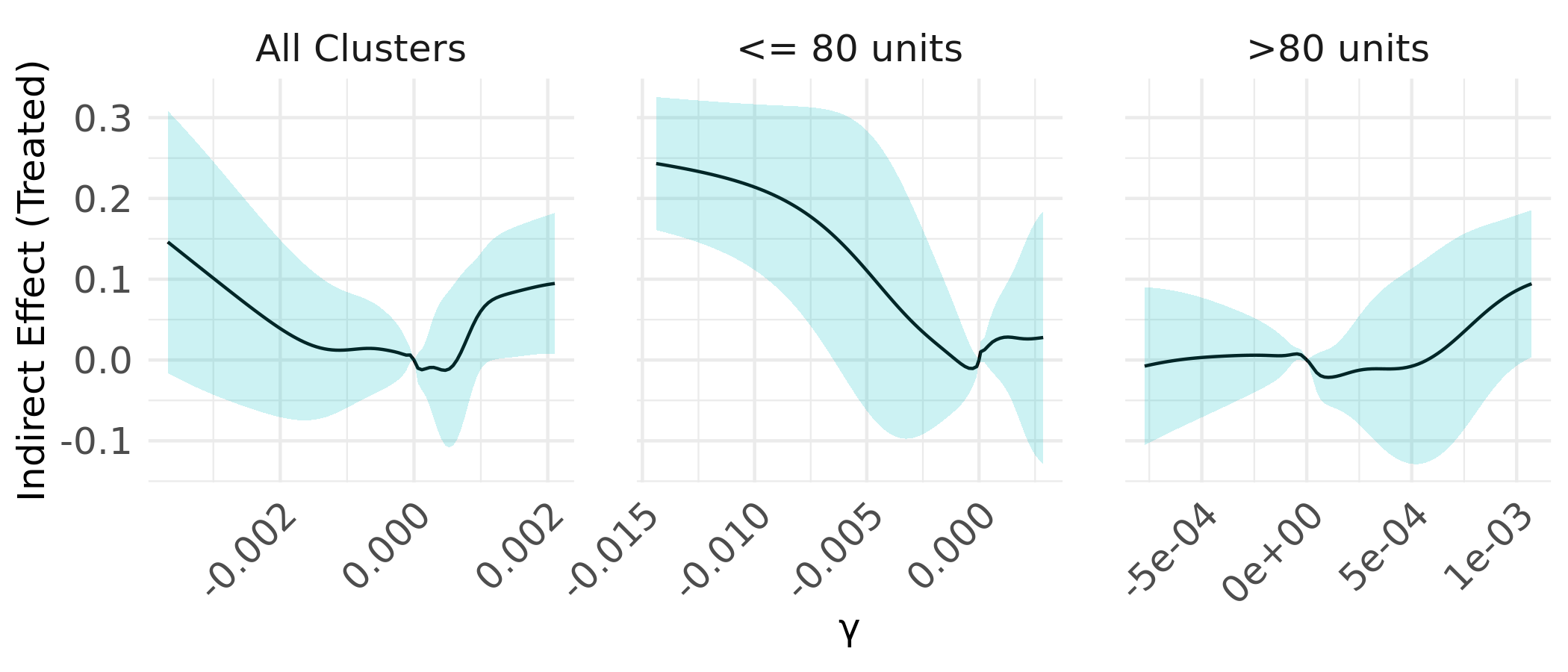}}
\caption{The indirect effect on treated individuals when targeting treatment based on betweenness. Larger values of $\gamma$ correspond to strategies that assign a larger probability of treatment to individuals with a larger value of betweenness. Results are stratified by cluster size.}
\label{fig:cai_btwn_ie1}
\end{figure}

\begin{figure}[H]
\centerline{\includegraphics[width=0.9\textwidth]{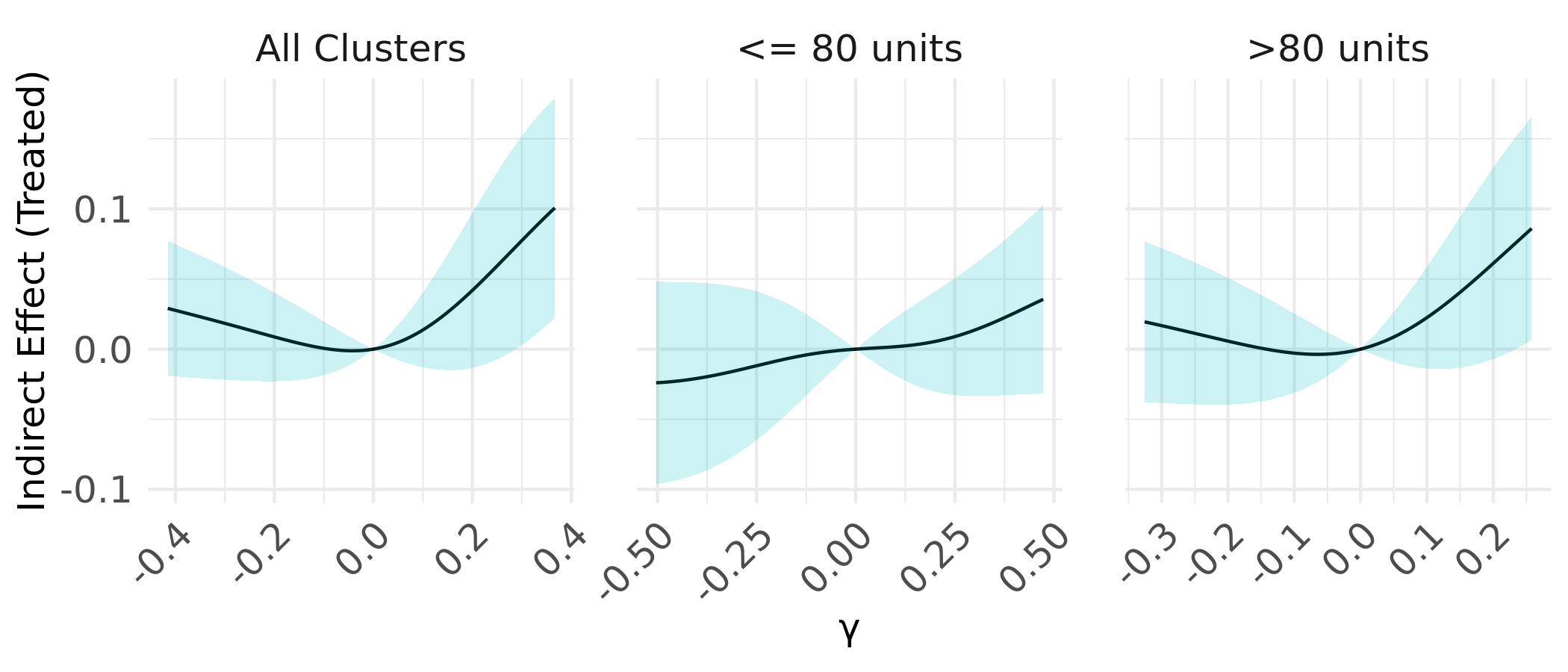}}
\caption{The indirect effect on treated individuals when targeting treatment based on the perceived probability of a future disaster. Larger values of $\gamma$ correspond to strategies that assign a larger probability of treatment to individuals with a larger perceived probability of a future disaster. Results are stratified by cluster size.}
\label{fig:cai_disaster_ie1}
\end{figure}

\begin{figure}[H]
\centerline{\includegraphics[width=0.9\textwidth]{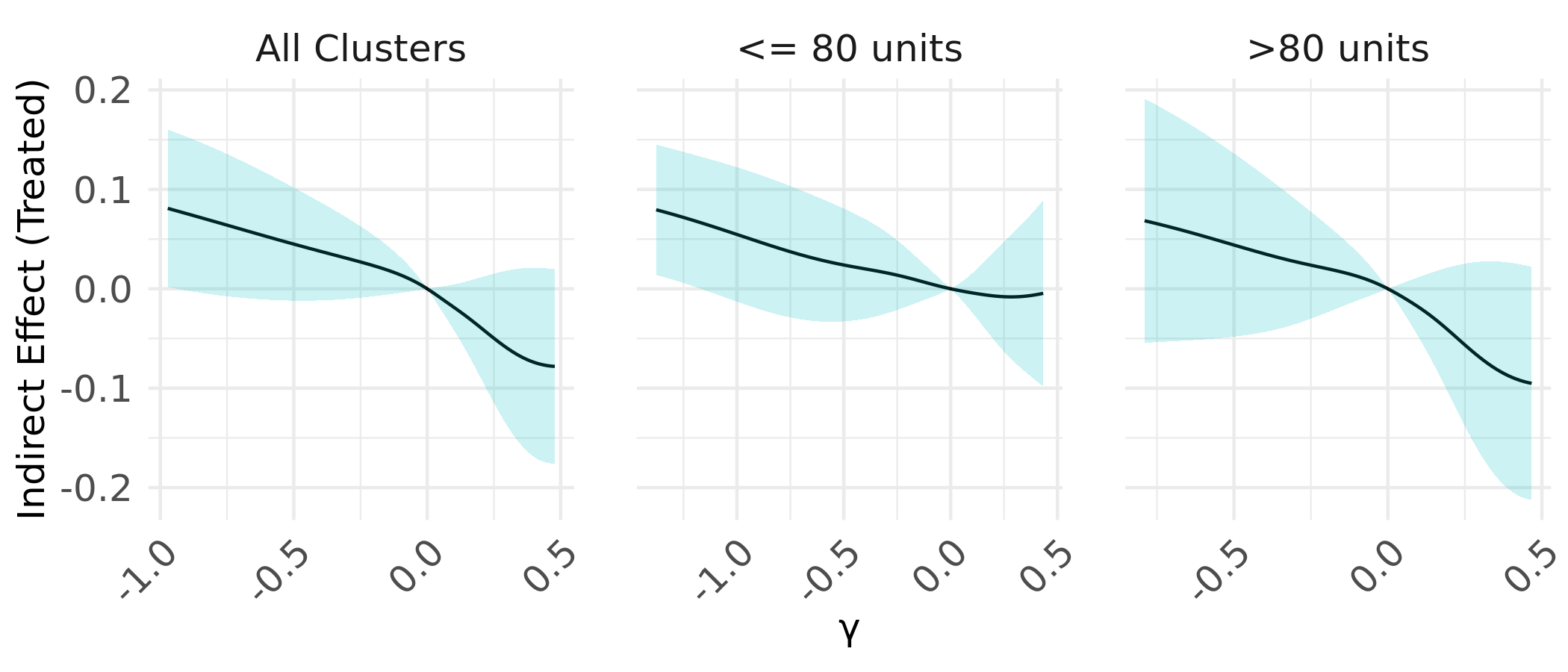}}
\caption{The indirect effect on treated individuals when targeting treatment based on the area of rice cultivation. Larger values of $\gamma$ correspond to strategies that assign a larger probability of treatment to individuals with a larger area of rice cultivation. Results are stratified by cluster size.}
\label{fig:cai_rice_ie1}
\end{figure}

This appendix includes additional analyses on the weather insurance dataset described in the main manuscript. We present additional information on the distribution of network characteristics in the application dataset (\cref{fig:betweenness}, \cref{fig:degree}). We then include figures for the direct effect (\cref{fig:cai_dgr_de}, \cref{fig:cai_btwn_de}, \cref{fig:cai_rice_de}, \cref{fig:cai_disaster_de}) and indirect effects (\cref{fig:cai_dgr_ie0}, \cref{fig:cai_btwn_ie0}, \cref{fig:cai_rice_ie0}, \cref{fig:cai_disaster_ie0},
\cref{fig:cai_dgr_ie1}, \cref{fig:cai_btwn_ie1}, \cref{fig:cai_rice_ie1}, \cref{fig:cai_disaster_ie1}) of interventions targeting individuals dependent on degree, betweenness, area of rice cultivated, and perceived probability of future disaster. 

%%%% DIRECT EFFECT FIGURES %%%%%%%%%%%%

Recall, the direct effect contrasts average individual outcomes when the individual is treated versus untreated, and the rest of the cluster has a fixed treatment assignment mechanism. We looked at how changing the cluster's treatment mechanism, assigned through $\bm \gamma$, changes the direct effect. The direct effect may depend on the treatment allocation strategy defined by $\bm \gamma$ when the effect of receiving the treatment depends on who else is treated. 
%Direct effect interacts with spillover effects in that if spillover effects are large, individual treatment effects may be less important. 

When treatment propensities depend on network degree, there is no significant change in DE across different treatment strategies in either pooled or stratified analyses (p = 0.306 for all clusters, 0.09 for clusters with $\le80$ units, and 0.423 for clusters with $>80$ units) (\cref{fig:cai_dgr_de}).

When treatment propensities depend on network betweenness, there is a significant change in DE across different treatment strategies in smaller clusters (p = 0.211 for all clusters, 0.009 for clusters with $\le80$ units, and 0.406 for clusters with $>80$ units) (\cref{fig:cai_btwn_de}). In clusters of $\le80$ units, there is a larger DE when, in the rest of the cluster, individuals with low betweenness have a higher treatment propensity. Correspondingly, the DE is lower when individuals with a high betweenness have a higher treatment propensity. A possible explanation is that an individual is more impacted by spillover effects when high-betweenness individuals are treated, attenuating the direct effect of their own treatment. 

When treatment propensities depend on the perceived probability of a future disaster, there is a significant change in DE across different treatment strategies in all clusters and larger clusters (p = 0.018 for all clusters, 0.247 for clusters with $\le80$ units, and 0.036 for clusters with $>80$ units)(\cref{fig:cai_disaster_de}). Consistent across cluster sizes, when the rest of the cluster assigns a higher treatment propensity to individuals with a large perceived probability of future disaster, the direct effect is larger. This suggests that an individual is more affected by spillover when their treated neighbors are individuals with a low perceived probability of future disaster, and again a higher spillover may attenuate the direct effect of individual treatment.

When treatment propensities depend on the area of rice cultivated, there are no significant changes in DE across different treatment strategies in pooled or stratified analyses(p = 0.173 for all clusters, 0.614 for clusters with $\le80$ units, and 0.069 for clusters with $>80$ units)(\cref{fig:cai_rice_de}).

%%%%%% INDIRECT EFFECT ON UNTREATED FIGURES %%%%%%%%%%%%%%

Let us now turn to the indirect effects. In this context, an indirect effect on untreated, i.e., $IE(0, \alpha, \bm \gamma, \bm 0)$, is the change in average outcome for an untreated unit when the surrounding units have a covariate dependent treatment propensity compared to a covariate agnostic treatment propensity. Then, the indirect effect would be non-zero if a unit's outcome depend on who else is treated among its neighbors.

When treatment propensities depend on network degree, there is a significant change in $IE(0, \alpha, \bm \gamma, \bm 0)$ both across all clusters and when stratified into larger and smaller clusters (p $<0.001$ for all clusters, p=0.028 for clusters with $\le80$ units, and p$<0.001$ for clusters with $>80$ units)(\cref{fig:cai_dgr_ie0}). Pooled across all clusters and in larger clusters, interventions assigning treatment with higher probabilities to high degree individuals appear to decrease the average outcome among the untreated. In smaller clusters, the same strategies appear to increase their outcome. A possible explanation for this discrepancy is the differences in network structure of small versus large clusters, as discussed in the main text for similar OE results. 

When treatment propensities depend on network betweenness, there is no statistically significant change in $IE(0, \alpha, \bm \gamma, \bm 0)$ across strategies (p = 0.368 for all clusters, p=0.379 for clusters with $\le80$ units, and p=0.446 for clusters with $>80$ units) (\cref{fig:cai_btwn_ie0}).

When treatment propensities depend on perceived probability of future disaster, there is a significant change in $IE(0, \alpha, \bm \gamma, \bm 0)$ only for analyses pooled across all clusters (p = 0.047 for all clusters, p=0.556 for clusters with $\le80$ units, and p=0.148 for clusters with $>80$ units) (\cref{fig:cai_disaster_ie0}). Pooled across all clusters, interventions assigning higher treatment propensity to individuals with a larger perceived probability of future disaster have a smaller $IE(0, \alpha, \bm \gamma, \bm 0)$.

When treatment propensities depend on the area of rice cultivated, there is no statistically significant change in $IE(0, \alpha, \bm \gamma, \bm 0)$ across strategies (p = 0.0058 for all clusters, p=0.336 for clusters with $\le80$ units, and p=0.383 for clusters with $>80$ units) (\cref{fig:cai_btwn_de})

%%%%%% INDIRECT EFFECT ON TREATED FIGURES %%%%%%%%%%%%%%%%

The indirect effect on treated, $IE(1, \alpha, \bm \gamma, \bm 0)$, is the change in average outcome for a treated unit when the surrounding units have a covariate dependent treatment propensity compared to a covariate agnostic treatment propensity. 

When treatment propensities depend on network degree, there is a significant change in $IE(1, \alpha, \bm \gamma, \bm 0)$ pooled across all clusters and in larger clusters (p = 0.004 for all clusters, p=0.265 for clusters with $\le80$ units, and p=0.039 for clusters with $>80$ units) (\cref{fig:cai_dgr_ie1}). In the significant strata, strategies that assign a higher treatment propensity to individuals with a large network degree have a smaller $IE(1, \alpha, \bm \gamma, \bm 0)$, that is, the average outcome for the treated units decrease. 

When treatment propensities depend on network betweenness, there is no significant change in $IE(1, \alpha, \bm \gamma, \bm 0)$ across strata (p = 0.178 for all clusters, p=0.065 for clusters with $\le80$ units, and p=0.242 for clusters with $>80$ units) (\cref{fig:cai_btwn_ie1}).

When treatment propensities depend on perceived probability of future disaster, there is no significant change in $IE(1, \alpha, \bm \gamma, \bm 0)$ across strata (p = 0.061 for all clusters, p=0.384 for clusters with $\le80$ units, and p=0.142 for clusters with $>80$ units) (\cref{fig:cai_disaster_ie1})

Finally, when treatment propensities depend on area of rice cultivation, there is a significant change in $IE(1, \alpha, \bm \gamma, \bm 0)$ pooled across all clusters (p = 0.038 for all clusters, p=0.239 for clusters with $\le80$ units, and p=0.109 for clusters with $>80$ units) (\cref{fig:cai_rice_ie1}). In the pooled analyses, strategies that assign a higher treatment propensity to individuals with a smaller area of rice cultivation yield a higher $IE(1, \alpha, \bm \gamma, \bm 0)$, that is, a higher outcome among the treated.


\newpage
\subsection*{Web appendix G - calculating true causal effects for Scenario 1}
In order to calculate the bias and coverage of the proposed estimators, we used simulations to calculate the true causal effects under each set of parameters $\boldsymbol{\beta}$ and for each set of treatment allocation parameters $\gamma$, $\alpha$ we consider 
%. We simulated these quantities through the following 
The simulation procedure is as follows: 
\begin{itemize}
\item Simulate a dataset with 200 clusters 
    \item For each cluster generate $\mathbf{X}^{(1)}$ and $\mathbf{X}^{(2)}$
    \item For each cluster, generate a vector of counterfactual treatment probabilities using $\gamma$, $\alpha$, $\mathbf{X}^{(1)}$, and $\mathbf{X}^{(2)}$
    \item Using the vector of treatment probabilities, draw a new treatment vector, $\mathbf{A}$ 
    \item Using the observed covariates and the newly generated treatment vectors, calculate $(T_{ij}, U_{ij}, V_{ij})$ for each unit.
    \item Use the linear model of Scenario 1 with fixed $\boldsymbol{\beta}$ to simulate potential outcome, ${Y}_{ij}(\mathbf{A}_i)$, using the observed covariates, the random treatment vector, and the calculated $(T_{ij}, U_{ij}, V_{ij})$
    \item Calculate causal effects by contrasting appropriate of potential outcomes
\end{itemize}
We repeat this process 2000 times for each set of parameters, and then take the mean of the calculated effects as the true value.

\subsection*{Web appendix H - calculating true causal effects for Scenario 2}
To calculate the "true" causal effects for the diffusion simulations we use the following procedure:
\begin{itemize}
    \item Simulate a dataset with 200 5-unit star network clusters 
    \item For each cluster fix $\mathbf{X}^{(1)} = 1$ for the central node, and otherwise $\mathbf{X}^{(1)} = 0$
    \item Simulate $\mathbf{X}^{(2)}$ either randomly or correlated with $\mathbf{X}^{(1)}$, depending on concordance parameter
  %  \item For each cluster, draw 200 possible $\mathbf{X}^{(2)}$ vectors
    \item 
    %For each of the 2000*200 possible clusters, 
    Use $\alpha$, $\gamma$, and covariates $\mathbf{X}^{(1)}$ and $\mathbf{X}^{(1)}$ to calculate a predicted probability of treatment for each individual
    \item Draw a random vector of treatments, $\mathbf{A}$, from the probabilities in the prior step
    \item Simulate the  diffusion process along each network edge using the parameter for the probability of diffusion occurring
    \item Set outcome $Y_{ij} = 1$ for individuals where $A_{ij} = 1$ or diffusion occurred from at least one edge connected to the unit
    \item Calculate average outcomes and contrasts to calculate true causal effects
 \end{itemize}
 We repeat this process 2000 times for each set of parameters, and then take the mean of the calculated effects as the true value.

\subsection*{Web appendix I - statistical test}
In the main manuscript, we described in details the statistical test for overall effects. For detecting differences in overall effect, we used the test statistic $T = \max_{\bm \gamma_1, \bm \gamma_2 \in \bm \Gamma(\mathcal{K}_P)} \{OE(\alpha, \bm \gamma_1, \bm 0) - OE(\alpha, \bm \gamma_2, \bm 0)\}$. When detecting differences in direct effect, we use the test statistic $T = \max_{\bm \gamma_1, \bm \gamma_2 \in \bm \Gamma(\mathcal{K}_P)} \{DE(\alpha, \bm \gamma_1) - DE(\alpha, \bm \gamma_2)\}$. Similarly, when detecting differences in indirect effect, we use the test statistic  $T = \max_{\bm \gamma_1, \bm \gamma_2 \in \bm \Gamma(\mathcal{K}_P)} \{IE(\alpha, \bm \gamma_1, \bm 0, a) - IE(\alpha, \bm \gamma_2, \bm 0, a)\}$ where a = 0 for indirect effect on the untreated and a = 1 for indirect effect on the treated. 

\subsubsection*{Simulation study for the evaluation of the statistical test}
We assessed the power of the proposed statistical test by applying the test to the simulated scenarios with a linear outcome model as in Scenario 1. This includes settings with no interference, homogeneous interference, heterogeneous interference through a single variable, and heterogeneous interference through two variables. We stratified results by $\beta_3$, $\beta_4$, and $\beta_5$ values and averaged across concordance and $\gamma$ values.  The parameter $\beta_3$ determines homogeneous interference, $\beta_4$ determines heterogeneous interference by $\mathbf{X}^{(1)}$, and $\beta_5$ determines heterogeneous interference by $\mathbf{X}^{(2)}$. For each set of parameters, we examined power and level when applying our statistical test to overall effect estimators (\cref{tab:powertable}), direct effect, and indirect effects (\cref{supp_table:test_DE_IE})

\begin{table}[t]
\centering
%For each set of parameters we tested in our simulation study, we calculated the power of the proposed statistical test applied to overall effect estimators. 
%The parameter $\beta_3$ determines homogeneous interference, $\beta_4$ determines heterogeneous interference by $\mathbf{X}^{(1)}$, and $\beta_5$ determines heterogeneous interference by $\mathbf{X}^{(2)}$. 
%The test statistic corresponds to the largest absolute difference in overall effect between all the considered interventions. A test statistic of 0.42 can be detected with $\ge$ 75\% power, and a test statistic of 0.67 has nearly 100\% power.
%\textbf{Power of Novel Statistical Test for Heterogeneous Interference}

\begin{tabular}{|c|c|c|c|c|c|c|}
  \hline
$\beta_3$ & $\beta_4$ & $\beta_5$ & Concordance & Test Statistic & OE Level & OE Power \\ 
  \hline
0 & 0 & 0 & 0 & 0.17 & 0.08 &   \\ 
  0 & 0 & 0 & 0.65 & 0.17 & 0.1 &   \\ 
  1 & 0 & 0 & 0 & 0.20 & 0.06 &   \\ 
  1 & 0 & 0 & 0.65 & 0.20 & 0.07 &   \\ 
  0 & 0 & 1 & 0 & 0.38 &   & 0.87 \\ 
  0 & 0 & 1 & 0.65 & 0.39 &   & 0.88 \\ 
  1 & 0 & 1 & 0 & 0.40 &   & 0.7 \\ 
  1 & 0 & 1 & 0.65 & 0.42 &   & 0.75 \\ 
  0 & 1 & 0 & 0 & 0.39 &   & 0.89 \\ 
  0 & 1 & 0 & 0.65 & 0.41 &   & 0.89 \\ 
  1 & 1 & 0 & 0 & 0.40 &   & 0.74 \\ 
  1 & 1 & 0 & 0.65 & 0.41 &   & 0.71 \\ 
  0 & 1 & 1 & 0 & 0.56 &   & 0.99 \\ 
  0 & 1 & 1 & 0.65 & 0.68 &   & 1 \\ 
  1 & 1 & 1 & 0 & 0.56 &   & 0.97 \\ 
  1 & 1 & 1 & 0.65 & 0.68 &   & 0.98 \\ 
  0 & 2 & 0 & 0 & 0.70 &   & 1 \\ 
  0 & 2 & 0 & 0.65 & 0.74 &   & 1 \\ 
  1 & 2 & 0 & 0 & 0.71 &   & 0.99 \\ 
  1 & 2 & 0 & 0.65 & 0.74 &   & 1 \\ 
  0 & 2 & 1 & 0 & 0.83 &   & 1 \\ 
  0 & 2 & 1 & 0.65 & 1.04 &   & 1 \\ 
  1 & 2 & 1 & 0 & 0.84 &   & 1 \\ 
  1 & 2 & 1 & 0.65 & 1.03 &   & 1 \\ 
   \hline
\end{tabular}

\caption{
Level and power of our proposed  statistical test for overall effects.}
\label{tab:powertable}
\end{table} 
In \cref{tab:powertable} for overall effects, the first 4 rows correspond to scenarios where the null hypothesis is true, and thus, we report the level of the test. For all other scenarios, where the null hypothesis is false, we report the power of the test.
To detect differences in overall effect across treatment allocation strategies, the statistical test had a level of 6-10\% percent and a power of $\ge$75\% for differences in overall effect of at least as large as 0.42 and nearly 100\% for differences at least as large as 0.67. 

We also examined power and level for indirect and direct effect estimators.  For direct effect, we only calculate level because there were no scenarios where we expected there to be heterogeneity in DE. Results are shown in \cref{supp_table:test_DE_IE}. We see that the level of the proposed test is close to nominal and that the power is above 90\% throughout.

% latex table generated in R 4.2.0 by xtable 1.8-4 package
% Sun Oct 27 12:32:12 2024
\begin{table}[t]
\centering
\begin{tabular}{|c|c|c|c|c|c|c|c|c|}
  \hline
$\beta_3$ & $\beta_4$ & $\beta_5$ & Concordance & DE Level & IE(0) Level & IE(0) Power & IE(1) Level & IE(1) Power \\ 
  \hline
0 & 0 & 0 & 0 & 0.07 & 0.07 &   & 0.09 &   \\ 
  0 & 0 & 0 & 0.65 & 0.10 & 0.07 &   & 0.06 &   \\ 
  1 & 0 & 0 & 0 & 0.06 & 0.05 &   & 0.06 &   \\ 
  0 & 0 & 1 & 0 & 0.05 &   & 0.93 &   & 0.93 \\ 
  0 & 0 & 1 & 0.65 & 0.07 &   & 0.94 &   & 0.94 \\ 
  1 & 0 & 1 & 0 & 0.07 &   & 0.93 &   & 0.92 \\ 
  0 & 1 & 0 & 0 & 0.09 &   & 0.94 &   & 0.95 \\ 
  0 & 1 & 0 & 0.65 & 0.08 &   & 0.95 &   & 0.94 \\ 
  0 & 1 & 1 & 0 & 0.05 &   & 0.99 &   & 1 \\ 
  0 & 1 & 1 & 0.65 & 0.08 &   & 1 &   & 1 \\ 
  0 & 2 & 0 & 0 & 0.08 &   & 1 &   & 1 \\ 
  0 & 2 & 0 & 0.65 & 0.07 &   & 1 &   & 1 \\ 
  0 & 2 & 1 & 0 & 0.09 &   & 1 &   & 1 \\ 
  0 & 2 & 1 & 0.65 & 0.08 &   & 1 &   & 1 \\ 
   \hline
\end{tabular}
\caption{
Level and power of our proposed statistical test to detect heterogeneity in direct effect (DE) and indirect effects (IE(0) and IE(1)).}
\label{supp_table:test_DE_IE}
\end{table}

\subsubsection*{Note on proposed inferential procedure}
Rather than a single statistical test that incorporates information from all estimated causal effects (e.g., OE's), we could generate a series of tests for each contrast. For instance, for the overall effect, to test the composite null hypothesis $H_0: OE(\alpha, \bm \gamma, \bm 0)=0$ for all $\bm \gamma \in \Gamma(\mathcal{K}_P)$, we would conduct a test for each $\bm \gamma \in \Gamma(\mathcal{K}_P)$ with the test statistic  $T = OE(\alpha, \bm \gamma, \bm 0)$. However, compared to our proposed test, this would be a less powerful approach, because we would need to adjust for multiple testing and we would not be able to take advantage of correlations between test statistics.

%\end{document}

\end{document}